\documentclass[trackchanges,twocolumn]{aastex701}

\usepackage{graphicx}	
\usepackage{amsmath}    

\def\subinrm#1{\sb{\mathrm{#1}}}
{\catcode`\_=13 \global\let_=\subinrm}
\mathcode`_="8000
\def\supinrm#1{\sp{\mathrm{#1}}}
{\catcode`\^=13 \global\let^=\supinrm}
\mathcode`^="8000
\def\upsubscripts{\catcode`\_=12 } 
\def\upsupscripts{\catcode`\^=12 }

\begin{document}
\title{How Distance Affects GRB Prompt Emission Measurements}

\author[orcid=0000-0002-1103-7082,sname='Moss',gname='Michael']{Michael J. Moss}
\altaffiliation{NASA Postdoctoral Fellow}
\affiliation{Astrophysics Science Division, NASA Goddard Space Flight Center, Greenbelt, MD 20771, USA}
\email[show]{mikejmoss3@gmail.com}  

\author[orcid=0000-0002-7851-9756]{Amy Y. Lien}
\affiliation{University of Tampa, Department of Chemistry, Biochemistry, and Physics, 401 W. Kennedy Blvd, Tampa, FL 33606, USA}
\email{alien@ut.edu}

\author[orcid=0000-0003-1673-970X]{S. Bradley Cenko}
\affiliation{Astrophysics Science Division, NASA Goddard Space Flight Center, Greenbelt, MD 20771, USA}
\affiliation{Joint Space-Science Institute, University of Maryland, College Park, MD 20742, USA}
\email{brad.cenko@nasa.gov}

\author[orcid=0000-0001-5780-8770]{Sylvain Guiriec}
\affiliation{Astrophysics Science Division, NASA Goddard Space Flight Center, Greenbelt, MD 20771, USA}
\affiliation{The Department of Physics, The George Washington University, 725 21st NW, Washington, DC 20052, USA}
\email{sguiriec@gwu.edu}

\author[orcid=0000-0001-9803-3879]{Craig B. Markwardt}
\affiliation{Astrophysics Science Division, NASA Goddard Space Flight Center, Greenbelt, MD 20771, USA}
\email{craig.b.markwardt@nasa.gov}

\upsubscripts
\upsupscripts

\begin{abstract}
We investigated how Gamma-Ray Burst (GRB) prompt emission measurements are affected by increasing distance to the source. We selected a sample of 26 bright GRBs with measured redshifts $z<1$ observed by the Burst Alert Telescope (BAT) on board the Neil Gehrels Swift Observatory (Swift) and simulated what BAT would have observed if the GRBs were at larger redshifts. We measured the durations of the simulated gamma-ray signals using a Bayesian block approach and calculated the enclosed fluences and peak fluxes. As expected, we found that almost all durations (fluences) measured for simulated high-$z$ GRBs were shorter (less) than their true durations (energies) due to low signal-to-noise ratio emission becoming completely dominated by background, i.e., the ``tip-of-the-iceberg'' effect. This effect strongly depends on the profile and intensity of the source light curve. Due to the uniqueness of GRB light curves, there is no common behavior in the evolution of measured durations with redshift. We compared our synthetic high-$z$ (i.e., $z>3$) GRBs to a sample of 72 observed high-$z$ bursts and found that the two samples were not inconsistent with being drawn from the same underlying population. We conclude that: (i) prompt emission durations (fluences) of high-$z$ GRBs observed by Swift/BAT are most likely underestimations, sometimes by factors of $\sim$several tens ($\sim2$), and (ii) changes in the average GRB prompt emission duration and fluence with increasing redshift are consistent with the tip-of-the-iceberg effect.
\end{abstract}


\section{Introduction}  

A Gamma-ray Burst (GRB) is characterized by a bright highly-variable flash of gamma-rays that can last for a fraction of a second or tens of seconds, known as prompt emission. This initial phase is followed by a period of decaying emission that can be observed from radio frequencies to TeV energies and that lasts for hours, days, and, occasionally, years, known as afterglow emission. These catastrophic events are the signatures of highly relativistic material being jetted away from a compact object \citep{1978MNRAS.183..359C,1992MNRAS.258P..41R} newly formed by either the core-collapse of a massive star \citep{1998Natur.395..670G,2003Natur.423..847H,2003ApJ...591L..17S} or the merger of a compact binary system \citep{2013Natur.500..547T,2017ApJ...848L..13A}. The prompt emission is often associated with particle acceleration and energy dissipation mechanisms occurring within the outflow \citep{2014IJMPD..2330002Z}, while afterglow emission is often adequately modeled as a decelerating relativistic blast wave continually sweeping up material from the medium surrounding the burst \citep{1976PhFl...19.1130B,1997MNRAS.288L..51W,1998ApJ...497L..17S,2000ApJ...537..191F}. 

\begin{figure}
	\centering
	\includegraphics[width=\linewidth]{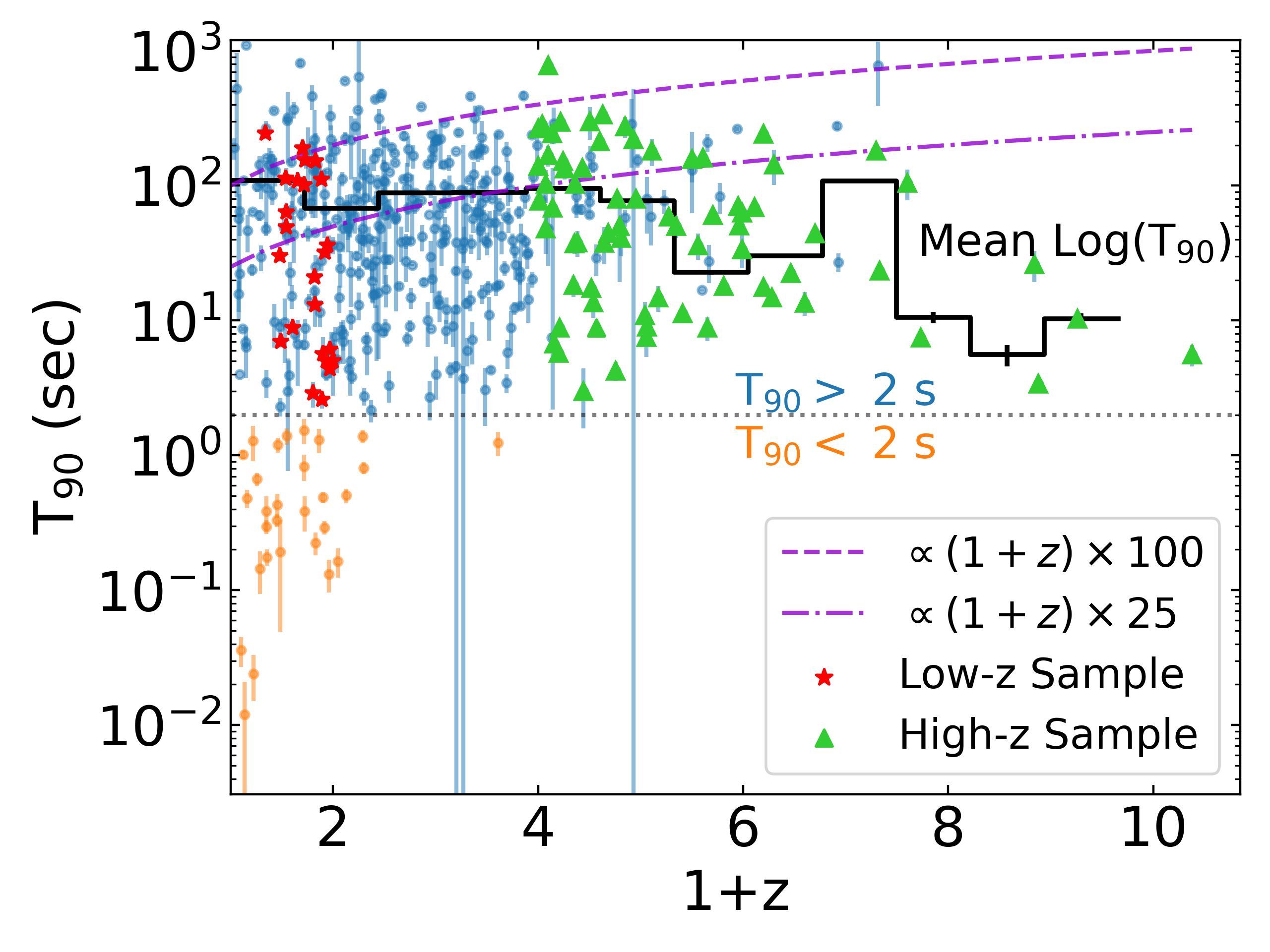}
	\caption{T$_{90}$ measurements of Swift/BAT GRBs as a function of their measured redshifts. The blue and orange points are GRBs with a measured prompt duration of T$_{90}>2$ s and T$_{90}<2$ s, i.e., long and short GRBs, respectively. The lower and upper violet dashed lines are $\propto(1+z) T$ for rest frame durations of T = 25 s and 100 s, respectively. The black line is the weighted mean of the log durations of LGRBs in each sequential redshift bin, i.e., $\mu = \sum x_i\sigma_i^{-2} / \sum \sigma_i^{-2}$, where $x_i = \log_{10}(\text{T}_{90, i})$. The red stars and green triangles indicate GRBs in the low-$z$ and high-$z$ samples of this work, respectively (see Tables \ref{tab: low-z} and \ref{tab: high-z}). \label{fig: grb-t90-vs-z}}
\end{figure}

Due to their incredible luminosities, GRBs can be detected across most of the observable universe. They have been detected from redshifts as low as $z = 0.0085$ \citep{1999A&AS..138..465G} and out to $z\sim9.4$ \citep{2009GCN..9281....1U,2011ApJ...736....7C}, making them excellent tools to study cosmological aspects of our universe, such as star formation rates \citep{1997ApJ...486L..71T,1998MNRAS.294L..13W,1998A&A...339L...1M,2001ApJ...548..522P} and the re-ionization phase of the early Universe \citep{1998ApJ...501...15M,2006PASJ...58..485T,2019MNRAS.483.5380T,2024arXiv240313126F}.

The duration of GRB prompt emission is typically defined using the T$_{90}$ metric, i.e., the time interval which encompasses 5$\%$ to 95$\%$ of the observed burst count fluence in the 50-300 keV energy band \citep{1993ApJ...413L.101K}. By defining the spectral hardness of a GRB as the ratio between the fluence in the 100-300 keV and 50-100 keV energy bands, there is a bimodality in the observed GRB population separated into short-hard GRBs ( SGRBs; i.e., T$_{90}<2$ s and a higher hardness ratio) and long-soft GRBs (LGRBs; i.e., T$_{90}>2$ s and a lower hardness ratio). Exactly where the separation between the two populations occurs is unclear and has been found to depend on which instrument was used to perform the measurements \citep{2012ApJ...749..110B}.

The coincidental detections of LGRBs with core-collapse supernova \citep{1998Natur.395..670G,2003Natur.423..847H} and SGRBs with binary-merger events \citep{2013Natur.500..547T,2017ApJ...848L..13A} have provided a naive explanation for the observed bimodality. However, there is a significant overlap between the durations and hardness ratios of the two populations and several GRBs have called into question the simple association between a GRB's duration and spectral properties with its progenitor system. For example, GRB 200826A has a T$_{90} \sim 1.14$ s, placing it into the SGRB population, but was later found to have a supernova counterpart \citep{2021NatAs...5..917A,2022ApJ...932....1R}, linking it to a collapsar progenitor. Conversely, GRBs 211211A and 230307A were both clearly LGRBs (i.e., T$_{90} \sim$ 122 s and 35 s, respectively), but both were associated with counterparts consistent with kilonovae, which are signatures of compact binary mergers \citep{2022Natur.612..223R,2022Natur.612..228T,2024Natur.626..737L}. 

Naively, the average measured duration of LGRBs should increase with redshift simply due to time dilation, i.e., $\overline{\text{T}}_{90} \propto (1+z)$. However, the average duration of LGRBs seems to remain constant out to high redshifts ($z\sim7$), at which point the average duration begins to sharply decrease (see Fig. \ref{fig: grb-t90-vs-z}). 

Low SNR observations lead to underestimations of GRB durations and fluences due purely to instrumental biases; this is known as the ``tip-of-the-iceberg'' effect \citep{1997ApJ...490...79B,2013ApJ...765..116K,2013MNRAS.436.3640L,2022ApJ...927..157M,2024arXiv240303266L}. Low SNR observations can be caused by either the observing conditions during an observation (e.g., background level and incident angle of the source relative to the detector plane; \citealt{2022ApJ...927..157M,2024arXiv240303266L}) or the physical conditions of the source itself (e.g., intrinsic brightness and distance; \citealt{1997ApJ...490...79B,2013ApJ...765..116K,2013MNRAS.436.3640L,2024arXiv240303266L}). Furthermore, the duration of GRB prompt emission is energy dependent. Higher energy emission is typically shorter than the emission at lower energies \citep{1995ApJ...448L.101F}. As the distance to a GRB increases, the rest-frame spectra is redshifted such that higher energy parts of the intrinsic GRB spectrum will fall in the instrument energy band, meaning we should expect the duration of high-redshift GRBs to be shorter.

\citet{1997ApJ...490...79B} compared the duration measurements of bright GRBs detected by the \textit{Burst And Transient Source Experiment} on board the Compton Gamma-Ray Observatory (CGRO/BATSE) and found that a brightness bias exists for GRB duration measurements. This bias was sufficient to obscure a time-dilation factor of order $\sim$2. \citet{2013ApJ...765..116K} simulated a set of synthetic Fast-Rise Exponential-Decay (FRED) light curves each with different brightnesses at increasing distances, folded them through the instrument response matrix of CGRO/BATSE, added background signal, and then measured their durations. The authors found that at lower redshifts, the burst durations evolved as $\propto(1+z)$, however, at higher redshifts the burst durations stopped following this trend and began to decrease due to decreasing signal-to-noise ratios (SNRs). \citet{2013MNRAS.436.3640L} took a sample of low-redshift bursts observed by Swift/BAT and simulated the burst light curves at higher redshifts with the purpose of comparing the durations measured for the simulated bursts to the durations measured for the farthest GRBs observed by Swift/BAT. The authors found that their simulated light curves had average duration measurements higher than the average duration of the observed high-redshift bursts by a factor $\gtrsim3$ and that the duration measurement distribution between the two samples only had a $\sim1\%$ probability of being consistent with being drawn from the same underlying population.

In this work, we further investigated the tip-of-the-iceberg effect to understand how duration and fluence estimates for GRB prompt emission observed by Swift/BAT evolve with redshift, expanding upon previous studies by taking into account proper distance corrections and folding observed spectra with the instrument response of Swift/BAT to properly account for signal loss due to instrument sensitivity. In Section \ref{sec: Methods}, we outline the procedures and methods we used to simulate Swift/BAT observations of GRBs. In Section \ref{sec: Results}, we display our results with discussion in Section \ref{sec: comparisons}. We conclude in Section \ref{sec: Conclusion}. Throughout this work we assume typical parameter values for standard $\Lambda$CDM cosmology models, i.e., $\Omega_m = 0.3$, $\Omega_{\Lambda} = 0.7$, and $H_0 = 67.4$ km s$^{-1}$ Mpc $^{-1}$ \citep{2020A&A...641A...6P}. 

\section{Methods} \label{sec: Methods}

\begin{figure*}[ht!]
	\centering
	\includegraphics[width=0.85\linewidth]{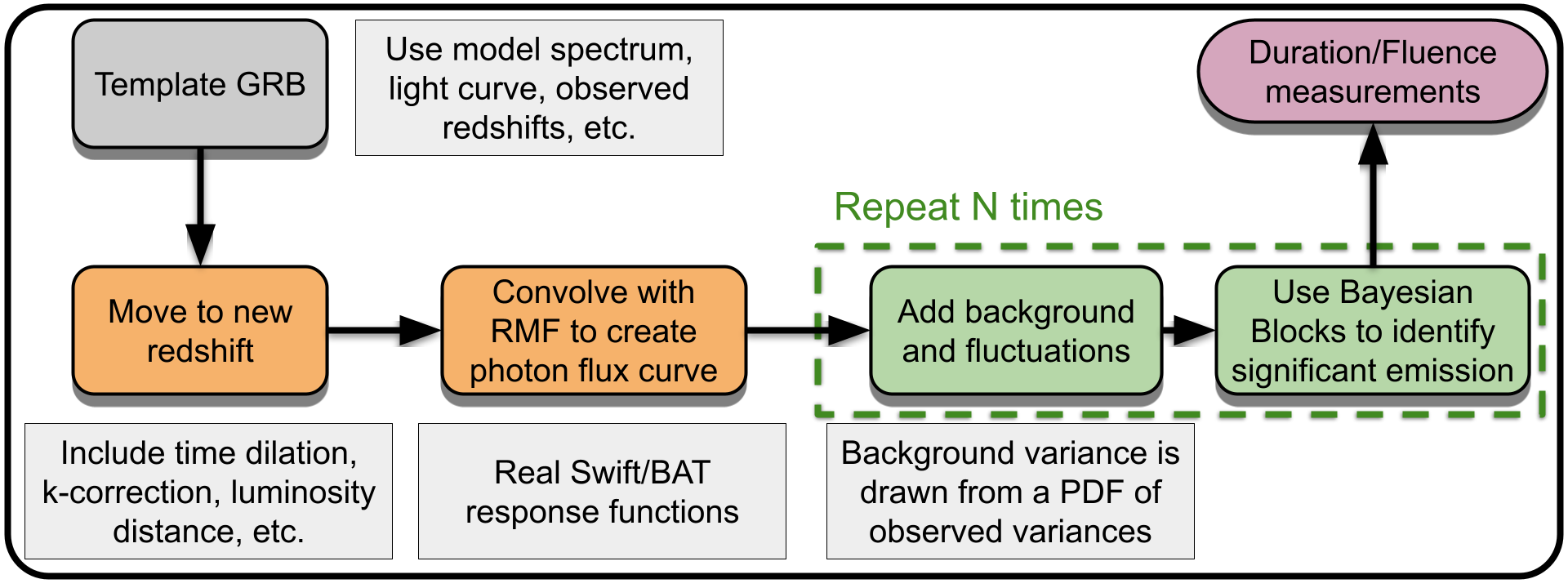}
	\caption{Work flow schematic of the simulations and measurements performed in this work. \label{fig: schematic}}
\end{figure*}

Our goal was to study how the distance to a GRB impacts the duration and fluences estimated for Swift/BAT observations. To do this, we selected a sample of GRBs observed with Swift/BAT at low redshift ($z<1$) and simulated them at increasing distances up to $z=15$ or  until the bursts could no longer be detected by Swift/BAT. We used the Bayesian block algorithm to estimate the T$_{90}$ and fluences of the simulated bursts and compared these results to the input burst parameters to assess the impact of the distance to the source. All simulations in this work were completed using the publicly available \texttt{simmes} package
\footnote{\url{https://github.com/mikemoss3/simmes}}. Finally, we compared the measurements of our simulated bursts to a sample of GRBs observed with Swift/BAT at redshifts of $z\geq3$. A schematic of the methodology used in this work is shown in Figure \ref{fig: schematic}.

We compiled our sample of GRBs by first selecting bursts observed by Swift/BAT \citep{2005SSRv..120..143B} between December 2004 - June 2025 that had redshift measurements. The automated analysis of Swift/BAT GRBs includes a spectral fit to the 1-second peak flux emission in the 15-150 keV range using both power law and cut-off power law models. For the GRBs in our sample, we assume their spectra are well represented by the best-fit cut-off power law
\footnote{A spectrum with a peak energy is required so that the total energy of a simulated GRB does not tend toward infinity as it is simulated out to increasing redshifts.} 
(CPL)\textbf{, regardless of whether the automated analysis preferred a power law or CPL}. We required each burst to have a CPL peak energy to be $E_p \leq 550$ keV. This removed any bursts with unconstrained peak energies far outside the BAT energy band
\footnote{Since these bursts will be used as mock GRBs for our simulations, we do not require the peak energies to be strongly constrained measurements, only that they are in reasonable agreement with measurements made for the entire GRB population}. 

Typically, GRB spectra peak at lower energies near the start and end of the burst emission. Therefore, using time-resolved spectra in our simulations generally lead to burst signals being lost into the background noise faster than when using a single time-integrated spectra. We found that using time-resolved spectra did not lead to significant changes in our final interpretations, but can increase computation time significantly. Therefore, we decided to use the best-fit 1-second peak spectral model found for each burst as the spectrum across the entire burst interval. As a result, our estimations on the loss of burst signal are conservative.

\begin{figure}[ht!]
	\centering
	\includegraphics[width=\linewidth]{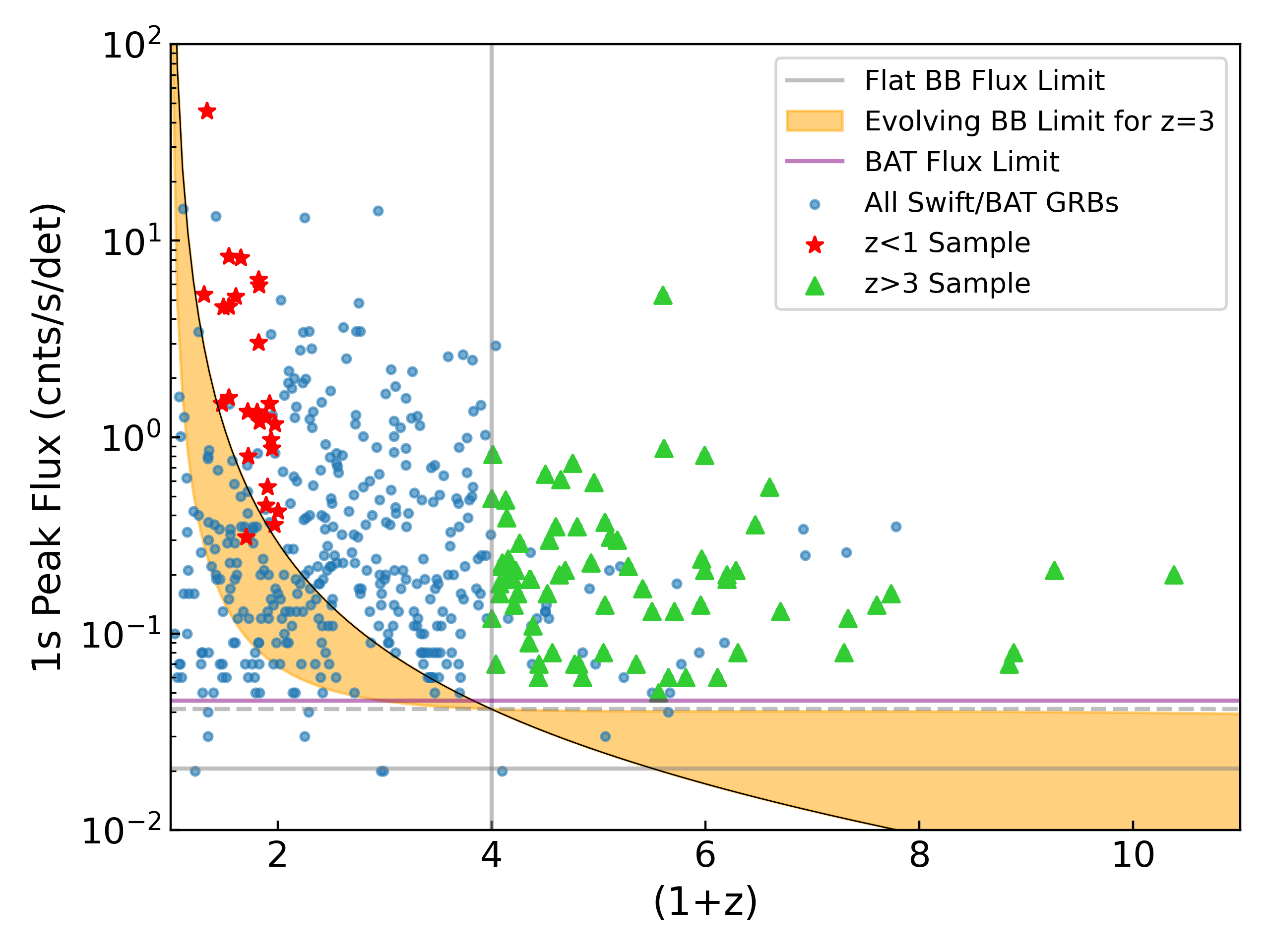}
	\caption{Peak flux measurements of Swift/BAT GRBs as a function of their measured redshifts. Blue points are all Swift/BAT GRBs. The red stars and green triangles indicate GRBs in the low-$z$ and high-$z$ samples of this work, respectively (see Tables \ref{tab: low-z} and \ref{tab: high-z}). The horizontal gray line indicates the theoretical Bayesian block sensitivity limit, the dashed line is twice that limit \textbf{(see text for details)}. For reference, the horizontal purple line indicates the BAT flux sensitivity assuming a 1 second exposure \citep{2013ApJS..207...19B,2016ApJ...829....7L}. The thin black line represents a 1s peak flux value equal to twice Bayesian block sensitivity limit at $z=3$ calculated at other redshifts between $z = 0 - 10$ assuming a spectrum $dN/dE \propto (E/50.\text{ keV})^\alpha e^{-E(2+\alpha) / E_p}$, where the photon index is assumed to be $\alpha = -1.5$ and the peak energy to be $E_p = 50$ keV. The region shaded in yellow \textbf{defines the same extrapolated flux value but assuming different spectral shapes (see text for details)}. \label{fig: grb-peak-f-vs-z}}
\end{figure}

\begin{figure}[ht!]
	\centering
	\includegraphics[width=\linewidth]{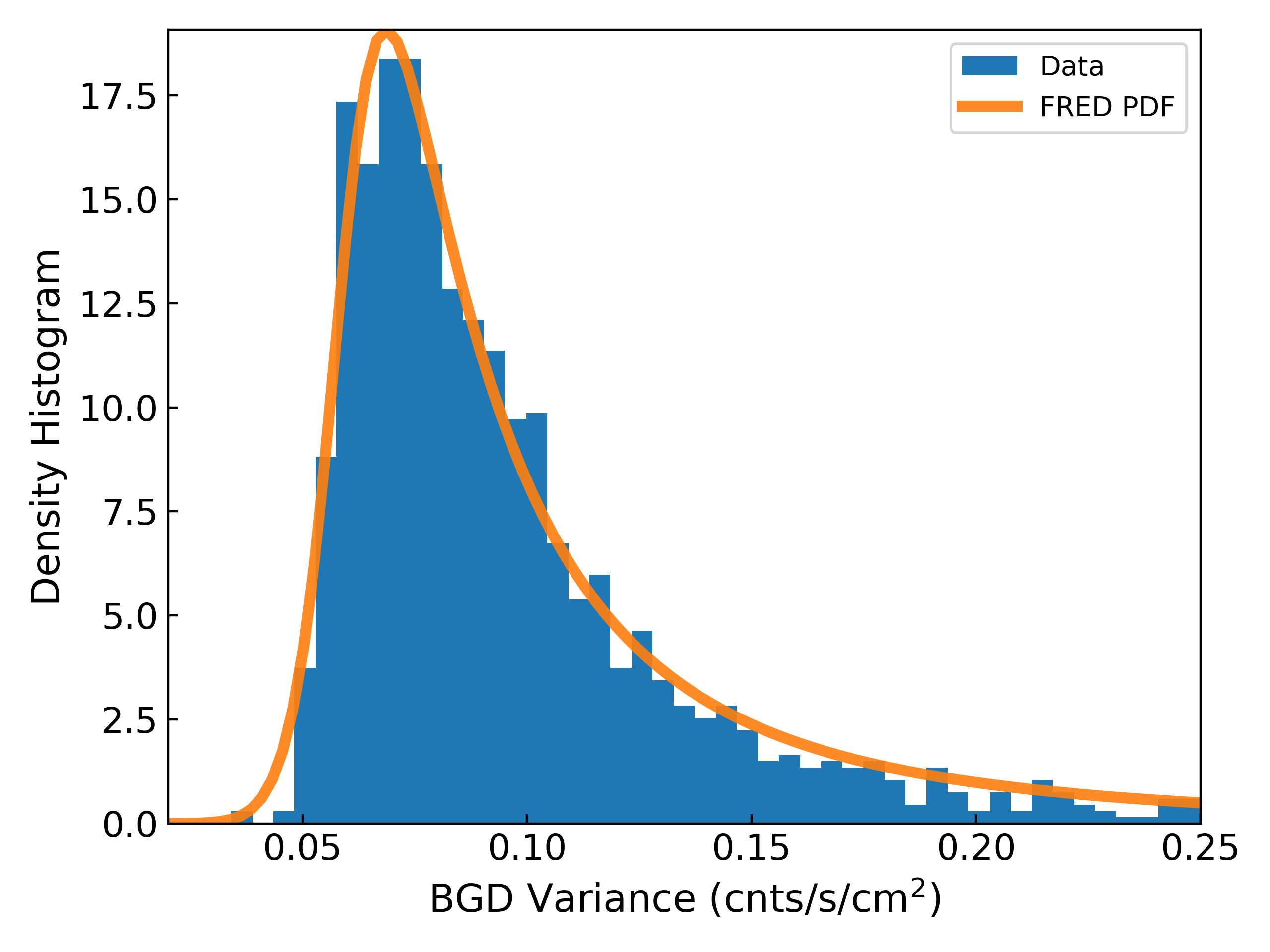}
	\caption{We compiled the 1-second time-bin background variances from the mask-weighted light curves of all Swift/BAT GRBs (blue) and fit a FRED function to the distribution to create a probability distribution function (orange). The background variances were calculated within the 50 second intervals preceding and following the emission of each burst. We sampled this distribution during our simulations to determine the background variance of each simulated light curve.}
	\label{fig: background variances}
\end{figure}

Lastly, we applied a distance-dependent peak-flux cut by assuming a mock peak flux value at $z=3$ and calculating what the peak flux would be as a function of redshifts (see Eq. \ref{eq:flux_calc} below) and for a range of spectral shapes (i.e., assuming a CPL spectrum defined as $dN/dE = N_0 (E/50.\text{ keV})^\alpha e^{-E(2+\alpha) / E_p}$, where the photon index ranged from $\alpha = -1.5$ to $-0.5$, peak energies ranged from $50$ to $550$ keV, and $N_0=1$ is the flux normalization ). The peak flux value we assumed is equal to twice the theoretical Bayesian block sensitivity limit, $A$, given by $A = \bar{\sigma}\Delta t \sqrt{2 \log{n}}$ \citep{2013ApJ...764..167S}, where $\bar{\sigma}$ is average background variance of BAT mask-weighted light curves (see Fig \ref{fig: background variances}), $\Delta t$ is the time bin size, and $n$ is the number of bins in the light curve. We assume $\Delta t = 1$ sec and $n=50$. \textbf{This flux value was extrapolated to all redshifts between $z=0 - 10$ assuming spectral defined by $dN/dE \propto (E/50.\text{ keV})^\alpha e^{-E(2+\alpha) / E_p}$, where the photon index was varied between $\alpha = -1.5$ to $-0.5$ and peak energies varied between from $50$ to $550$ keV.} This flux cut ensured that each burst in our sample would be measurable out to at least $z=3$ (see yellow region in Fig. \ref{fig: grb-peak-f-vs-z}).

We then separated the observed sample into low-$z$ GRBs (i.e., 26 GRBs with $z<1$; see red stars in Fig. \ref{fig: grb-t90-vs-z} and \ref{fig: grb-peak-f-vs-z}) and high-$z$ GRBs (i.e., 72 GRBs with $z>3$; see green triangles in Fig. \ref{fig: grb-t90-vs-z} and \ref{fig: grb-peak-f-vs-z}). Tables \ref{tab: low-z} and \ref{tab: high-z} in Appendix \ref{app:data_tables} list the GRBs included in this work.  We now describe how the observed low-$z$ GRBs were simulated out to higher redshifts and then compared to the observed high-$z$ sample.

For each GRB in our low-$z$ sample, we download the observed Swift/BAT mask-weighted 15 - 350 keV light curve
\footnote{Mask-weighting is a standard procedure for coded-mask aperture instruments such as Swift/BAT \citep{2005SSRv..120..143B,2022ApJ...941..169D}.} 
and best-fit spectral model parameters
\footnote{See automated GRB data products at \url{https://swift.gsfc.nasa.gov/results/batgrbcat/}}. 
For bursts in our sample with T$_{90}\leq10$ sec, we used 64 ms time binned light curves and 1 second time binned light curves for bursts with T$_{90}>10$ sec. Because we assumed the burst spectrum is equal to the best-fit spectrum model found for the 1-second peak interval, we normalized the light curves by the fluence within the 1-second peak interval, leaving us with a dimensionless light curve that realistically mimics GRB light curve behavior. We did not apply any smoothing to the light curves. These normalized light curves were used to indicate the relative fluxes in each bin while the best-fit spectral model were used to determine the actual count rates in each bin. For each light curve, we isolate the burst signal and remove background by omitting any signal outside of the estimated T$_{100}$\footnote{In addition to the T$_{90}$, the Swift/BAT team reports the duration which encompasses 100\% of the significantly detected burst emission, i.e., the T$_{100}$.}.

We simulated GRBs at higher redshifts by taking into account the cosmological effects on the observed light curve profile and spectra (i.e., time dilation, $k$ correction, luminosity distance). A burst observed in the energy band $[e_1, e_2]$ at a redshift $z_1$ with a measured flux $F_{z_1, [e_1, e_2]}$ and spectrum $\phi_1(E)$ will have a \textbf{flux} given by

\begin{align}
	F_{z_2, [e_1, e_2]} = & \frac{D^2_{L, z_1} (1+z_2)}{D^2_{L, z_2}(1+z_1)} F_{z_1, [e_1, e_2]} \times \nonumber \\ 
	& \frac{k[e_1, e_2, E_1, E_2, z_1, \phi_{z_1}(E)]}{k[e_1, e_2, E_1, E_2, z_2, \phi_{z_2}(E)]} \label{eq:flux_calc}
\end{align}
if it were instead at $z_2$, where $\phi_{z_2}(E)$ is the spectrum that would be observed at $z_2$, $D_{L,z_i}$ is the luminosity distance at a redshift $z_i$, and $[E_1, E_2] = [0.1, 10,000]$ keV adequately approximates the bolometric energy band. For this work, we use the complete Swift/BAT energy band, i.e., $[e_1, e_2] = [15, 350]$ keV. $k[e_1, e_2, E_1, E_2, z, \phi(E)]$ is the $k$-correction factor that occurs as the observed spectra is redshifted into the observing band \citep{2001AJ....121.2879B},
\begin{align}
	k[e_1, e_2, E_1, E_2, z, \phi(E)] &= \frac{F_{[E_1/(1+z), E_2/(1+z)]}}{F_{[e_1, e_2]}} \\
	&= \frac{\int_{E_1/(1+z)}^{E_2/(1+z)} E \phi(E) dE}{\int_{e_1}^{e_2} E \phi(E) dE},
\end{align}

The redshifted spectra (units photons / s / keV / cm$^2$) are then convolved with the Swift/BAT instrument response function to produce mock folded spectra (units counts / sec / keV / on-axis fully-illuminated detector
\footnote{The unit ``on-axis fully-illuminated detector'' is unique to coded-mask aperture instruments, a definition of what each term signifies can be found here: \url{https://swift.gsfc.nasa.gov/analysis/threads/batfluxunitsthread.html}}
). For this work, we assumed that all simulated bursts were observed on-axis compared to the detector bore-sight (i.e., $\theta_{inc} = 0$ and PCODE=1, where PCODE is the partial coding fraction which defines the average fraction of each detector illuminated by a source through the coded mask accounting for both the source incident angle and instrument geometry). Furthermore, we assumed that all 32,768 detectors on the Swift/BAT detector plane were enabled during each mock observation. We found that we needed to scale the folded spectra by a factor of 2 to ensure that the simulations would remain self consistent with the input observations. Taking the integral of the folded spectra across the most-sensitive interval of the BAT  energy band (i.e., 15 - 150 keV) gave an observed count flux (units of counts / sec / on-axis fully-illuminated detector), which differs from the emitted count flux due to instrumental effects and sensitivity. We then multiplied the normalized light curves by their respective observed count fluxes to obtain mock Swift/BAT mask-weighted light curves for each burst.

Mask-weighted light curves are inherently background subtracted, meaning that outside of any source emission interval the average count rate will be $\sim 0$ cnts/s. The fluctuations around zero are determined by the raw background count rate during observation (e.g., a larger raw background count level leads to larger variations in the mask-weighted light curve). To mimic these background fluctuations, we compiled a distribution of 1-second time-bin background variances for all GRBs observed with Swift/BAT (see Fig. \ref{fig: background variances}) and fit a FRED function \citep{2003ApJ...596..389K} to the distribution to create a probability distribution function (PDF). For each simulation, we randomly selected a background variance value from this PDF and then added mock background fluctuations to the light curve.

Finally, the mock mask-weighted light curves were passed through a Bayesian-block algorithm\footnote{\url{https://docs.astropy.org/en/stable/api/astropy.stats.bayesian_blocks.html}} to obtain duration measurements \citep{2013ApJ...764..167S}. 
For this work, we did not use the Swift/BAT trigger algorithms to trigger on simulated light curves and, instead, directly applied the Bayesian block algorithm to obtain duration measurements. This methodology is more similar to ground-based analyses of Swift/BAT data when searching for sub-threshold bursts. The Bayesian-block algorithm is sensitive to the duration of the background interval it uses to test for significant emission. However, increasing the light curve length also increases Bayesian block computation time. For that reason, when we added mock background emission, we use a background interval that was at least $\geq$T$_{100}$ onto either side of the source emission interval.

For every burst in our low-$z$ sample, we simulated the burst across 50 different redshifts, starting at the burst's observed redshift and increase to $z=15$ or until the burst was no longer measurable using the Bayesian-block algorithm. We simulate mock GRBs 1000 times at each redshift, varying the count flux for each simulation as described above.

\section{Impact of Distance on GRB Measurements} \label{sec: Results}

\begin{figure}[ht!]
	\centering
	\includegraphics[width=0.7\linewidth]{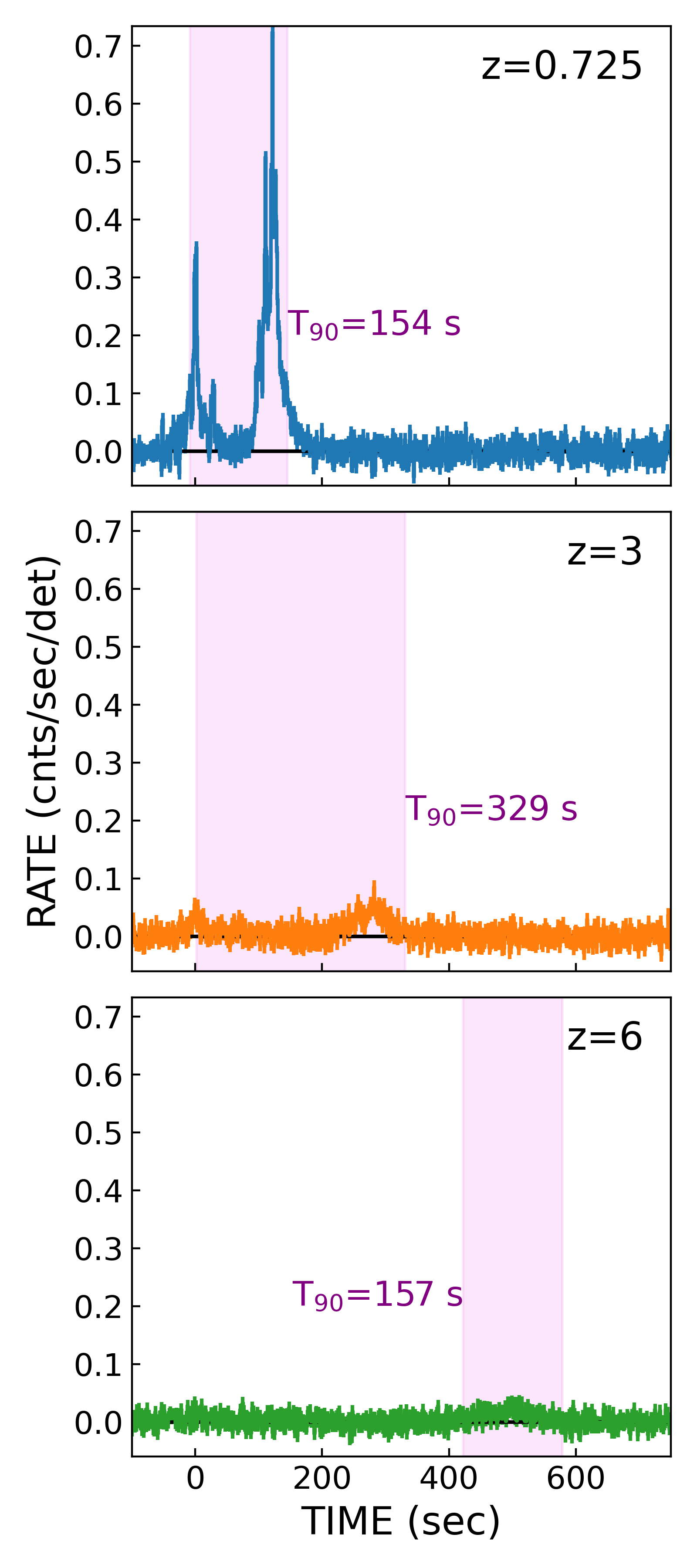}
	\caption{\textit{Top}: The light curve of GRB 140512A, observed at a redshift of $z=0.725$, displays multiple pulses of emission (T$_{100}\sim$154 seconds). Shaded regions indicate the measured T$_{90}$. \textit{Center} and \textit{Bottom}: Using our simulation tool, we show what the same light curve looks like when shifted to $z=3$ and $z=6$, respectively. With increasing redshift, the luminosity of the burst is decreased and the signal is stretched due to time dilation. At $z=3$, the early time emission is still visible and could be interpreted as a precursor. At $z=6$, only the main emission period remains significant for Swift/BAT.}
	\label{fig: example light curve}
\end{figure}

To illustrate the tip-of-the-iceberg effect, we show an example GRB 140512A simulated at different redshifts in Figure \ref{fig: example light curve}. At the burst's observed redshift, $z=0.725$, the light curve shows multiple pulses of varying brightnesses lasting for T$_{90, rest}\sim154$ sec. When the burst is moved to $z=3$, only two dim pulses separated by a quiescent period are visible for a total duration of T$_{90}\sim329$ sec. In this case, the early emission episode may even be interpreted as a GRB precursor \citep{1991Natur.350..592M,1995ApJ...452..145K,2010ApJ...723.1711T}. At $z=6$, only the main emission period remains significant above background and the duration becomes T$_{90}\sim157$ sec, similar to the duration measured at $z=0.725$.

\begin{figure}[ht!]
	\centering
	\includegraphics[width=0.65\linewidth]{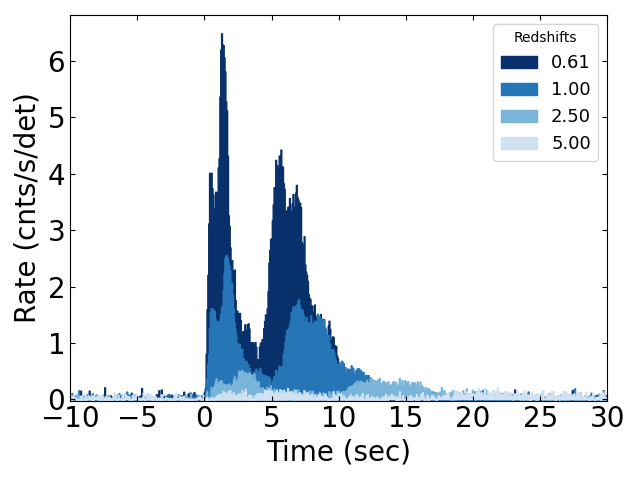}
	\includegraphics[width=0.85\linewidth]{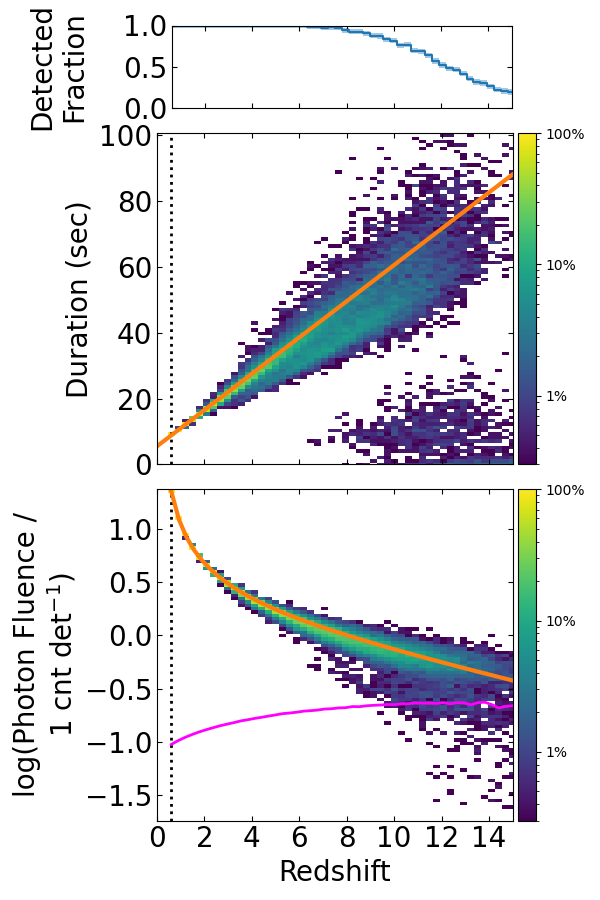}
	\caption{\textit{a}: The 15 - 150 keV Light curve of GRB 050525A simulated at increasing redshifts (indicated by lighter shades of blue). \textit{b}: The fraction of simulations able to be measured by the Bayesian block algorithm in each redshift bin ($1,000$ simulations were performed in each $z$ bin). \textit{c}: Density plot of the measured T$_{90}$ at increasing redshifts for mock GRBs generated from GRB 050525A data. The black dotted line indicates the measured redshift of the burst, $z_{obs}=0.61$. The orange line indicates $\propto$T$_{90} \times (1+z)/(1+z_{obs})$. Bins with $<3$ detections were excluded to remove false positives. \textit{d}: Density plot of the measured fluence as a function of increasing redshift. The orange line indicates the analytically expected fluence, i.e., $\propto k(z)/D^2_L(z)$. The magenta line indicates the Bayesian block detection threshold defined $A = \sigma \sqrt{2\log(N) T}$ (see text for parameter definitions). \textbf{The top and bottom color bars indicate the percentage of simulations performed at the same redshift that obtain the same T$_{90}$ or fluence measurement, respectively.} \label{fig: 050525A}}
\end{figure}

In Figure \ref{fig: 050525A}, we show the simulation results for GRB 050525A, a burst with an observed light curve that exhibits two bright and well-defined FRED pulses with a measured duration of T$_{90,rest}=8.836$ sec. The durations and fluences measured for all simulations are displayed in the two density plots and show how the T$_{90}$ and fluence measured for the simulations evolve with redshift. As the redshift increases from the measured redshift $z=0.61$ out to higher redshifts, the measured durations stay proportional to $\propto$T$_{90,rest}\times(1+z)$. However, above $z\sim8$, duration measurements begin to drop far below the expected time-dilation line, indicating that the tip-of-the-iceberg effect is starting to have an effect on the duration measurements. Duration measurements above T$_{90,rest}\times(1+z)$ are purely due to random fluctuations in the background noise. The fluence measurements are proportional to $\propto k(z)/D_L^2(z)$ but quickly become underestimates of the intrinsic fluence.

\begin{figure}[ht!]
	\centering
	\includegraphics[width=0.65\linewidth]{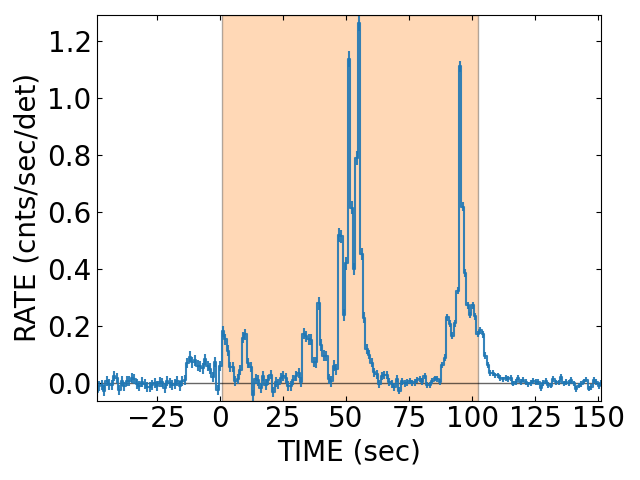}
	\includegraphics[width=0.85\linewidth]{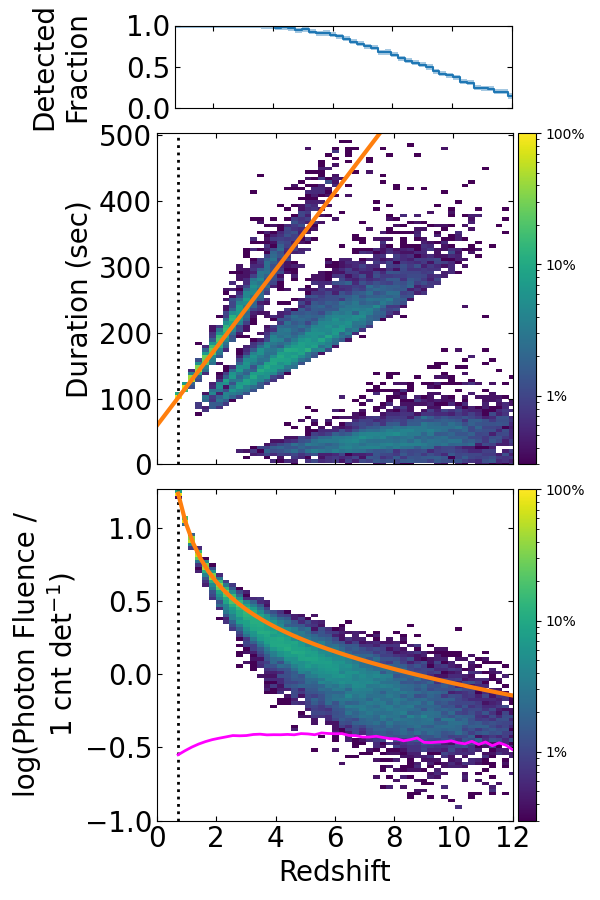}
	\digitalasset
	\caption{\textit{a}: The 15 - 150 keV light curve for GRB 111228A. The orange shaded region indicates the T$_{90}$ measured by the automated BAT analysis pipeline. Panels \textit{b}, \textit{c}, and \textit{d} are described in Figure \ref{fig: 050525A}. The complex multi-pulsed light curve of GRB 111228A leads to discontinuities in the T$_{90}$ evolution. At a redshift of $\sim3$, this burst could have had a measured T$_{90}\approx20$ sec, an order of magnitude shorter than the true signal duration. The fluence evolution does not show the same strong discontinuities because a majority of the fluence occurs during the brightest pulse. Figures of the simulation results for the remaining GRBs in our low-$z$ sample (24 figures) are available in the online version of the article. \label{fig: 111228A}}
\end{figure}

Another example of our simulation results is shown for GRB 111228A in Figure \ref{fig: 111228A}. Starting at $z\sim2$ the duration measurements begin to have a bimodal or even a trimodal distribution. This behavior occurs when the dimmer pulses at the beginning and end of GRB 111228A become so dim they are no longer significant above background and, therefore, are not included in the duration estimate found by the Bayesian block algorithm. At $z\sim2$, GRB 111228A would have had an equally likely chance to have a measured duration of $\sim 175$ sec and $\sim75$ sec. Depending on the structure of the input light curve, it is common to see such abrupt jumps or discontinuities in the evolution of the T$_{90}$ measurements. However, these abrupt changes in T$_{90}$ measurements are not necessarily present in the measured fluence, since the majority of the fluence is within the brightest pulse(s) of the light curve.

Once source signal becomes dim enough that it is no longer distinguishable from background noise it becomes unrecoverable without prior knowledge of the underlying light curve. And since prompt emission light curves of GRBs vary so strongly between bursts, there is no common behavior in the evolution of the measured durations with increasing redshift. However, for every GRB in our sample, duration measurements at higher redshifts significantly underestimated the time-dilation corrected duration, indicating that the tip-of-the-iceberg effect plays a significant role in the measured duration of the simulated light curves. Furthermore, some similar features occurred in the duration measurement evolution of several bursts due to similarities in their light curve profiles, e.g., dim pulses towards the beginning or end of the emission interval may lead to disconnected regions of probable duration measurements, as was the case for GRB 111228A.

In some cases, the measured durations seem to show a trend of rising above $T = (1+z)T_{rest}$ more than just random fluctuations would cause, e.g., \textbf{see the measurements for GRB 111228A that rise above the $(1+z)$ line at $z\sim5$} (see Fig. \ref{fig: 111228A}). \textbf{This behavior isn't possible in simple analytical cases, but can occur} when GRBs have dim pulses or tails of emission on either side of their main emission period that were not completely included in the original T$_{90}$ interval (since the first 5$\%$ and last 95$\%$ are not included). With increasing redshift, the signal-to-noise ratio of both the main emission period and the dim pulses decrease but, occasionally, background fluctuations will make the dim signal significant enough that it becomes measurable by the Bayesian block algorithm again, and leads to a duration measurement higher than expected. \textbf{Considering the light curve of GRB 111228A, the emission that occurs between -20 to 0 seconds will become dimmer at higher redshifts, but will occasionally become significant enough to be included in the T$_{90}$ measurements due to background fluctuations, leading to a T$_{90}$ measurement rising above the $(1+z)$ line}. The complete figure set including the input light curves and simulation results for the remaining power law and cut-off power law models GRBs in our low-$z$ sample (24 figures) is available in the online version of the article.

\subsection{Long then Short GRBs} \label{sec: long to short}

\begin{figure}[ht!]
	\centering
	\includegraphics[width=0.75\linewidth]{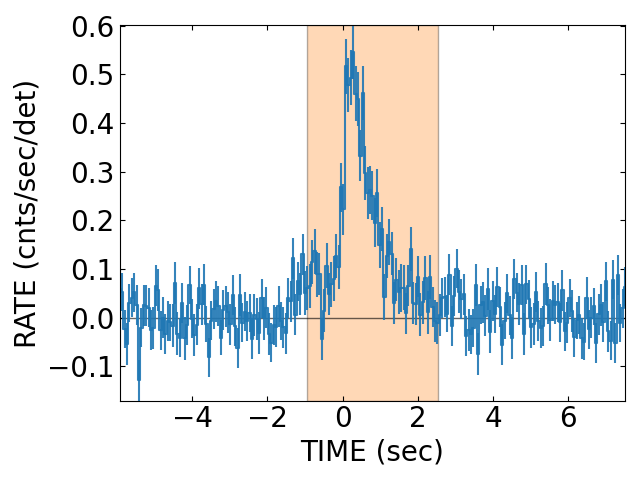}
	\includegraphics[width=0.75\linewidth]{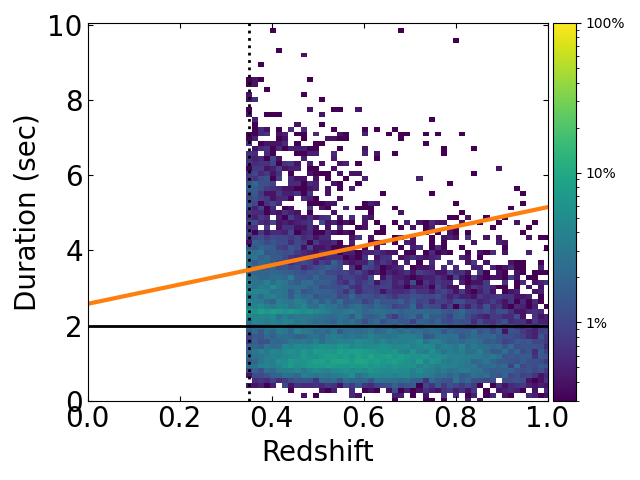}
	\includegraphics[width=0.75\linewidth]{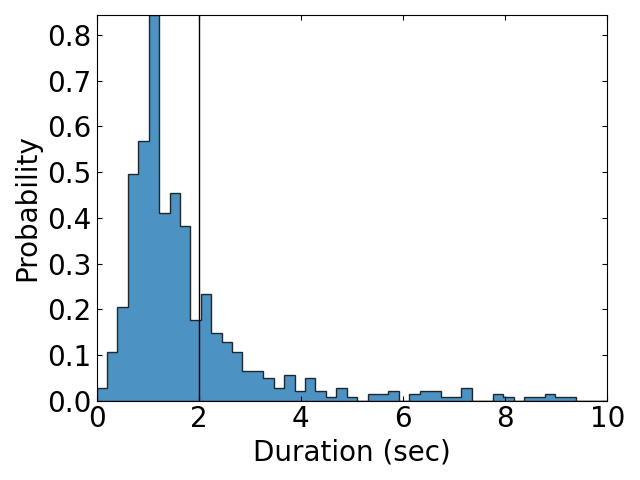}
	\caption{\textit{Top}: The 15 - 150 keV light curve for GRB 120311A as observed by Swift/BAT. The orange shaded region indicates the T$_{90}$ measured by the automated BAT analysis pipeline. \textit{Middle}: Same description as Figure \ref{fig: 050525A}\textit{b}, but for simulations using GRB 120311A as the input GRB. The horizontal, black line indicates the traditional separation between long and short GRBs (i.e., 2 seconds). This is an example of a long GRB becoming a short GRB, simply due to observational bias. \text{Bottom}: A slice of the duration measurements for GRB 120311A at $z\sim0.7$.}
	\label{fig: 120311A}
\end{figure}

In some cases, we found that long GRBs could be measured as short GRBs had they been at higher redshifts (e.g., see GRB 050525 in Fig. \ref{fig: 050525A}). GRB 120311A is not part of our low-$z$ sample, however we simulated it as an exemplary case of this long-to-short behavior. GRB 120311A has a measured redshift of $z=0.35$ \citep{2019A&A...623A..92S} displaying a FRED-like light curve with a T$_{90}=3.480$ sec. The spectrum of GRB 120311A observed by Swift/BAT was best fit by a power law model (with spectral index $\alpha = -2.176$ and normalization $N_0 = 0.0197$ cnts s$^{-1}$ cm$^{-2}$ keV$^{-1}$) and, furthermore, was not observed by other gamma-ray instruments, such as Konus instrument onboard the Wind observatory or the Gamma-ray Burst Monitor onboard the \textit{Fermi} Gamma-ray Space Telescope, so we could not constrain the peak energy for a cut-off power-law spectral model. For our simulations, we assumed the measured spectral index and normalization and a peak energy of 550 keV, typical for short-hard GRBs. From our simulations, we found that even at the burst's observed redshift, $z=0.35$, by simply varying the background, the T$_{90}$ measurement for this burst can change to values ranging from $1 \lesssim $T$_{90} \lesssim 7$ sec. This is due to the dim tail in the FRED-like pulse of the light curve as parts of the tail become more or less significant above background due to noise fluctuations. However, if instead this burst was observed at a redshift of $z=0.7$, we would most likely have measured a T$_{90}\sim1$ sec (see Fig. \ref{fig: 120311A}). 

We have found that this phenomenon is possible, but does not happen commonly for our sample. It is much more common that a long GRB will fade completely into the noise and become undetectable before becoming a short GRB. However, the sample of bursts we simulated is small compared to the total number of GRBs and, furthermore, selected with criteria in mind to create a bright sample of bursts that could be comparable to high-$z$ bursts. A study on how frequently GRBs with intrinsic durations $>2$ sec are observed as SGRBs with T$_{90}<2$ sec must include a much larger sample of GRBs, including those with unique light curve structures like extended emission, and will be completed in future work. 

\section{Comparing observed and simulated high-$z$ GRBs} \label{sec: comparisons}

\begin{figure*}[ht!]
	\centering
	\includegraphics[width=0.3\linewidth]{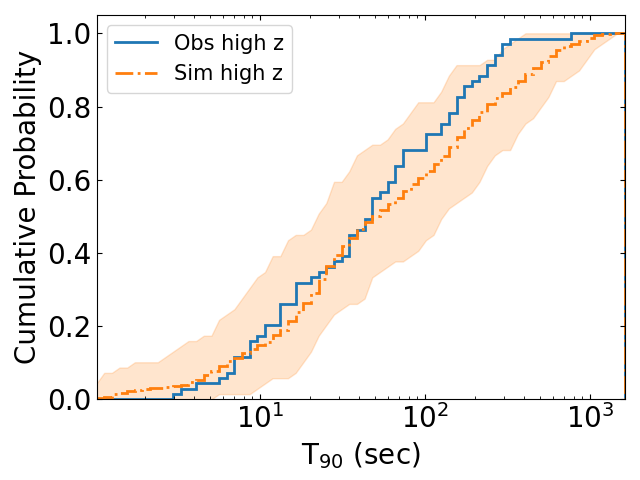}
	\includegraphics[width=0.3\linewidth]{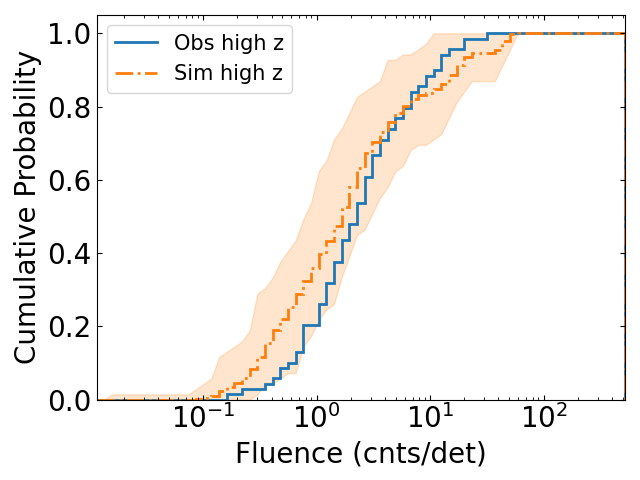}
	\includegraphics[width=0.3\linewidth]{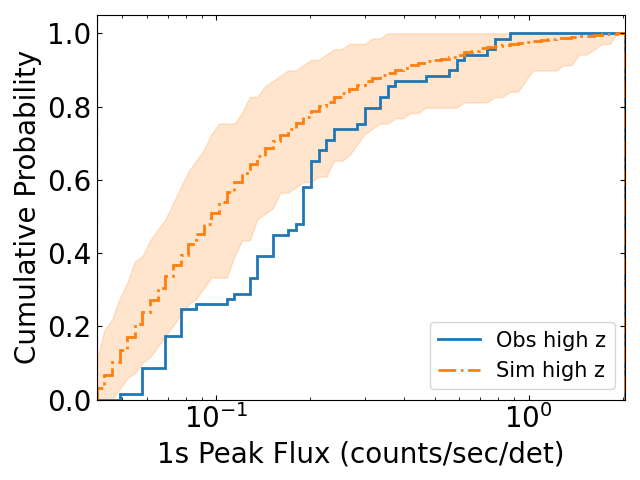}
	\includegraphics[width=0.3\linewidth]{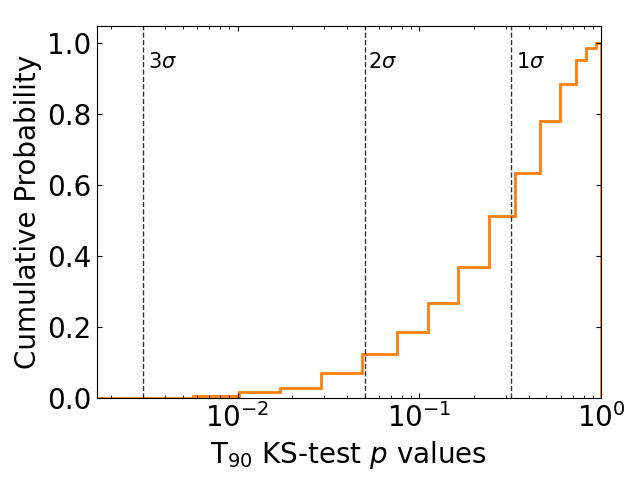}
	\includegraphics[width=0.3\linewidth]{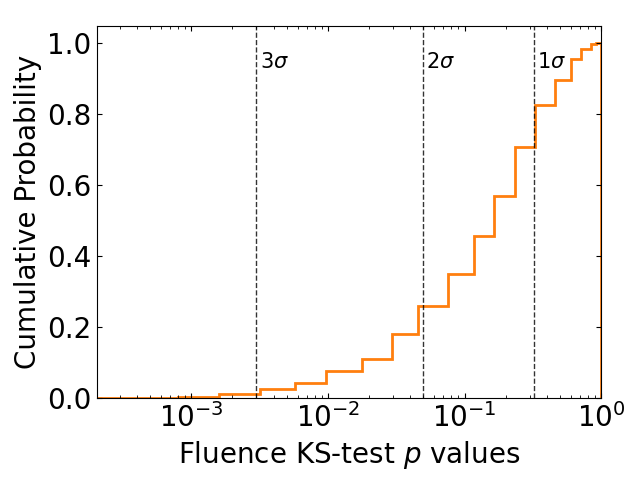}
	\includegraphics[width=0.3\linewidth]{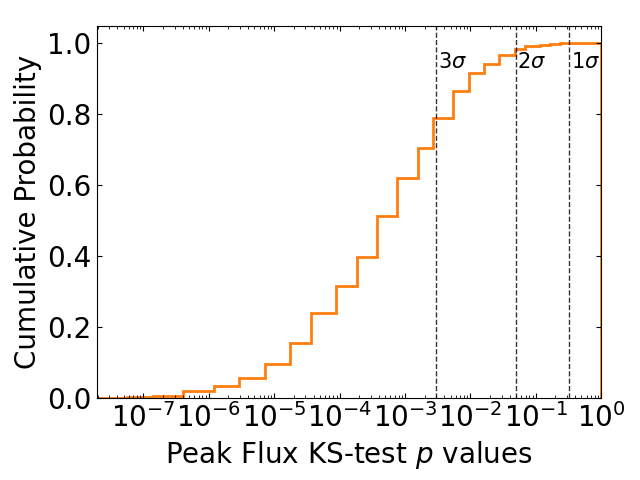}
	\caption{\textit{Top Row}: Cumulative distributions of the T$_{90}$ (\textit{left}), fluence (\textit{center}), and peak flux (\textit{right}) measurements for the observed high-$z$ sample (solid, blue) and the simulated high-$z$ sample (dashed orange). The orange shaded regions indicate the range of cumulative distributions generated from 1,000 trials of randomly sampling 73 simulation results from our entire simulated high-$z$ sample (i.e., the same number as in our observed high-$z$ sample). \textit{Bottom Row}: Cumulative distributions of KS-test $p$-values calculated between the 1,000 randomly-drawn simulation-subsample distributions and the observed distributions. The vertical dashed lines indicate 1, 2, and 3 $\sigma$ cut-offs.}
	\label{fig: cum dists}
\end{figure*}

In the top panels of Figure \ref{fig: cum dists} we display the T$_{90}$, fluence, and 1s peak flux cumulative probability distributions for the observed high-$z$ sample (solid, blue) and the simulated high-$z$ sample (dashed, orange). To test the compatibility of our simulations with the observations we randomly selected 73 simulation results from our entire sample of high-$z$ simulations (i.e., 73 being the same number of bursts in our observed high-$z$ sample) to create 1,000 subsamples. These subsample distributions were used to generate the orange shaded regions in the top panels of Figure \ref{fig: cum dists}. We performed a two-sample Kolmogorov-Smirnov test (KS-test) between each subsample and the observed high-$z$ bursts. The cumulative distributions of the KS-test $p$-values are displayed in the bottom panels of Figure \ref{fig: cum dists}. We find that when comparing the durations of the two samples, $52.2\%$ ($99.6\%$) of $p$-values are above $0.32$ ($0.003$), indicating that $52.2\%$ ($99.6\%$) of the random subsamples are within 1$\sigma$ (3$\sigma$) of the observed distribution. Similarly, when comparing the fluence distributions, $29.4\%$ ($99.4\%$) of $p$-values are found to be greater than $0.32$ ($0.003$). However, when we compared the peak-flux distributions, we find no random subsamples are within 1$\sigma$ of the observed distribution and only $28.4\%$ are within 3$\sigma$ (see Tab. \ref{tab:p_vals_all}).

If only the durations of the samples are compared, the simulations and the observations are not incompatible with being drawn from the same underlying distribution (i.e., we cannot reject the null hypothesis of the KS-test) whether a 1, 2, or 3$\sigma$ agreement threshold is used. When comparing the fluence measurements, a 2, or 3$\sigma$ threshold still leads to the same conclusion. However, the peak-flux distributions of the two samples appear to differ significantly, both in shape (see Fig. \ref{fig: cum dists} \textit{Bottom Right}) and according to the $p$-value distribution (see Tab. \ref{tab:p_vals_all}), implying that they were not drawn from the same underlying distribution. On average, the peak fluxes of the observed GRB sample are higher than simulations.

\begin{deluxetable}{llll}
	\tablewidth{0pt}
	\tablecaption{Percentage of KS-Test $p$-values calculated between the high-$z$ simulation subsamples and all observed high-$z$ GRBs that fall within 1, 2, or 3$\sigma$ \label{tab:p_vals_all}}
	\tablehead
	{
		\colhead{Metric} & \colhead{1$\sigma$} & \colhead{2$\sigma$} & \colhead{3$\sigma$}
	}
	\startdata
		Durations & 52.2$\%$ & 90.4$\%$ & 99.6$\%$ \\ 
		Fluences & 29.4$\%$ & 75.8$\%$ & 99.4$\%$ \\ 
		Peak Flux & 0.0$\%$ & 1.2$\%$ & 28.4$\%$ \\ 
	\enddata
\end{deluxetable}

\begin{deluxetable}{llll}
	\tablewidth{0pt}
	\tablecaption{Percentage of KS-Test $p$-values calculated between the high-$z$ simulation subsamples and high-$z$ GRBs observed before 2012 that fall within 1, 2, or 3$\sigma$ \label{tab:p_vals_cut}}
	\tablehead
	{
		\colhead{Metric} & \colhead{1$\sigma$} & \colhead{2$\sigma$} & \colhead{3$\sigma$}
	}
	\startdata
		Durations & 75.9$\%$ & 98.3$\%$ & 100$\%$ \\ 
		Fluences & 74.8$\%$ & 97.5$\%$ & 100$\%$ \\ 
		Peak Fluxes & 15.3$\%$ & 54.0$\%$ & 94.1$\%$ \\ 
	\enddata
\end{deluxetable}

Motivated by the agreement between the duration and fluence measurements, we investigated whether there may be some bias that has arisen in Swift/BAT GRBs over the mission's lifetime. Splitting the observed sample in roughly half leads to 39 GRBs observed before 2012-01-01 ($\sim35\%$ of the missions life at the time of this publication) and 33 GRBs observed after. In Appendix \ref{sec:excluding_late_mission_grbs}, we discuss possible indications that there may be an evolving bias in Swift/BAT GRBs and possible explanations.

\begin{figure*}[ht!]
	\centering
	\includegraphics[width=0.3\linewidth]{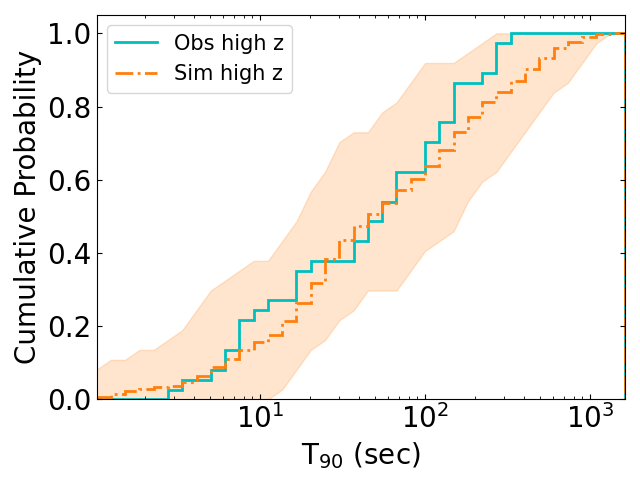}
	\includegraphics[width=0.3\linewidth]{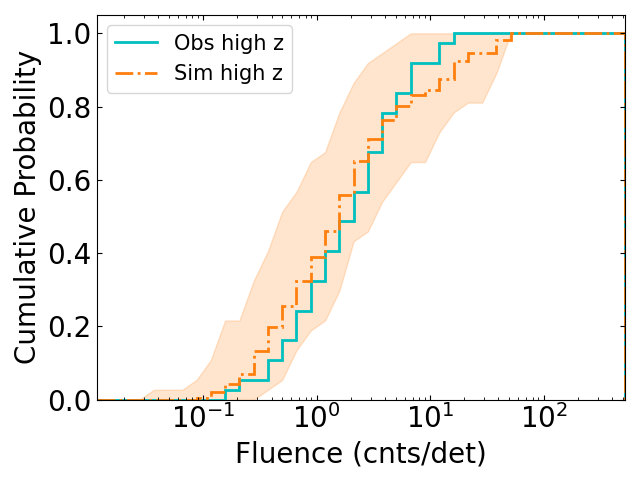}
	\includegraphics[width=0.3\linewidth]{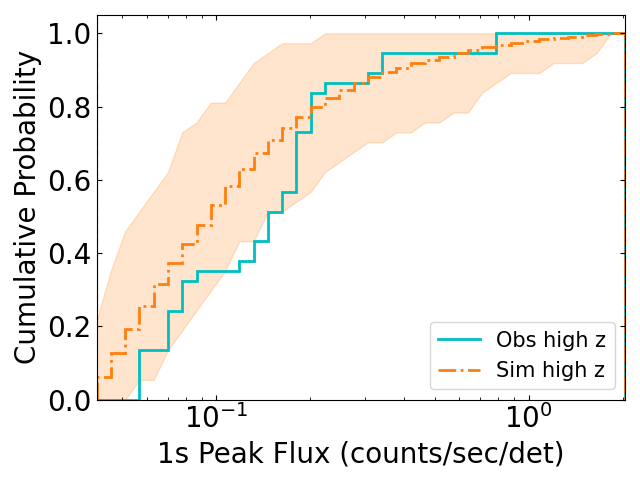}
	\includegraphics[width=0.3\linewidth]{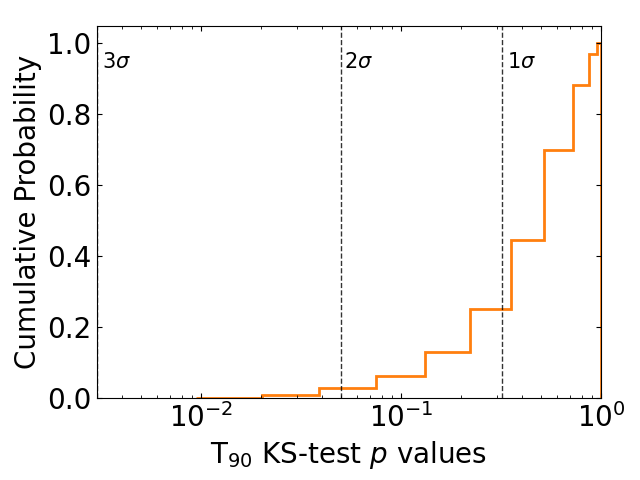}
	\includegraphics[width=0.3\linewidth]{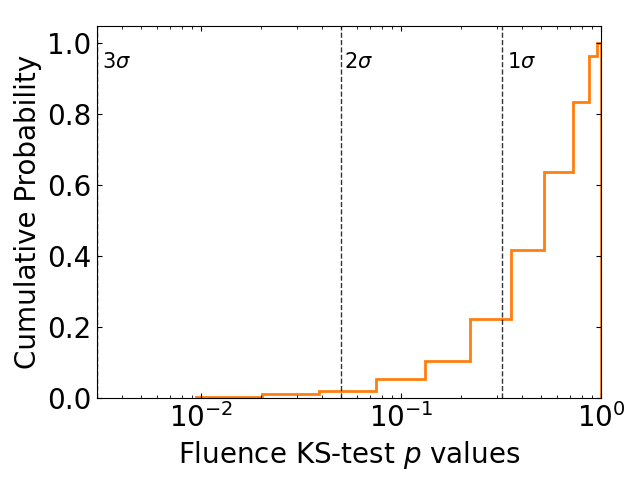}
	\includegraphics[width=0.3\linewidth]{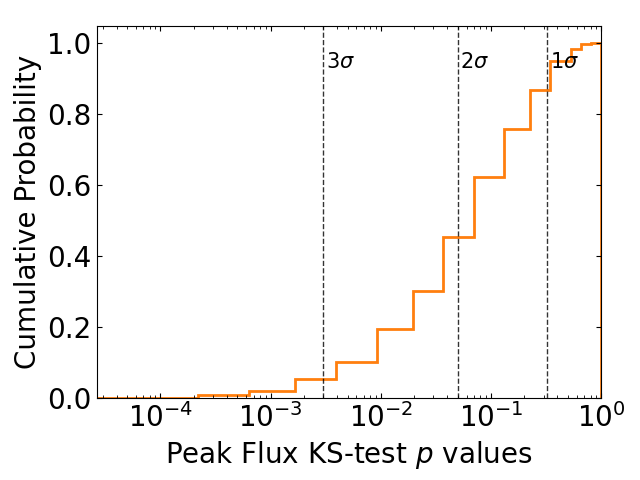}
	\caption{Same description as for Figure \ref{fig: cum dists}, but only using high-$z$ bursts observed by Swift/BAT before January 1, 2012.}
	\label{fig: cum dists cut}
\end{figure*}

The pre-2012 observations and simulation measurements are in a much stronger agreement than the total observed sample (see Fig. \ref{fig: cum dists cut} and Tab. \ref{tab:p_vals_cut}). Both the duration and fluence distributions find $\sim75\%$ of the random subsamples to be within 1$\sigma$ of the pre-2012 observations. The peak-flux measurements have also become more consistent, i.e., $15\%$ ($94.1\%$) of the random subsamples are within 1$\sigma$ (3$\sigma$). The agreement between the distributions across each metric implies that the pre-2012 GRB measurements and our mock GRB measurements are not inconsistent with coming from the same underlying distributions. 

One possible source of this difference is our requirement that all bursts in both the low-$z$ and high-$z$ samples have observed peak energies constrained to be $E_p\leq550$ keV. For a burst at a redshift of $z=0.1$, this observational limit would correspond to a rest-frame $E_{p,0} = 605$ keV, while for a burst at $z=6$, this would correspond to $E_{p,0} = 3,850$ keV. According to the empirical Yonetoku relation, there exists a positive relation between a GRB's observed luminosity and its peak energy \citep{2004ApJ...609..935Y}. So, because of the observer-frame $E_p$ limit, our simulated high-$z$ sample will likely have lower peak energies and, therefore, possibly have lower luminosities compared to the observed high-$z$ sample (which is demonstrated in Fig. \ref{fig: cum dists} \textit{Top Right}). However, this effect is not time dependent and cannot explain why the simulations are in such stronger agreement with the pre-2012 sample of observed bursts than the post-2012 sample.

The fact that the low-$z$ and pre-2012 high-$z$ samples are consistent simply by taking into account cosmological effects and proper observational biases indicates that (i) the significant underestimations of duration and fluence measurements at high $z$ that are found in the simulations should also be assumed to exist in observed high-$z$ GRBs and (ii) no additional physics is needed to explain differences in low-$z$ and high-$z$ GRB prompt duration and fluence measurements.

Since we find that the tip-of-the-iceberg effect leads to dim emission becoming undetectable for our mock GRBs located at high redshifts, we should also expect high-$z$ GRBs to experience the same effect. Consequently, we should not expect the average durations of GRBs to evolve exactly according to cosmological effects. These findings are in agreement with the observed average durations of Swift/BAT GRBs (see Fig. \ref{fig: grb-t90-vs-z}). Furthermore, in some cases we found that the measured burst fluence was a factor $\gtrsim 2$ lower than expected from simple distance corrections. These differences imply there may be a systematic error when estimating the isotropic gamma-ray energy values, $E_{\gamma, iso}$, of GRBs and the distribution of $E_{\gamma, iso}$ may extend to higher energies than previously thought \citep{2008ApJ...680..531K}.

\section{Conclusion} \label{sec: Conclusion}

In this work, we took a sample of bright GRBs observed at redshifts $z<1$ and simulated them out to $z>3$. We then compared these simulated GRBs to a sample of GRBs observed at $z>3$. From this work we found that:
\begin{enumerate}
	\item due to the tip-of-the-iceberg effect, the duration and fluence measurements of the simulated GRBs are underestimations of their true durations. Sometimes leading to T$_{90}$ measurements that are an order of magnitude below the intrinsic duration.
	\item This bias is also true for fluence measurements to a lesser degree, but can be up to a factor of $\sim2$ at high $z$ for some bursts.
	\item This same effect can occasionally turn long GRBs into short GRBs. 
	\item Lastly, the simulated high-$z$ sample is consistent with being drawn from the same population as the observed high-$z$ sample, implying that observed high-$z$ GRBs also suffer from underestimated duration and fluence measurements \textbf{that are found to occur in our simulations.}
\end{enumerate}

In contrast to our results, \citet{2013MNRAS.436.3640L} found a $\sim1\%$ probability that the high-$z$ GRB sample observed by Swift/BAT is drawn from the same population as the bright low-$z$ sample, implying their simulated and observed samples are, at best, marginally consistent. One possibility is the difference in selected samples. \citet{2013MNRAS.436.3640L} used a sample of 114 low-redshift GRBs for which the \citet{2010MNRAS.403.1296W} pulse-fitting methodology had been applied, instead of the bright sample of bursts used in this work. Furthermore, there are differences in the methodologies used to perform the GRB simulations that could have lead to the inconsistency between the results. For example, \citet{2013MNRAS.436.3640L} simulated a sample of bursts with measured redshifts $z<4$ out to a single redshift value equal to that of the average redshift of their high-$z$ sample, i.e., $\bar{z}_{high} = 7.66$. Further comparisons between the two methodologies will be needed to determine the cause of the inconsistent results.

Over its lifetime, Swift's average pointing interval has decreased, i.e., from $\sim$990 sec in 2005 to $\sim$650 sec in 2025. Recent studies of a potential class of ultra-long GRBs (i.e., GRBs with T$_{90}$s of a few thousand seconds or more; \citealt{2011A&A...528A..15G,2013ApJ...766...30G,2014ApJ...781...13L,2020ExA....50...91D}) have proposed that Swift's pointing strategy may bias against the discovery of GRBs with durations similar to or longer that the average pointing interval. From this work, we find that typical LGRBs are not observed as ultra-long GRBs at high redshifts. The only burst in our sample with simulation measurements of T$_{90}>1,000$ seconds at $z=9$ is GRB 130427A, which was an exceptionally bright GRB observed at a redshift of $z=0.34$ and shows significant emission in BAT $>900$ seconds after the initial trigger time. Therefore, if a population of ultra-long GRBs does exist at moderate- to high- redshifts, it would require an evolution in the $\gamma$-ray emission duration with redshift or, potentially, a unique progenitor system only present at those redshifts.

Studies of the tip-of-the-iceberg effect have investigated its effect on various duration measurement algorithm, e.g., Bayesian blocks, T$_{90}$, and T$_{50}$ methods. Similar to the results of this work, it has been found that each duration estimator is affected by underestimation biases. Duration estimation algorithms that attempt to estimate the total length of emission, e.g., the Bayesian blocks and T$_{100}$ methods, tend to be more strongly affected that more conservative duration estimations, e.g., the T$_{50}$ \citep{1997ApJ...490...79B,2002A&A...393L..29M,2013ApJ...765..116K}. Using alternative duration measurement techniques, like only measuring the T$_{90}$ of each isolated pulse in a GRB light curve, may lead to more consistent estimates of central engine lifetimes.

Our results imply that the measured duration of gamma-ray emission is not always an accurate estimation of the lifetime of GRB central engine activity due to instrumental bias (a result similarly found by \citealt{2020MNRAS.496.2910P}). This claim is further supported by recent observations made by the Einstein Probe (EP), which has seen the X-ray emission of some GRBs lasting much longer than the gamma-ray emission observed by gamma-ray detectors such as Swift/BAT and Konus/Wind \citep{2025NatAs.tmp...34L,2025ApJ...988L..34J,2025arXiv250925877L}. The difference in these duration measurements may not be physical but, instead, may simply be a reflection of the fact that the EP is a more sensitive all-sky monitor compared to our current soft gamma-ray instruments (see Extended Data Fig. 1 in \citet{2025NatAs.tmp...34L}; \citealt{2022hxga.book...86Y}). More sensitive all-sky monitors that can detect the onset of the soft gamma-ray emission will be needed to truly constrain the lifetime of GRB central engines.

\begin{acknowledgments}
MM’s research was supported by an appointment to the NASA Postdoctoral Program at the NASA Goddard Space Flight Center, administered by Oak Ridge Associated Universities under contract with NASA. 
\end{acknowledgments}

\facilities{Swift(BAT)}

\software{numpy \citep{harris2020array},
			astropy \citep{2013A&A...558A..33A,2018AJ....156..123A,2022ApJ...935..167A},
			scipy \citep{2020SciPy-NMeth}
			}

\section*{Data Availability}

All Swift/BAT GRB observations are publicly available in an online catalog \url{https://swift.gsfc.nasa.gov/results/batgrbcat/}. The data were promptly and automatically analyzed using the publicly available HEASOFT and XSPEC tools. The analysis and simulations performed in this work were completed using the publicly available \texttt{simmes} package built upon on the Numpy, Scipy, Astropy and Fitsio packages and is available for download at \url{https://github.com/mikemoss3/simmes} or installed via \texttt{pip}. The data is available in a machine readable format alongside the article. 


\bibliographystyle{aasjournalv7}
\bibliography{bibliography}

@ARTICLE{1976PhFl...19.1130B,
       author = {{Blandford}, R.~D. and {McKee}, C.~F.},
        title = "{Fluid dynamics of relativistic blast waves}",
      journal = {Physics of Fluids},
         year = 1976,
        month = aug,
       volume = {19},
        pages = {1130-1138},
          doi = {10.1063/1.861619},
       adsurl = {https://ui.adsabs.harvard.edu/abs/1976PhFl...19.1130B}
}

@ARTICLE{1978MNRAS.183..359C,
       author = {{Cavallo}, G. and {Rees}, M.~J.},
        title = "{A qualitative study of cosmic fireballs and gamma -ray bursts.}",
      journal = {\mnras},
         year = 1978,
        month = may,
       volume = {183},
        pages = {359-365},
          doi = {10.1093/mnras/183.3.359},
       adsurl = {https://ui.adsabs.harvard.edu/abs/1978MNRAS.183..359C}
}

@ARTICLE{1991Natur.350..592M,
       author = {{Murakami}, T. and {Inoue}, H. and {Nishimura}, J. and {van Paradijs}, J. and {Fenimore}, E.~E.},
        title = "{A {\ensuremath{\gamma}}-ray burst preceded by X-ray activity}",
      journal = {\nat},
         year = 1991,
        month = apr,
       volume = {350},
       number = {6319},
        pages = {592-594},
          doi = {10.1038/350592a0},
       adsurl = {https://ui.adsabs.harvard.edu/abs/1991Natur.350..592M}
}

@ARTICLE{1992MNRAS.258P..41R,
       author = {{Rees}, M.~J. and {Meszaros}, P.},
        title = "{Relativistic fireballs - Energy conversion and time-scales.}",
      journal = {\mnras},
         year = 1992,
        month = sep,
       volume = {258},
        pages = {41},
          doi = {10.1093/mnras/258.1.41P},
       adsurl = {https://ui.adsabs.harvard.edu/abs/1992MNRAS.258P..41R}
}

@ARTICLE{1993ApJ...413L.101K,
       author = {{Kouveliotou}, Chryssa and {Meegan}, Charles A. and {Fishman}, Gerald J. and {Bhat}, Narayana P. and {Briggs}, Michael S. and {Koshut}, Thomas M. and {Paciesas}, William S. and {Pendleton}, Geoffrey N.},
        title = "{Identification of Two Classes of Gamma-Ray Bursts}",
      journal = {\apjl},
         year = 1993,
        month = aug,
       volume = {413},
        pages = {L101},
          doi = {10.1086/186969},
       adsurl = {https://ui.adsabs.harvard.edu/abs/1993ApJ...413L.101K}
}

@ARTICLE{1995ApJ...448L.101F,
       author = {{Fenimore}, E.~E. and {in 't Zand}, J.~J.~M. and {Norris}, J.~P. and {Bonnell}, J.~T. and {Nemiroff}, R.~J.},
        title = "{Gamma-Ray Burst Peak Duration as a Function of Energy}",
      journal = {\apjl},
         year = 1995,
        month = aug,
       volume = {448},
        pages = {L101},
          doi = {10.1086/309603},
archivePrefix = {arXiv},
       eprint = {astro-ph/9504075},
 primaryClass = {astro-ph},
       adsurl = {https://ui.adsabs.harvard.edu/abs/1995ApJ...448L.101F}
}

@ARTICLE{1995ApJ...452..145K,
       author = {{Koshut}, Thomas M. and {Kouveliotou}, Chryssa and {Paciesas}, William S. and {van Paradijs}, Jan and {Pendleton}, Geoffrey N. and {Briggs}, Michael S. and {Fishman}, Gerald J. and {Meegan}, Charles A.},
        title = "{Gamma-Ray Burst Precursor Activity as Observed with BATSE}",
      journal = {\apj},
         year = 1995,
        month = oct,
       volume = {452},
        pages = {145},
          doi = {10.1086/176286},
       adsurl = {https://ui.adsabs.harvard.edu/abs/1995ApJ...452..145K}
}

@ARTICLE{1997MNRAS.288L..51W,
       author = {{Wijers}, Ralph A.~M.~J. and {Rees}, Martin J. and {Meszaros}, Peter},
        title = "{Shocked by GRB 970228: the afterglow of a cosmological fireball}",
      journal = {\mnras},
         year = 1997,
        month = jul,
       volume = {288},
       number = {4},
        pages = {L51-L56},
          doi = {10.1093/mnras/288.4.L51},
archivePrefix = {arXiv},
       eprint = {astro-ph/9704153},
 primaryClass = {astro-ph},
       adsurl = {https://ui.adsabs.harvard.edu/abs/1997MNRAS.288L..51W}
}

@ARTICLE{1997ApJ...486L..71T,
       author = {{Totani}, Tomonori},
        title = "{Cosmological Gamma-Ray Bursts and Evolution of Galaxies}",
      journal = {\apjl},
         year = 1997,
        month = sep,
       volume = {486},
       number = {2},
        pages = {L71-L74},
          doi = {10.1086/310853},
archivePrefix = {arXiv},
       eprint = {astro-ph/9707051},
 primaryClass = {astro-ph},
       adsurl = {https://ui.adsabs.harvard.edu/abs/1997ApJ...486L..71T}
}

@ARTICLE{1997ApJ...490...79B,
       author = {{Bonnell}, J.~T. and {Norris}, J.~P. and {Nemiroff}, R.~J. and {Scargle}, J.~D.},
        title = "{Brightness-independent Measurements of Gamma-Ray Burst Durations}",
      journal = {\apj},
         year = 1997,
        month = nov,
       volume = {490},
       number = {1},
        pages = {79-91},
          doi = {10.1086/304841},
       adsurl = {https://ui.adsabs.harvard.edu/abs/1997ApJ...490...79B}
}

@ARTICLE{1998MNRAS.294L..13W,
       author = {{Wijers}, Ralph A.~M.~J. and {Bloom}, Joshua S. and {Bagla}, Jasjeet S. and {Natarajan}, Priyamvada},
        title = "{Gamma-ray bursts from stellar remnants: probing the Universe at high redshift}",
      journal = {\mnras},
         year = 1998,
        month = feb,
       volume = {294},
       number = {1},
        pages = {L13-L17},
          doi = {10.1046/j.1365-8711.1998.01328.x10.1111/j.1365-8711.1998.01328.x},
archivePrefix = {arXiv},
       eprint = {astro-ph/9708183},
 primaryClass = {astro-ph},
       adsurl = {https://ui.adsabs.harvard.edu/abs/1998MNRAS.294L..13W}
}

@ARTICLE{1998ApJ...497L..17S,
       author = {{Sari}, Re'em and {Piran}, Tsvi and {Narayan}, Ramesh},
        title = "{Spectra and Light Curves of Gamma-Ray Burst Afterglows}",
      journal = {\apjl},
         year = 1998,
        month = apr,
       volume = {497},
       number = {1},
        pages = {L17-L20},
          doi = {10.1086/311269},
archivePrefix = {arXiv},
       eprint = {astro-ph/9712005},
 primaryClass = {astro-ph},
       adsurl = {https://ui.adsabs.harvard.edu/abs/1998ApJ...497L..17S}
}

@ARTICLE{1998ApJ...501...15M,
       author = {{Miralda-Escud{\'e}}, Jordi},
        title = "{Reionization of the Intergalactic Medium and the Damping Wing of the Gunn-Peterson Trough}",
      journal = {\apj},
         year = 1998,
        month = jul,
       volume = {501},
       number = {1},
        pages = {15-22},
          doi = {10.1086/305799},
archivePrefix = {arXiv},
       eprint = {astro-ph/9708253},
 primaryClass = {astro-ph},
       adsurl = {https://ui.adsabs.harvard.edu/abs/1998ApJ...501...15M}
}

@ARTICLE{1998Natur.395..670G,
       author = {{Galama}, T.~J. and {Vreeswijk}, P.~M. and {van Paradijs}, J. and {Kouveliotou}, C. and {Augusteijn}, T. and {B{\"o}hnhardt}, H. and {Brewer}, J.~P. and {Doublier}, V. and {Gonzalez}, J. -F. and {Leibundgut}, B. and {Lidman}, C. and {Hainaut}, O.~R. and {Patat}, F. and {Heise}, J. and {in't Zand}, J. and {Hurley}, K. and {Groot}, P.~J. and {Strom}, R.~G. and {Mazzali}, P.~A. and {Iwamoto}, K. and {Nomoto}, K. and {Umeda}, H. and {Nakamura}, T. and {Young}, T.~R. and {Suzuki}, T. and {Shigeyama}, T. and {Koshut}, T. and {Kippen}, M. and {Robinson}, C. and {de Wildt}, P. and {Wijers}, R.~A.~M.~J. and {Tanvir}, N. and {Greiner}, J. and {Pian}, E. and {Palazzi}, E. and {Frontera}, F. and {Masetti}, N. and {Nicastro}, L. and {Feroci}, M. and {Costa}, E. and {Piro}, L. and {Peterson}, B.~A. and {Tinney}, C. and {Boyle}, B. and {Cannon}, R. and {Stathakis}, R. and {Sadler}, E. and {Begam}, M.~C. and {Ianna}, P.},
        title = "{An unusual supernova in the error box of the {\ensuremath{\gamma}}-ray burst of 25 April 1998}",
      journal = {\nat},
         year = 1998,
        month = oct,
       volume = {395},
       number = {6703},
        pages = {670-672},
          doi = {10.1038/27150},
archivePrefix = {arXiv},
       eprint = {astro-ph/9806175},
 primaryClass = {astro-ph},
       adsurl = {https://ui.adsabs.harvard.edu/abs/1998Natur.395..670G}
}

@ARTICLE{1998A&A...339L...1M,
       author = {{Mao}, Shude and {Mo}, H.~J.},
        title = "{The nature of the host galaxies for gamma-ray bursts}",
      journal = {\aap},
         year = 1998,
        month = nov,
       volume = {339},
        pages = {L1-L4},
          doi = {10.48550/arXiv.astro-ph/9808342},
archivePrefix = {arXiv},
       eprint = {astro-ph/9808342},
 primaryClass = {astro-ph},
       adsurl = {https://ui.adsabs.harvard.edu/abs/1998A&A...339L...1M}
}

@ARTICLE{1999A&AS..138..465G,
       author = {{Galama}, T.~J. and {Vreeswijk}, P.~M. and {van Paradijs}, J. and {Kouveliotou}, C. and {Augusteijn}, T. and {Patat}, F. and {Heise}, J. and {in 't Zand}, J. and {Groot}, P.~J. and {Wijers}, R.~A.~M.~J. and {Pian}, E. and {Palazzi}, E. and {Frontera}, F. and {Masetti}, N.},
        title = "{On the possible association of SN 1998bw and GRB 980425}",
      journal = {\aaps},
         year = 1999,
        month = sep,
       volume = {138},
        pages = {465-466},
          doi = {10.1051/aas:1999311},
       adsurl = {https://ui.adsabs.harvard.edu/abs/1999A&AS..138..465G}
}

@ARTICLE{2000ApJ...537..191F,
       author = {{Frail}, D.~A. and {Waxman}, E. and {Kulkarni}, S.~R.},
        title = "{A 450 Day Light Curve of the Radio Afterglow of GRB 970508: Fireball Calorimetry}",
      journal = {\apj},
         year = 2000,
        month = jul,
       volume = {537},
       number = {1},
        pages = {191-204},
          doi = {10.1086/309024},
archivePrefix = {arXiv},
       eprint = {astro-ph/9910319},
 primaryClass = {astro-ph},
       adsurl = {https://ui.adsabs.harvard.edu/abs/2000ApJ...537..191F}
}

@ARTICLE{2001ApJ...548..522P,
       author = {{Porciani}, Cristiano and {Madau}, Piero},
        title = "{On the Association of Gamma-Ray Bursts with Massive Stars: Implications for Number Counts and Lensing Statistics}",
      journal = {\apj},
         year = 2001,
        month = feb,
       volume = {548},
       number = {2},
        pages = {522-531},
          doi = {10.1086/319027},
archivePrefix = {arXiv},
       eprint = {astro-ph/0008294},
 primaryClass = {astro-ph},
       adsurl = {https://ui.adsabs.harvard.edu/abs/2001ApJ...548..522P}
}

@ARTICLE{2001AJ....121.2879B,
       author = {{Bloom}, Joshua S. and {Frail}, Dale A. and {Sari}, Re'em},
        title = "{The Prompt Energy Release of Gamma-Ray Bursts using a Cosmological k-Correction}",
      journal = {\aj},
         year = 2001,
        month = jun,
       volume = {121},
       number = {6},
        pages = {2879-2888},
          doi = {10.1086/321093},
archivePrefix = {arXiv},
       eprint = {astro-ph/0102371},
 primaryClass = {astro-ph},
       adsurl = {https://ui.adsabs.harvard.edu/abs/2001AJ....121.2879B}
}

@ARTICLE{2002A&A...393L..29M,
       author = {{McBreen}, S. and {McBreen}, B. and {Hanlon}, L. and {Quilligan}, F.},
        title = "{Cumulative light curves of gamma-ray bursts and relaxation systems}",
      journal = {\aap},
         year = 2002,
        month = oct,
       volume = {393},
        pages = {L29-L32},
          doi = {10.1051/0004-6361:20021073},
archivePrefix = {arXiv},
       eprint = {astro-ph/0208347},
 primaryClass = {astro-ph},
       adsurl = {https://ui.adsabs.harvard.edu/abs/2002A&A...393L..29M}
}

@ARTICLE{2003Natur.423..847H,
       author = {{Hjorth}, Jens and {Sollerman}, Jesper and {M{\o}ller}, Palle and {Fynbo}, Johan P.~U. and {Woosley}, Stan E. and {Kouveliotou}, Chryssa and {Tanvir}, Nial R. and {Greiner}, Jochen and {Andersen}, Michael I. and {Castro-Tirado}, Alberto J. and {Castro Cer{\'o}n}, Jos{\'e} Mar{\'\i}a and {Fruchter}, Andrew S. and {Gorosabel}, Javier and {Jakobsson}, P{\'a}ll and {Kaper}, Lex and {Klose}, Sylvio and {Masetti}, Nicola and {Pedersen}, Holger and {Pedersen}, Kristian and {Pian}, Elena and {Palazzi}, Eliana and {Rhoads}, James E. and {Rol}, Evert and {van den Heuvel}, Edward P.~J. and {Vreeswijk}, Paul M. and {Watson}, Darach and {Wijers}, Ralph A.~M.~J.},
        title = "{A very energetic supernova associated with the {\ensuremath{\gamma}}-ray burst of 29 March 2003}",
      journal = {\nat},
         year = 2003,
        month = jun,
       volume = {423},
       number = {6942},
        pages = {847-850},
          doi = {10.1038/nature01750},
archivePrefix = {arXiv},
       eprint = {astro-ph/0306347},
 primaryClass = {astro-ph},
       adsurl = {https://ui.adsabs.harvard.edu/abs/2003Natur.423..847H}
}

@ARTICLE{2003ApJ...591L..17S,
       author = {{Stanek}, K.~Z. and {Matheson}, T. and {Garnavich}, P.~M. and {Martini}, P. and {Berlind}, P. and {Caldwell}, N. and {Challis}, P. and {Brown}, W.~R. and {Schild}, R. and {Krisciunas}, K. and {Calkins}, M.~L. and {Lee}, J.~C. and {Hathi}, N. and {Jansen}, R.~A. and {Windhorst}, R. and {Echevarria}, L. and {Eisenstein}, D.~J. and {Pindor}, B. and {Olszewski}, E.~W. and {Harding}, P. and {Holland}, S.~T. and {Bersier}, D.},
        title = "{Spectroscopic Discovery of the Supernova 2003dh Associated with GRB 030329}",
      journal = {\apjl},
         year = 2003,
        month = jul,
       volume = {591},
       number = {1},
        pages = {L17-L20},
          doi = {10.1086/376976},
archivePrefix = {arXiv},
       eprint = {astro-ph/0304173},
 primaryClass = {astro-ph},
       adsurl = {https://ui.adsabs.harvard.edu/abs/2003ApJ...591L..17S}
}

@ARTICLE{2003ApJ...596..389K,
       author = {{Kocevski}, Dan and {Ryde}, Felix and {Liang}, Edison},
        title = "{Search for Relativistic Curvature Effects in Gamma-Ray Burst Pulses}",
      journal = {\apj},
         year = 2003,
        month = oct,
       volume = {596},
       number = {1},
        pages = {389-400},
          doi = {10.1086/377707},
archivePrefix = {arXiv},
       eprint = {astro-ph/0303556},
 primaryClass = {astro-ph},
       adsurl = {https://ui.adsabs.harvard.edu/abs/2003ApJ...596..389K}
}

@ARTICLE{2004ApJ...609..935Y,
       author = {{Yonetoku}, D. and {Murakami}, T. and {Nakamura}, T. and {Yamazaki}, R. and {Inoue}, A.~K. and {Ioka}, K.},
        title = "{Gamma-Ray Burst Formation Rate Inferred from the Spectral Peak Energy-Peak Luminosity Relation}",
      journal = {\apj},
         year = 2004,
        month = jul,
       volume = {609},
       number = {2},
        pages = {935-951},
          doi = {10.1086/421285},
archivePrefix = {arXiv},
       eprint = {astro-ph/0309217},
 primaryClass = {astro-ph},
       adsurl = {https://ui.adsabs.harvard.edu/abs/2004ApJ...609..935Y}
}

@ARTICLE{2005SSRv..120..143B,
       author = {{Barthelmy}, Scott D. and {Barbier}, Louis M. and {Cummings}, Jay R. and {Fenimore}, Ed E. and {Gehrels}, Neil and {Hullinger}, Derek and {Krimm}, Hans A. and {Markwardt}, Craig B. and {Palmer}, David M. and {Parsons}, Ann and {Sato}, Goro and {Suzuki}, Masaya and {Takahashi}, Tadayuki and {Tashiro}, Makota and {Tueller}, Jack},
        title = "{The Burst Alert Telescope (BAT) on the SWIFT Midex Mission}",
      journal = {\ssr},
         year = 2005,
        month = oct,
       volume = {120},
       number = {3-4},
        pages = {143-164},
          doi = {10.1007/s11214-005-5096-3},
archivePrefix = {arXiv},
       eprint = {astro-ph/0507410},
 primaryClass = {astro-ph},
       adsurl = {https://ui.adsabs.harvard.edu/abs/2005SSRv..120..143B}
}

@ARTICLE{2006PASJ...58..485T,
       author = {{Totani}, Tomonori and {Kawai}, Nobuyuki and {Kosugi}, George and {Aoki}, Kentaro and {Yamada}, Toru and {Iye}, Masanori and {Ohta}, Kouji and {Hattori}, Takashi},
        title = "{Implications for Cosmic Reionization from the Optical Afterglow Spectrum of the Gamma-Ray Burst 050904 at z = 6.3$^{*}$}",
      journal = {\pasj},
         year = 2006,
        month = jun,
       volume = {58},
       number = {3},
        pages = {485-498},
          doi = {10.1093/pasj/58.3.485},
archivePrefix = {arXiv},
       eprint = {astro-ph/0512154},
 primaryClass = {astro-ph},
       adsurl = {https://ui.adsabs.harvard.edu/abs/2006PASJ...58..485T}
}

@ARTICLE{2008ApOpt..47.2739S,
       author = {{Skinner}, Gerald K.},
        title = "{Sensitivity of coded mask telescopes}",
      journal = {Applied Optics},
         year = 2008,
        month = may,
       volume = {47},
       number = {15},
        pages = {2739-2749},
          doi = {10.1364/AO.47.002739},
archivePrefix = {arXiv},
       eprint = {0804.1089},
 primaryClass = {astro-ph},
       adsurl = {https://ui.adsabs.harvard.edu/abs/2008ApOpt..47.2739S}
}

@ARTICLE{2008ApJ...680..531K,
       author = {{Kocevski}, Daniel and {Butler}, Nathaniel},
        title = "{Gamma-Ray Burst Energetics in the Swift Era}",
      journal = {\apj},
         year = 2008,
        month = jun,
       volume = {680},
       number = {1},
        pages = {531-538},
          doi = {10.1086/586693},
archivePrefix = {arXiv},
       eprint = {0707.4478},
 primaryClass = {astro-ph},
       adsurl = {https://ui.adsabs.harvard.edu/abs/2008ApJ...680..531K}
}

@ARTICLE{2009GCN..9281....1U,
       author = {{Ukwatta}, T.~N. and {Barthelmy}, S.~D. and {Evans}, P.~A. and {Gehrels}, N. and {Markwardt}, C.~B. and {Page}, K.~L. and {Palmer}, D.~M. and {Rowlinson}, B.~A. and {Siegel}, M.~H. and {Stamatikos}, M. and {Vetere}, L.},
        title = "{GRB 090429B: Swift detection of a burst.}",
      journal = {GRB Coordinates Network},
         year = 2009,
        month = jan,
       volume = {9281},
        pages = {1},
       adsurl = {https://ui.adsabs.harvard.edu/abs/2009GCN..9281....1U}
}

@ARTICLE{2009ApJ...701..824N,
       author = {{Nysewander}, M. and {Fruchter}, A.~S. and {Pe'er}, A.},
        title = "{A Comparison of the Afterglows of Short- and Long-duration Gamma-ray Bursts}",
      journal = {\apj},
         year = 2009,
        month = aug,
       volume = {701},
       number = {1},
        pages = {824-836},
          doi = {10.1088/0004-637X/701/1/824},
archivePrefix = {arXiv},
       eprint = {0806.3607},
 primaryClass = {astro-ph},
       adsurl = {https://ui.adsabs.harvard.edu/abs/2009ApJ...701..824N}
}

@ARTICLE{2010MNRAS.403.1296W,
       author = {{Willingale}, R. and {Genet}, F. and {Granot}, J. and {O'Brien}, P.~T.},
        title = "{The spectral-temporal properties of the prompt pulses and rapid decay phase of gamma-ray bursts}",
      journal = {\mnras},
         year = 2010,
        month = apr,
       volume = {403},
       number = {3},
        pages = {1296-1316},
          doi = {10.1111/j.1365-2966.2009.16187.x},
archivePrefix = {arXiv},
       eprint = {0912.1759},
 primaryClass = {astro-ph.HE},
       adsurl = {https://ui.adsabs.harvard.edu/abs/2010MNRAS.403.1296W}
}

@ARTICLE{2010ApJ...723.1711T,
       author = {{Troja}, E. and {Rosswog}, S. and {Gehrels}, N.},
        title = "{Precursors of Short Gamma-ray Bursts}",
      journal = {\apj},
         year = 2010,
        month = nov,
       volume = {723},
       number = {2},
        pages = {1711-1717},
          doi = {10.1088/0004-637X/723/2/1711},
archivePrefix = {arXiv},
       eprint = {1009.1385},
 primaryClass = {astro-ph.HE},
       adsurl = {https://ui.adsabs.harvard.edu/abs/2010ApJ...723.1711T}
}

@ARTICLE{2011A&A...528A..15G,
       author = {{Gruber}, D. and {Kr{\"u}hler}, T. and {Foley}, S. and {Nardini}, M. and {Burlon}, D. and {Rau}, A. and {Bissaldi}, E. and {von Kienlin}, A. and {McBreen}, S. and {Greiner}, J. and et al.},
        title = "{Fermi/GBM observations of the ultra-long GRB 091024. A burst with an optical flash}",
      journal = {\aap},
         year = 2011,
        month = apr,
       volume = {528},
          eid = {A15},
        pages = {A15},
          doi = {10.1051/0004-6361/201015891},
archivePrefix = {arXiv},
       eprint = {1101.1099},
 primaryClass = {astro-ph.HE},
       adsurl = {https://ui.adsabs.harvard.edu/abs/2011A&A...528A..15G}
}

@ARTICLE{2011ApJ...736....7C,
       author = {{Cucchiara}, A. and {Levan}, A.~J. and {Fox}, D.~B. and {Tanvir}, N.~R. and {Ukwatta}, T.~N. and {Berger}, E. and {Kr{\"u}hler}, T. and {K{\"u}pc{\"u} Yolda{\c{s}}}, A. and {Wu}, X.~F. and {Toma}, K. and {Greiner}, J. and {Olivares}, F.~E. and {Rowlinson}, A. and {Amati}, L. and {Sakamoto}, T. and {Roth}, K. and {Stephens}, A. and {Fritz}, Alexander and {Fynbo}, J.~P.~U. and {Hjorth}, J. and {Malesani}, D. and {Jakobsson}, P. and {Wiersema}, K. and {O'Brien}, P.~T. and {Soderberg}, A.~M. and {Foley}, R.~J. and {Fruchter}, A.~S. and {Rhoads}, J. and {Rutledge}, R.~E. and {Schmidt}, B.~P. and {Dopita}, M.~A. and {Podsiadlowski}, P. and {Willingale}, R. and {Wolf}, C. and {Kulkarni}, S.~R. and {D'Avanzo}, P.},
        title = "{A Photometric Redshift of z \raisebox{-0.5ex}\textasciitilde 9.4 for GRB 090429B}",
      journal = {\apj},
         year = 2011,
        month = jul,
       volume = {736},
       number = {1},
          eid = {7},
        pages = {7},
          doi = {10.1088/0004-637X/736/1/7},
archivePrefix = {arXiv},
       eprint = {1105.4915},
 primaryClass = {astro-ph.CO},
       adsurl = {https://ui.adsabs.harvard.edu/abs/2011ApJ...736....7C}
}

@ARTICLE{2012ApJ...749...68S,
       author = {{Salvaterra}, R. and {Campana}, S. and {Vergani}, S.~D. and {Covino}, S. and {D'Avanzo}, P. and {Fugazza}, D. and {Ghirlanda}, G. and {Ghisellini}, G. and {Melandri}, A. and {Nava}, L. and {Sbarufatti}, B. and {Flores}, H. and {Piranomonte}, S. and {Tagliaferri}, G.},
        title = "{A Complete Sample of Bright Swift Long Gamma-Ray Bursts. I. Sample Presentation, Luminosity Function and Evolution}",
      journal = {\apj},
         year = 2012,
        month = apr,
       volume = {749},
       number = {1},
          eid = {68},
        pages = {68},
          doi = {10.1088/0004-637X/749/1/68},
archivePrefix = {arXiv},
       eprint = {1112.1700},
 primaryClass = {astro-ph.CO},
       adsurl = {https://ui.adsabs.harvard.edu/abs/2012ApJ...749...68S}
}

@ARTICLE{2012ApJ...749..110B,
       author = {{Bromberg}, Omer and {Nakar}, Ehud and {Piran}, Tsvi and {Sari}, Re'em},
        title = "{An Observational Imprint of the Collapsar Model of Long Gamma-Ray Bursts}",
      journal = {\apj},
         year = 2012,
        month = apr,
       volume = {749},
       number = {2},
          eid = {110},
        pages = {110},
          doi = {10.1088/0004-637X/749/2/110},
archivePrefix = {arXiv},
       eprint = {1111.2990},
 primaryClass = {astro-ph.HE},
       adsurl = {https://ui.adsabs.harvard.edu/abs/2012ApJ...749..110B}
}

@ARTICLE{2013ApJ...764..167S,
       author = {{Scargle}, Jeffrey D. and {Norris}, Jay P. and {Jackson}, Brad and {Chiang}, James},
        title = "{Studies in Astronomical Time Series Analysis. VI. Bayesian Block Representations}",
      journal = {\apj},
         year = 2013,
        month = feb,
       volume = {764},
       number = {2},
          eid = {167},
        pages = {167},
          doi = {10.1088/0004-637X/764/2/167},
archivePrefix = {arXiv},
       eprint = {1207.5578},
 primaryClass = {astro-ph.IM},
       adsurl = {https://ui.adsabs.harvard.edu/abs/2013ApJ...764..167S}
}

@ARTICLE{2013ApJ...765..116K,
       author = {{Kocevski}, Daniel and {Petrosian}, Vahe},
        title = "{On the Lack of Time Dilation Signatures in Gamma-Ray Burst Light Curves}",
      journal = {\apj},
         year = 2013,
        month = mar,
       volume = {765},
       number = {2},
          eid = {116},
        pages = {116},
          doi = {10.1088/0004-637X/765/2/116},
       adsurl = {https://ui.adsabs.harvard.edu/abs/2013ApJ...765..116K}
}

@ARTICLE{2013ApJ...766...30G,
       author = {{Gendre}, B. and {Stratta}, G. and {Atteia}, J.~L. and {Basa}, S. and {Bo{\"e}r}, M. and {Coward}, D.~M. and {Cutini}, S. and {D'Elia}, V. and {Howell}, E.~J. and {Klotz}, A. and et al.},
        title = "{The Ultra-long Gamma-Ray Burst 111209A: The Collapse of a Blue Supergiant?}",
      journal = {\apj},
         year = 2013,
        month = mar,
       volume = {766},
       number = {1},
          eid = {30},
        pages = {30},
          doi = {10.1088/0004-637X/766/1/30},
archivePrefix = {arXiv},
       eprint = {1212.2392},
 primaryClass = {astro-ph.HE},
       adsurl = {https://ui.adsabs.harvard.edu/abs/2013ApJ...766...30G}
}

@ARTICLE{2013Natur.500..547T,
       author = {{Tanvir}, N.~R. and {Levan}, A.~J. and {Fruchter}, A.~S. and {Hjorth}, J. and {Hounsell}, R.~A. and {Wiersema}, K. and {Tunnicliffe}, R.~L.},
        title = "{A `kilonova' associated with the short-duration {\ensuremath{\gamma}}-ray burst GRB 130603B}",
      journal = {\nat},
         year = 2013,
        month = aug,
       volume = {500},
       number = {7464},
        pages = {547-549},
          doi = {10.1038/nature12505},
archivePrefix = {arXiv},
       eprint = {1306.4971},
 primaryClass = {astro-ph.HE},
       adsurl = {https://ui.adsabs.harvard.edu/abs/2013Natur.500..547T}
}

@ARTICLE{2013ApJS..207...19B,
       author = {{Baumgartner}, W.~H. and {Tueller}, J. and {Markwardt}, C.~B. and {Skinner}, G.~K. and {Barthelmy}, S. and {Mushotzky}, R.~F. and {Evans}, P.~A. and {Gehrels}, N.},
        title = "{The 70 Month Swift-BAT All-sky Hard X-Ray Survey}",
      journal = {\apjs},
         year = 2013,
        month = aug,
       volume = {207},
       number = {2},
          eid = {19},
        pages = {19},
          doi = {10.1088/0067-0049/207/2/19},
archivePrefix = {arXiv},
       eprint = {1212.3336},
 primaryClass = {astro-ph.HE},
       adsurl = {https://ui.adsabs.harvard.edu/abs/2013ApJS..207...19B}
}

@ARTICLE{2013A&A...558A..33A,
       author = {{Astropy Collaboration} and {Robitaille}, Thomas P. and {Tollerud}, Erik J. and {Greenfield}, Perry and {Droettboom}, Michael and {Bray}, Erik and {Aldcroft}, Tom and {Davis}, Matt and {Ginsburg}, Adam and {Price-Whelan}, Adrian M. and et al.},
        title = "{Astropy: A community Python package for astronomy}",
      journal = {\aap},
         year = 2013,
        month = oct,
       volume = {558},
          eid = {A33},
        pages = {A33},
          doi = {10.1051/0004-6361/201322068},
archivePrefix = {arXiv},
       eprint = {1307.6212},
 primaryClass = {astro-ph.IM},
       adsurl = {https://ui.adsabs.harvard.edu/abs/2013A&A...558A..33A}
}

@ARTICLE{2013MNRAS.436.3640L,
       author = {{Littlejohns}, O.~M. and {Tanvir}, N.~R. and {Willingale}, R. and {Evans}, P.~A. and {O'Brien}, P.~T. and {Levan}, A.~J.},
        title = "{Are gamma-ray bursts the same at high redshift and low redshift?}",
      journal = {\mnras},
         year = 2013,
        month = dec,
       volume = {436},
       number = {4},
        pages = {3640-3655},
          doi = {10.1093/mnras/stt1841},
archivePrefix = {arXiv},
       eprint = {1309.7045},
 primaryClass = {astro-ph.HE},
       adsurl = {https://ui.adsabs.harvard.edu/abs/2013MNRAS.436.3640L}
}

@ARTICLE{2014ApJ...781...13L,
       author = {{Levan}, A.~J. and {Tanvir}, N.~R. and {Starling}, R.~L.~C. and {Wiersema}, K. and {Page}, K.~L. and {Perley}, D.~A. and {Schulze}, S. and {Wynn}, G.~A. and {Chornock}, R. and {Hjorth}, J. and {Cenko}, S.~B. and {Fruchter}, A.~S. and {O'Brien}, P.~T. and {Brown}, G.~C. and {Tunnicliffe}, R.~L. and {Malesani}, D. and {Jakobsson}, P. and {Watson}, D. and {Berger}, E. and {Bersier}, D. and {Cobb}, B.~E. and {Covino}, S. and {Cucchiara}, A. and {de Ugarte Postigo}, A. and {Fox}, D.~B. and {Gal-Yam}, A. and {Goldoni}, P. and {Gorosabel}, J. and {Kaper}, L. and {Kr{\"u}hler}, T. and {Karjalainen}, R. and {Osborne}, J.~P. and {Pian}, E. and {S{\'a}nchez-Ram{\'\i}rez}, R. and {Schmidt}, B. and {Skillen}, I. and {Tagliaferri}, G. and {Th{\"o}ne}, C. and {Vaduvescu}, O. and {Wijers}, R.~A.~M.~J. and {Zauderer}, B.~A.},
        title = "{A New Population of Ultra-long Duration Gamma-Ray Bursts}",
      journal = {\apj},
         year = 2014,
        month = jan,
       volume = {781},
       number = {1},
          eid = {13},
        pages = {13},
          doi = {10.1088/0004-637X/781/1/13},
archivePrefix = {arXiv},
       eprint = {1302.2352},
 primaryClass = {astro-ph.HE},
       adsurl = {https://ui.adsabs.harvard.edu/abs/2014ApJ...781...13L}
}

@ARTICLE{2014IJMPD..2330002Z,
       author = {{Zhang}, Bing},
        title = "{Gamma-Ray Burst Prompt Emission}",
      journal = {International Journal of Modern Physics D},
         year = 2014,
        month = dec,
       volume = {23},
       number = {2},
          eid = {1430002},
        pages = {1430002},
          doi = {10.1142/S021827181430002X},
archivePrefix = {arXiv},
       eprint = {1402.7022},
 primaryClass = {astro-ph.HE},
       adsurl = {https://ui.adsabs.harvard.edu/abs/2014IJMPD..2330002Z}
}

@ARTICLE{2016ApJ...829....7L,
       author = {{Lien}, Amy and {Sakamoto}, Takanori and {Barthelmy}, Scott D. and {Baumgartner}, Wayne H. and {Cannizzo}, John K. and {Chen}, Kevin and {Collins}, Nicholas R. and {Cummings}, Jay R. and {Gehrels}, Neil and {Krimm}, Hans A. and et al.},
        title = "{The Third Swift Burst Alert Telescope Gamma-Ray Burst Catalog}",
      journal = {\apj},
         year = 2016,
        month = sep,
       volume = {829},
       number = {1},
          eid = {7},
        pages = {7},
          doi = {10.3847/0004-637X/829/1/7},
archivePrefix = {arXiv},
       eprint = {1606.01956},
 primaryClass = {astro-ph.HE},
       adsurl = {https://ui.adsabs.harvard.edu/abs/2016ApJ...829....7L}
}

@ARTICLE{2017ApJ...848L..13A,
       author = {{Abbott}, B.~P. and {Abbott}, R. and {Abbott}, T.~D. and {Acernese}, F. and {Ackley}, K. and {Adams}, C. and {Adams}, T. and {Addesso}, P. and {Adhikari}, R.~X. and {Adya}, V.~B. and {Affeldt}, C. and {Afrough}, M. and {Agarwal}, B. and {Agathos}, M. and {Agatsuma}, K. and {Aggarwal}, N. and {Aguiar}, O.~D. and {Aiello}, L. and {Ain}, A. and {Ajith}, P. and {Allen}, B. and {Allen}, G. and {Allocca}, A. and {Aloy}, M.~A. and {Altin}, P.~A. and {Amato}, A. and {Ananyeva}, A. and {Anderson}, S.~B. and {Anderson}, W.~G. and {Angelova}, S.~V. and {Antier}, S. and {Appert}, S. and {Arai}, K. and {Araya}, M.~C. and {Areeda}, J.~S. and {Arnaud}, N. and {Arun}, K.~G. and {Ascenzi}, S. and {Ashton}, G. and {Ast}, M. and {Aston}, S.~M. and {Astone}, P. and {Atallah}, D.~V. and {Aufmuth}, P. and {Aulbert}, C. and {AultONeal}, K. and {Austin}, C. and {Avila-Alvarez}, A. and {Babak}, S. and {Bacon}, P. and {Bader}, M.~K.~M. and {Bae}, S. and {Baker}, P.~T. and {Baldaccini}, F. and {Ballardin}, G. and {Ballmer}, S.~W. and {Banagiri}, S. and {Barayoga}, J.~C. and {Barclay}, S.~E. and {Barish}, B.~C. and {Barker}, D. and {Barkett}, K. and {Barone}, F. and {Barr}, B. and {Barsotti}, L. and {Barsuglia}, M. and {Barta}, D. and {Bartlett}, J. and {Bartos}, I. and {Bassiri}, R. and {Basti}, A. and {Batch}, J.~C. and {Bawaj}, M. and {Bayley}, J.~C. and {Bazzan}, M. and {B{\'e}csy}, B. and {Beer}, C. and {Bejger}, M. and {Belahcene}, I. and {Bell}, A.~S. and {Berger}, B.~K. and {Bergmann}, G. and {Bero}, J.~J. and {Berry}, C.~P.~L. and {Bersanetti}, D. and {Bertolini}, A. and {Betzwieser}, J. and {Bhagwat}, S. and {Bhandare}, R. and {Bilenko}, I.~A. and {Billingsley}, G. and {Billman}, C.~R. and {Birch}, J. and {Birney}, R. and {Birnholtz}, O. and {Biscans}, S. and {Biscoveanu}, S. and {Bisht}, A. and {Bitossi}, M. and {Biwer}, C. and {Bizouard}, M.~A. and {Blackburn}, J.~K. and {Blackman}, J. and {Blair}, C.~D. and {Blair}, D.~G. and {Blair}, R.~M. and {Bloemen}, S. and {Bock}, O. and {Bode}, N. and {Boer}, M. and {Bogaert}, G. and {Bohe}, A. and {Bondu}, F. and {Bonilla}, E. and {Bonnand}, R. and {Boom}, B.~A. and {Bork}, R. and {Boschi}, V. and {Bose}, S. and {Bossie}, K. and {Bouffanais}, Y. and {Bozzi}, A. and {Bradaschia}, C. and {Brady}, P.~R. and {Branchesi}, M. and {Brau}, J.~E. and {Briant}, T. and {Brillet}, A. and {Brinkmann}, M. and {Brisson}, V. and {Brockill}, P. and {Broida}, J.~E. and {Brooks}, A.~F. and {Brown}, D.~A. and {Brown}, D.~D. and {Brunett}, S. and {Buchanan}, C.~C. and {Buikema}, A. and {Bulik}, T. and {Bulten}, H.~J. and {Buonanno}, A. and {Buskulic}, D. and {Buy}, C. and {Byer}, R.~L. and {Cabero}, M. and {Cadonati}, L. and {Cagnoli}, G. and {Cahillane}, C. and {Calder{\'o}n Bustillo}, J. and {Callister}, T.~A. and {Calloni}, E. and {Camp}, J.~B. and {Canepa}, M. and {Canizares}, P. and {Cannon}, K.~C. and {Cao}, H. and {Cao}, J. and {Capano}, C.~D. and {Capocasa}, E. and {Carbognani}, F. and {Caride}, S. and {Carney}, M.~F. and {Casanueva Diaz}, J. and {Casentini}, C. and {Caudill}, S. and {Cavagli{\`a}}, M. and {Cavalier}, F. and {Cavalieri}, R. and {Cella}, G. and {Cepeda}, C.~B. and {Cerd{\'a}-Dur{\'a}n}, P. and {Cerretani}, G. and {Cesarini}, E. and {Chamberlin}, S.~J. and {Chan}, M. and {Chao}, S. and {Charlton}, P. and {Chase}, E. and {Chassande-Mottin}, E. and {Chatterjee}, D. and {Chatziioannou}, K. and {Cheeseboro}, B.~D. and {Chen}, H.~Y. and {Chen}, X. and {Chen}, Y. and {Cheng}, H. -P. and {Chia}, H. and {Chincarini}, A. and {Chiummo}, A. and {Chmiel}, T. and {Cho}, H.~S. and {Cho}, M. and {Chow}, J.~H. and {Christensen}, N. and {Chu}, Q. and {Chua}, A.~J.~K. and {Chua}, S. and {Chung}, A.~K.~W. and {Chung}, S. and {Ciani}, G. and {Ciolfi}, R. and {Cirelli}, C.~E. and {Cirone}, A. and {Clara}, F. and {Clark}, J.~A. and {Clearwater}, P. and {Cleva}, F. and {Cocchieri}, C. and {Coccia}, E. and {Cohadon}, P. -F. and {Cohen}, D. and {Colla}, A. and {Collette}, C.~G. and {Cominsky}, L.~R. and {Constancio}, M., Jr. and {Conti}, L. and {Cooper}, S.~J. and {Corban}, P. and {Corbitt}, T.~R. and {Cordero-Carri{\'o}n}, I. and {Corley}, K.~R. and {Cornish}, N. and {Corsi}, A. and {Cortese}, S. and {Costa}, C.~A. and {Coughlin}, M.~W. and {Coughlin}, S.~B. and {Coulon}, J. -P. and {Countryman}, S.~T. and {Couvares}, P. and {Covas}, P.~B. and {Cowan}, E.~E. and {Coward}, D.~M. and {Cowart}, M.~J. and {Coyne}, D.~C. and {Coyne}, R. and {Creighton}, J.~D.~E. and {Creighton}, T.~D. and {Cripe}, J. and {Crowder}, S.~G. and {Cullen}, T.~J. and {Cumming}, A. and {Cunningham}, L. and {Cuoco}, E. and {Dal Canton}, T. and {D{\'a}lya}, G. and {Danilishin}, S.~L. and {D'Antonio}, S. and {Danzmann}, K. and {Dasgupta}, A. and {Da Silva Costa}, C.~F. and {Dattilo}, V. and {Dave}, I. and {Davier}, M. and {Davis}, D. and {Daw}, E.~J. and {Day}, B. and {De}, S. and {DeBra}, D. and {Degallaix}, J. and {De Laurentis}, M. and {Del{\'e}glise}, S. and {Del Pozzo}, W. and {Demos}, N. and {Denker}, T. and {Dent}, T. and {De Pietri}, R. and {Dergachev}, V. and {De Rosa}, R. and {DeRosa}, R.~T. and {De Rossi}, C. and {DeSalvo}, R. and {de Varona}, O. and {Devenson}, J. and {Dhurandhar}, S. and {D{\'\i}az}, M.~C. and {Di Fiore}, L. and {Di Giovanni}, M. and {Di Girolamo}, T. and {Di Lieto}, A. and {Di Pace}, S. and {Di Palma}, I. and {Di Renzo}, F. and {Doctor}, Z. and {Dolique}, V. and {Donovan}, F. and {Dooley}, K.~L. and {Doravari}, S. and {Dorrington}, I. and {Douglas}, R. and {Dovale {\'A}lvarez}, M. and {Downes}, T.~P. and {Drago}, M. and {Dreissigacker}, C. and {Driggers}, J.~C. and {Du}, Z. and {Ducrot}, M. and {Dupej}, P. and {Dwyer}, S.~E. and {Edo}, T.~B. and {Edwards}, M.~C. and {Effler}, A. and {Eggenstein}, H. -B. and {Ehrens}, P. and {Eichholz}, J. and {Eikenberry}, S.~S. and {Eisenstein}, R.~A. and {Essick}, R.~C. and {Estevez}, D. and {Etienne}, Z.~B. and {Etzel}, T. and {Evans}, M. and {Evans}, T.~M. and {Factourovich}, M. and {Fafone}, V. and {Fair}, H. and {Fairhurst}, S. and {Fan}, X. and {Farinon}, S. and {Farr}, B. and {Farr}, W.~M. and {Fauchon-Jones}, E.~J. and {Favata}, M. and {Fays}, M. and {Fee}, C. and {Fehrmann}, H. and {Feicht}, J. and {Fejer}, M.~M. and {Fernandez-Galiana}, A. and {Ferrante}, I. and {Ferreira}, E.~C. and {Ferrini}, F. and {Fidecaro}, F. and {Finstad}, D. and {Fiori}, I. and {Fiorucci}, D. and {Fishbach}, M. and {Fisher}, R.~P. and {Fitz-Axen}, M. and {Flaminio}, R. and {Fletcher}, M. and {Fong}, H. and {Font}, J.~A. and {Forsyth}, P.~W.~F. and {Forsyth}, S.~S. and {Fournier}, J. -D. and {Frasca}, S. and {Frasconi}, F. and {Frei}, Z. and {Freise}, A. and {Frey}, R. and {Frey}, V. and {Fries}, E.~M. and {Fritschel}, P. and {Frolov}, V.~V. and {Fulda}, P. and {Fyffe}, M. and {Gabbard}, H. and {Gadre}, B.~U. and {Gaebel}, S.~M. and {Gair}, J.~R. and {Gammaitoni}, L. and {Ganija}, M.~R. and {Gaonkar}, S.~G. and {Garcia-Quiros}, C. and {Garufi}, F. and {Gateley}, B. and {Gaudio}, S. and {Gaur}, G. and {Gayathri}, V. and {Gehrels}, N. and {Gemme}, G. and {Genin}, E. and {Gennai}, A. and {George}, D. and {George}, J. and {Gergely}, L. and {Germain}, V. and {Ghonge}, S. and {Ghosh}, Abhirup and {Ghosh}, Archisman and {Ghosh}, S. and {Giaime}, J.~A. and {Giardina}, K.~D. and {Giazotto}, A. and {Gill}, K. and {Glover}, L. and {Goetz}, E. and {Goetz}, R. and {Gomes}, S. and {Goncharov}, B. and {Gonz{\'a}lez}, G. and {Gonzalez Castro}, J.~M. and {Gopakumar}, A. and {Gorodetsky}, M.~L. and {Gossan}, S.~E. and {Gosselin}, M. and {Gouaty}, R. and {Grado}, A. and {Graef}, C. and {Granata}, M. and {Grant}, A. and {Gras}, S. and {Gray}, C. and {Greco}, G. and {Green}, A.~C. and {Gretarsson}, E.~M. and {Groot}, P. and {Grote}, H. and {Grunewald}, S. and {Gruning}, P. and {Guidi}, G.~M. and {Guo}, X. and {Gupta}, A. and {Gupta}, M.~K. and {Gushwa}, K.~E. and {Gustafson}, E.~K. and {Gustafson}, R. and {Halim}, O. and {Hall}, B.~R. and {Hall}, E.~D. and {Hamilton}, E.~Z. and {Hammond}, G. and {Haney}, M. and {Hanke}, M.~M. and {Hanks}, J. and {Hanna}, C. and {Hannam}, M.~D. and {Hannuksela}, O.~A. and {Hanson}, J. and {Hardwick}, T. and {Harms}, J. and {Harry}, G.~M. and {Harry}, I.~W. and {Hart}, M.~J. and {Haster}, C. -J. and {Haughian}, K. and {Healy}, J. and {Heidmann}, A. and {Heintze}, M.~C. and {Heitmann}, H. and {Hello}, P. and {Hemming}, G. and {Hendry}, M. and {Heng}, I.~S. and {Hennig}, J. and {Heptonstall}, A.~W. and {Heurs}, M. and {Hild}, S. and {Hinderer}, T. and {Hoak}, D. and {Hofman}, D. and {Holt}, K. and {Holz}, D.~E. and {Hopkins}, P. and {Horst}, C. and {Hough}, J. and {Houston}, E.~A. and {Howell}, E.~J. and {Hreibi}, A. and {Hu}, Y.~M. and {Huerta}, E.~A. and {Huet}, D. and {Hughey}, B. and {Husa}, S. and {Huttner}, S.~H. and {Huynh-Dinh}, T. and {Indik}, N. and {Inta}, R. and {Intini}, G. and {Isa}, H.~N. and {Isac}, J. -M. and {Isi}, M. and {Iyer}, B.~R. and {Izumi}, K. and {Jacqmin}, T. and {Jani}, K. and {Jaranowski}, P. and {Jawahar}, S. and {Jim{\'e}nez-Forteza}, F. and {Johnson}, W.~W. and {Johnson-McDaniel}, N.~K. and {Jones}, D.~I. and {Jones}, R. and {Jonker}, R.~J.~G. and {Ju}, L. and {Junker}, J. and {Kalaghatgi}, C.~V. and {Kalogera}, V. and {Kamai}, B. and {Kandhasamy}, S. and {Kang}, G. and {Kanner}, J.~B. and {Kapadia}, S.~J. and {Karki}, S. and {Karvinen}, K.~S. and {Kasprzack}, M. and {Kastaun}, W. and {Katolik}, M. and {Katsavounidis}, E. and {Katzman}, W. and {Kaufer}, S. and {Kawabe}, K. and {K{\'e}f{\'e}lian}, F. and {Keitel}, D. and {Kemball}, A.~J. and {Kennedy}, R. and {Kent}, C. and {Key}, J.~S. and {Khalili}, F.~Y. and {Khan}, I. and {Khan}, S. and {Khan}, Z. and {Khazanov}, E.~A. and {Kijbunchoo}, N. and {Kim}, Chunglee and {Kim}, J.~C. and {Kim}, K. and {Kim}, W. and {Kim}, W.~S. and {Kim}, Y. -M. and {Kimbrell}, S.~J. and {King}, E.~J. and {King}, P.~J. and {Kinley-Hanlon}, M. and {Kirchhoff}, R. and {Kissel}, J.~S. and {Kleybolte}, L. and {Klimenko}, S. and {Knowles}, T.~D. and {Koch}, P. and {Koehlenbeck}, S.~M. and {Koley}, S. and {Kondrashov}, V. and {Kontos}, A. and {Korobko}, M. and {Korth}, W.~Z. and {Kowalska}, I. and {Kozak}, D.~B. and {Kr{\"a}mer}, C. and {Kringel}, V. and {Krishnan}, B. and {Kr{\'o}lak}, A. and {Kuehn}, G. and {Kumar}, P. and {Kumar}, R. and {Kumar}, S. and {Kuo}, L. and {Kutynia}, A. and {Kwang}, S. and {Lackey}, B.~D. and {Lai}, K.~H. and {Landry}, M. and {Lang}, R.~N. and {Lange}, J. and {Lantz}, B. and {Lanza}, R.~K. and {Lartaux-Vollard}, A. and {Lasky}, P.~D. and {Laxen}, M. and {Lazzarini}, A. and {Lazzaro}, C. and {Leaci}, P. and {Leavey}, S. and {Lee}, C.~H. and {Lee}, H.~K. and {Lee}, H.~M. and {Lee}, H.~W. and {Lee}, K. and {Lehmann}, J. and {Lenon}, A. and {Leonardi}, M. and {Leroy}, N. and {Letendre}, N. and {Levin}, Y. and {Li}, T.~G.~F. and {Linker}, S.~D. and {Littenberg}, T.~B. and {Liu}, J. and {Lo}, R.~K.~L. and {Lockerbie}, N.~A. and {London}, L.~T. and {Lord}, J.~E. and {Lorenzini}, M. and {Loriette}, V. and {Lormand}, M. and {Losurdo}, G. and {Lough}, J.~D. and {Lousto}, C.~O. and {Lovelace}, G. and {L{\"u}ck}, H. and {Lumaca}, D. and {Lundgren}, A.~P. and {Lynch}, R. and {Ma}, Y. and {Macas}, R. and {Macfoy}, S. and {Machenschalk}, B. and {MacInnis}, M. and {Macleod}, D.~M. and {Maga{\~n}a Hernandez}, I. and {Maga{\~n}a-Sandoval}, F. and {Maga{\~n}a Zertuche}, L. and {Magee}, R.~M. and {Majorana}, E. and {Maksimovic}, I. and {Man}, N. and {Mandic}, V. and {Mangano}, V. and {Mansell}, G.~L. and {Manske}, M. and {Mantovani}, M. and {Marchesoni}, F. and {Marion}, F. and {M{\'a}rka}, S. and {M{\'a}rka}, Z. and {Markakis}, C. and {Markosyan}, A.~S. and {Markowitz}, A. and {Maros}, E. and {Marquina}, A. and {Martelli}, F. and {Martellini}, L. and {Martin}, I.~W. and {Martin}, R.~M. and {Martynov}, D.~V. and {Mason}, K. and {Massera}, E. and {Masserot}, A. and {Massinger}, T.~J. and {Masso-Reid}, M. and {Mastrogiovanni}, S. and {Matas}, A. and {Matichard}, F. and {Matone}, L. and {Mavalvala}, N. and {Mazumder}, N. and {McCarthy}, R. and {McClelland}, D.~E. and {McCormick}, S. and {McCuller}, L. and {McGuire}, S.~C. and {McIntyre}, G. and {McIver}, J. and {McManus}, D.~J. and {McNeill}, L. and {McRae}, T. and {McWilliams}, S.~T. and {Meacher}, D. and {Meadors}, G.~D. and {Mehmet}, M. and {Meidam}, J. and {Mejuto-Villa}, E. and {Melatos}, A. and {Mendell}, G. and {Mercer}, R.~A. and {Merilh}, E.~L. and {Merzougui}, M. and {Meshkov}, S. and {Messenger}, C. and {Messick}, C. and {Metzdorff}, R. and {Meyers}, P.~M. and {Miao}, H. and {Michel}, C. and {Middleton}, H. and {Mikhailov}, E.~E. and {Milano}, L. and {Miller}, A.~L. and {Miller}, B.~B. and {Miller}, J. and {Millhouse}, M. and {Milovich-Goff}, M.~C. and {Minazzoli}, O. and {Minenkov}, Y. and {Ming}, J. and {Mishra}, C. and {Mitra}, S. and {Mitrofanov}, V.~P. and {Mitselmakher}, G. and {Mittleman}, R. and {Moffa}, D. and {Moggi}, A. and {Mogushi}, K. and {Mohan}, M. and {Mohapatra}, S.~R.~P. and {Montani}, M. and {Moore}, C.~J. and {Moraru}, D. and {Moreno}, G. and {Morriss}, S.~R. and {Mours}, B. and {Mow-Lowry}, C.~M. and {Mueller}, G. and {Muir}, A.~W. and {Mukherjee}, Arunava and {Mukherjee}, D. and {Mukherjee}, S. and {Mukund}, N. and {Mullavey}, A. and {Munch}, J. and {Mu{\~n}iz}, E.~A. and {Muratore}, M. and {Murray}, P.~G. and {Napier}, K. and {Nardecchia}, I. and {Naticchioni}, L. and {Nayak}, R.~K. and {Neilson}, J. and {Nelemans}, G. and {Nelson}, T.~J.~N. and {Nery}, M. and {Neunzert}, A. and {Nevin}, L. and {Newport}, J.~M. and {Newton}, G. and {Ng}, K.~K.~Y. and {Nguyen}, T.~T. and {Nichols}, D. and {Nielsen}, A.~B. and {Nissanke}, S. and {Nitz}, A. and {Noack}, A. and {Nocera}, F. and {Nolting}, D. and {North}, C. and {Nuttall}, L.~K. and {Oberling}, J. and {O'Dea}, G.~D. and {Ogin}, G.~H. and {Oh}, J.~J. and {Oh}, S.~H. and {Ohme}, F. and {Okada}, M.~A. and {Oliver}, M. and {Oppermann}, P. and {Oram}, Richard J. and {O'Reilly}, B. and {Ormiston}, R. and {Ortega}, L.~F. and {O'Shaughnessy}, R. and {Ossokine}, S. and {Ottaway}, D.~J. and {Overmier}, H. and {Owen}, B.~J. and {Pace}, A.~E. and {Page}, J. and {Page}, M.~A. and {Pai}, A. and {Pai}, S.~A. and {Palamos}, J.~R. and {Palashov}, O. and {Palomba}, C. and {Pal-Singh}, A. and {Pan}, Howard and {Pan}, Huang-Wei and {Pang}, B. and {Pang}, P.~T.~H. and {Pankow}, C. and {Pannarale}, F. and {Pant}, B.~C. and {Paoletti}, F. and {Paoli}, A. and {Papa}, M.~A. and {Parida}, A. and {Parker}, W. and {Pascucci}, D. and {Pasqualetti}, A. and {Passaquieti}, R. and {Passuello}, D. and {Patil}, M. and {Patricelli}, B. and {Pearlstone}, B.~L. and {Pedraza}, M. and {Pedurand}, R. and {Pekowsky}, L. and {Pele}, A. and {Penn}, S. and {Perez}, C.~J. and {Perreca}, A. and {Perri}, L.~M. and {Pfeiffer}, H.~P. and {Phelps}, M. and {Piccinni}, O.~J. and {Pichot}, M. and {Piergiovanni}, F. and {Pierro}, V. and {Pillant}, G. and {Pinard}, L. and {Pinto}, I.~M. and {Pirello}, M. and {Pitkin}, M. and {Poe}, M. and {Poggiani}, R. and {Popolizio}, P. and {Porter}, E.~K. and {Post}, A. and {Powell}, J. and {Prasad}, J. and {Pratt}, J.~W.~W. and {Pratten}, G. and {Predoi}, V. and {Prestegard}, T. and {Prijatelj}, M. and {Principe}, M. and {Privitera}, S. and {Prodi}, G.~A. and {Prokhorov}, L.~G. and {Puncken}, O. and {Punturo}, M. and {Puppo}, P. and {P{\"u}rrer}, M. and {Qi}, H. and {Quetschke}, V. and {Quintero}, E.~A. and {Quitzow-James}, R. and {Raab}, F.~J. and {Rabeling}, D.~S. and {Radkins}, H. and {Raffai}, P. and {Raja}, S. and {Rajan}, C. and {Rajbhandari}, B. and {Rakhmanov}, M. and {Ramirez}, K.~E. and {Ramos-Buades}, A. and {Rapagnani}, P. and {Raymond}, V. and {Razzano}, M. and {Read}, J. and {Regimbau}, T. and {Rei}, L. and {Reid}, S. and {Reitze}, D.~H. and {Ren}, W. and {Reyes}, S.~D. and {Ricci}, F. and {Ricker}, P.~M. and {Rieger}, S. and {Riles}, K. and {Rizzo}, M. and {Robertson}, N.~A. and {Robie}, R. and {Robinet}, F. and {Rocchi}, A. and {Rolland}, L. and {Rollins}, J.~G. and {Roma}, V.~J. and {Romano}, R. and {Romel}, C.~L. and {Romie}, J.~H. and {Rosi{\'n}ska}, D. and {Ross}, M.~P. and {Rowan}, S. and {R{\"u}diger}, A. and {Ruggi}, P. and {Rutins}, G. and {Ryan}, K. and {Sachdev}, S. and {Sadecki}, T. and {Sadeghian}, L. and {Sakellariadou}, M. and {Salconi}, L. and {Saleem}, M. and {Salemi}, F. and {Samajdar}, A. and {Sammut}, L. and {Sampson}, L.~M. and {Sanchez}, E.~J. and {Sanchez}, L.~E. and {Sanchis-Gual}, N. and {Sandberg}, V. and {Sanders}, J.~R. and {Sassolas}, B. and {Sathyaprakash}, B.~S. and {Saulson}, P.~R. and {Sauter}, O. and {Savage}, R.~L. and {Sawadsky}, A. and {Schale}, P. and {Scheel}, M. and {Scheuer}, J. and {Schmidt}, J. and {Schmidt}, P. and {Schnabel}, R. and {Schofield}, R.~M.~S. and {Sch{\"o}nbeck}, A. and {Schreiber}, E. and {Schuette}, D. and {Schulte}, B.~W. and {Schutz}, B.~F. and {Schwalbe}, S.~G. and {Scott}, J. and {Scott}, S.~M. and {Seidel}, E. and {Sellers}, D. and {Sengupta}, A.~S. and {Sentenac}, D. and {Sequino}, V. and {Sergeev}, A. and {Shaddock}, D.~A. and {Shaffer}, T.~J. and {Shah}, A.~A. and {Shahriar}, M.~S. and {Shaner}, M.~B. and {Shao}, L. and {Shapiro}, B. and {Shawhan}, P. and {Sheperd}, A. and {Shoemaker}, D.~H. and {Shoemaker}, D.~M. and {Siellez}, K. and {Siemens}, X. and {Sieniawska}, M. and {Sigg}, D. and {Silva}, A.~D. and {Singer}, L.~P. and {Singh}, A. and {Singhal}, A. and {Sintes}, A.~M. and {Slagmolen}, B.~J.~J. and {Smith}, B. and {Smith}, J.~R. and {Smith}, R.~J.~E. and {Somala}, S. and {Son}, E.~J. and {Sonnenberg}, J.~A. and {Sorazu}, B. and {Sorrentino}, F. and {Souradeep}, T. and {Spencer}, A.~P. and {Srivastava}, A.~K. and {Staats}, K. and {Staley}, A. and {Steinke}, M. and {Steinlechner}, J. and {Steinlechner}, S. and {Steinmeyer}, D. and {Stevenson}, S.~P. and {Stone}, R. and {Stops}, D.~J. and {Strain}, K.~A. and {Stratta}, G. and {Strigin}, S.~E. and {Strunk}, A. and {Sturani}, R. and {Stuver}, A.~L. and {Summerscales}, T.~Z. and {Sun}, L. and {Sunil}, S. and {Suresh}, J. and {Sutton}, P.~J. and {Swinkels}, B.~L. and {Szczepa{\'n}czyk}, M.~J. and {Tacca}, M. and {Tait}, S.~C. and {Talbot}, C. and {Talukder}, D. and {Tanner}, D.~B. and {T{\'a}pai}, M. and {Taracchini}, A. and {Tasson}, J.~D. and {Taylor}, J.~A. and {Taylor}, R. and {Tewari}, S.~V. and {Theeg}, T. and {Thies}, F. and {Thomas}, E.~G. and {Thomas}, M. and {Thomas}, P. and {Thorne}, K.~A. and {Thorne}, K.~S. and {Thrane}, E. and {Tiwari}, S. and {Tiwari}, V. and {Tokmakov}, K.~V. and {Toland}, K. and {Tonelli}, M. and {Tornasi}, Z. and {Torres-Forn{\'e}}, A. and {Torrie}, C.~I. and {T{\"o}yr{\"a}}, D. and {Travasso}, F. and {Traylor}, G. and {Trinastic}, J. and {Tringali}, M.~C. and {Trozzo}, L. and {Tsang}, K.~W. and {Tse}, M. and {Tso}, R. and {Tsukada}, L. and {Tsuna}, D. and {Tuyenbayev}, D. and {Ueno}, K. and {Ugolini}, D. and {Unnikrishnan}, C.~S. and {Urban}, A.~L. and {Usman}, S.~A. and {Vahlbruch}, H. and {Vajente}, G. and {Valdes}, G. and {van Bakel}, N. and {van Beuzekom}, M. and {van den Brand}, J.~F.~J. and {Van Den Broeck}, C. and {Vander-Hyde}, D.~C. and {van der Schaaf}, L. and {van Heijningen}, J.~V. and {van Veggel}, A.~A. and {Vardaro}, M. and {Varma}, V. and {Vass}, S. and {Vas{\'u}th}, M. and {Vecchio}, A. and {Vedovato}, G. and {Veitch}, J. and {Veitch}, P.~J. and {Venkateswara}, K. and {Venugopalan}, G. and {Verkindt}, D. and {Vetrano}, F. and {Vicer{\'e}}, A. and {Viets}, A.~D. and {Vinciguerra}, S. and {Vine}, D.~J. and {Vinet}, J. -Y. and {Vitale}, S. and {Vo}, T. and {Vocca}, H. and {Vorvick}, C. and {Vyatchanin}, S.~P. and {Wade}, A.~R. and {Wade}, L.~E. and {Wade}, M. and {Walet}, R. and {Walker}, M. and {Wallace}, L. and {Walsh}, S. and {Wang}, G. and {Wang}, H. and {Wang}, J.~Z. and {Wang}, W.~H. and {Wang}, Y.~F. and {Ward}, R.~L. and {Warner}, J. and {Was}, M. and {Watchi}, J. and {Weaver}, B. and {Wei}, L. -W. and {Weinert}, M. and {Weinstein}, A.~J. and {Weiss}, R. and {Wen}, L. and {Wessel}, E.~K. and {We{\ss}els}, P. and {Westerweck}, J. and {Westphal}, T. and {Wette}, K. and {Whelan}, J.~T. and {Whitcomb}, S.~E. and {Whiting}, B.~F. and {Whittle}, C. and {Wilken}, D. and {Williams}, D. and {Williams}, R.~D. and {Williamson}, A.~R. and {Willis}, J.~L. and {Willke}, B. and {Wimmer}, M.~H. and {Winkler}, W. and {Wipf}, C.~C. and {Wittel}, H. and {Woan}, G. and {Woehler}, J. and {Wofford}, J. and {Wong}, K.~W.~K. and {Worden}, J. and {Wright}, J.~L. and {Wu}, D.~S. and {Wysocki}, D.~M. and {Xiao}, S. and {Yamamoto}, H. and {Yancey}, C.~C. and {Yang}, L. and {Yap}, M.~J. and {Yazback}, M. and {Yu}, Hang and {Yu}, Haocun and {Yvert}, M. and {Zadro{\.z}ny}, A. and {Zanolin}, M. and {Zelenova}, T. and {Zendri}, J. -P. and {Zevin}, M. and {Zhang}, L. and {Zhang}, M. and {Zhang}, T. and {Zhang}, Y. -H. and {Zhao}, C. and {Zhou}, M. and {Zhou}, Z. and {Zhu}, S.~J. and {Zhu}, X.~J. and {Zimmerman}, A.~B. and {Zucker}, M.~E. and {Zweizig}, J. and {(LIGO Scientific Collaboration} and {Virgo Collaboration} and {Burns}, E. and {Veres}, P. and {Kocevski}, D. and {Racusin}, J. and {Goldstein}, A. and {Connaughton}, V. and {Briggs}, M.~S. and {Blackburn}, L. and {Hamburg}, R. and {Hui}, C.~M. and {von Kienlin}, A. and {McEnery}, J. and {Preece}, R.~D. and {Wilson-Hodge}, C.~A. and {Bissaldi}, E. and {Cleveland}, W.~H. and {Gibby}, M.~H. and {Giles}, M.~M. and {Kippen}, R.~M. and {McBreen}, S. and {Meegan}, C.~A. and {Paciesas}, W.~S. and {Poolakkil}, S. and {Roberts}, O.~J. and {Stanbro}, M. and {Gamma-ray Burst Monitor}, (Fermi and {Savchenko}, V. and {Ferrigno}, C. and {Kuulkers}, E. and {Bazzano}, A. and {Bozzo}, E. and {Brandt}, S. and {Chenevez}, J. and {Courvoisier}, T.~J. -L. and {Diehl}, R. and {Domingo}, A. and {Hanlon}, L. and {Jourdain}, E. and {Laurent}, P. and {Lebrun}, F. and {Lutovinov}, A. and {Mereghetti}, S. and {Natalucci}, L. and {Rodi}, J. and {Roques}, J. -P. and {Sunyaev}, R. and {Ubertini}, P. and {(INTEGRAL}},
        title = "{Gravitational Waves and Gamma-Rays from a Binary Neutron Star Merger: GW170817 and GRB 170817A}",
      journal = {\apjl},
         year = 2017,
        month = oct,
       volume = {848},
       number = {2},
          eid = {L13},
        pages = {L13},
          doi = {10.3847/2041-8213/aa920c},
archivePrefix = {arXiv},
       eprint = {1710.05834},
 primaryClass = {astro-ph.HE},
       adsurl = {https://ui.adsabs.harvard.edu/abs/2017ApJ...848L..13A}
}

@ARTICLE{2018AJ....156..123A,
       author = {{Astropy Collaboration} and {Price-Whelan}, A.~M. and {Sip{\H{o}}cz}, B.~M. and {G{\"u}nther}, H.~M. and {Lim}, P.~L. and {Crawford}, S.~M. and {Conseil}, S. and {Shupe}, D.~L. and {Craig}, M.~W. and {Dencheva}, N. and et al.},
        title = "{The Astropy Project: Building an Open-science Project and Status of the v2.0 Core Package}",
      journal = {\aj},
         year = 2018,
        month = sep,
       volume = {156},
       number = {3},
          eid = {123},
        pages = {123},
          doi = {10.3847/1538-3881/aabc4f},
archivePrefix = {arXiv},
       eprint = {1801.02634},
 primaryClass = {astro-ph.IM},
       adsurl = {https://ui.adsabs.harvard.edu/abs/2018AJ....156..123A}
}

@ARTICLE{2019A&A...623A..92S,
       author = {{Selsing}, J. and {Malesani}, D. and {Goldoni}, P. and {Fynbo}, J.~P.~U. and {Kr{\"u}hler}, T. and {Antonelli}, L.~A. and {Arabsalmani}, M. and {Bolmer}, J. and {Cano}, Z. and {Christensen}, L. and {Covino}, S. and {D'Avanzo}, P. and {D'Elia}, V. and {De Cia}, A. and {de Ugarte Postigo}, A. and {Flores}, H. and {Friis}, M. and {Gomboc}, A. and {Greiner}, J. and {Groot}, P. and {Hammer}, F. and {Hartoog}, O.~E. and {Heintz}, K.~E. and {Hjorth}, J. and {Jakobsson}, P. and {Japelj}, J. and {Kann}, D.~A. and {Kaper}, L. and {Ledoux}, C. and {Leloudas}, G. and {Levan}, A.~J. and {Maiorano}, E. and {Melandri}, A. and {Milvang-Jensen}, B. and {Palazzi}, E. and {Palmerio}, J.~T. and {Perley}, D.~A. and {Pian}, E. and {Piranomonte}, S. and {Pugliese}, G. and {S{\'a}nchez-Ram{\'\i}rez}, R. and {Savaglio}, S. and {Schady}, P. and {Schulze}, S. and {Sollerman}, J. and {Sparre}, M. and {Tagliaferri}, G. and {Tanvir}, N.~R. and {Th{\"o}ne}, C.~C. and {Vergani}, S.~D. and {Vreeswijk}, P. and {Watson}, D. and {Wiersema}, K. and {Wijers}, R. and {Xu}, D. and {Zafar}, T.},
        title = "{The X-shooter GRB afterglow legacy sample (XS-GRB)}",
      journal = {\aap},
         year = 2019,
        month = mar,
       volume = {623},
          eid = {A92},
        pages = {A92},
          doi = {10.1051/0004-6361/201832835},
archivePrefix = {arXiv},
       eprint = {1802.07727},
 primaryClass = {astro-ph.HE},
       adsurl = {https://ui.adsabs.harvard.edu/abs/2019A&A...623A..92S}
}

@ARTICLE{2019MNRAS.483.5380T,
       author = {{Tanvir}, N.~R. and {Fynbo}, J.~P.~U. and {de Ugarte Postigo}, A. and {Japelj}, J. and {Wiersema}, K. and {Malesani}, D. and {Perley}, D.~A. and {Levan}, A.~J. and {Selsing}, J. and {Cenko}, S.~B. and {Kann}, D.~A. and {Milvang-Jensen}, B. and {Berger}, E. and {Cano}, Z. and {Chornock}, R. and {Covino}, S. and {Cucchiara}, A. and {D'Elia}, V. and {Gargiulo}, A. and {Goldoni}, P. and {Gomboc}, A. and {Heintz}, K.~E. and {Hjorth}, J. and {Izzo}, L. and {Jakobsson}, P. and {Kaper}, L. and {Kr{\"u}hler}, T. and {Laskar}, T. and {Myers}, M. and {Piranomonte}, S. and {Pugliese}, G. and {Rossi}, A. and {S{\'a}nchez-Ram{\'\i}rez}, R. and {Schulze}, S. and {Sparre}, M. and {Stanway}, E.~R. and {Tagliaferri}, G. and {Th{\"o}ne}, C.~C. and {Vergani}, S. and {Vreeswijk}, P.~M. and {Wijers}, R.~A.~M.~J. and {Watson}, D. and {Xu}, D.},
        title = "{The fraction of ionizing radiation from massive stars that escapes to the intergalactic medium}",
      journal = {\mnras},
         year = 2019,
        month = mar,
       volume = {483},
       number = {4},
        pages = {5380-5408},
          doi = {10.1093/mnras/sty3460},
archivePrefix = {arXiv},
       eprint = {1805.07318},
 primaryClass = {astro-ph.GA},
       adsurl = {https://ui.adsabs.harvard.edu/abs/2019MNRAS.483.5380T}
}

@ARTICLE{2019ApJ...878...52A,
       author = {{Ajello}, M. and {Arimoto}, M. and {Axelsson}, M. and {Baldini}, L. and {Barbiellini}, G. and {Bastieri}, D. and {Bellazzini}, R. and {Bhat}, P.~N. and {Bissaldi}, E. and {Blandford}, R.~D. and {Bonino}, R. and {Bonnell}, J. and {Bottacini}, E. and {Bregeon}, J. and {Bruel}, P. and {Buehler}, R. and {Cameron}, R.~A. and {Caputo}, R. and {Caraveo}, P.~A. and {Cavazzuti}, E. and {Chen}, S. and {Cheung}, C.~C. and {Chiaro}, G. and {Ciprini}, S. and {Costantin}, D. and {Crnogorcevic}, M. and {Cutini}, S. and {Dainotti}, M. and {D'Ammando}, F. and {de la Torre Luque}, P. and {de Palma}, F. and {Desai}, A. and {Desiante}, R. and {Di Lalla}, N. and {Di Venere}, L. and {Fana Dirirsa}, F. and {Fegan}, S.~J. and {Franckowiak}, A. and {Fukazawa}, Y. and {Funk}, S. and {Fusco}, P. and {Gargano}, F. and {Gasparrini}, D. and {Giglietto}, N. and {Giordano}, F. and {Giroletti}, M. and {Green}, D. and {Grenier}, I.~A. and {Grove}, J.~E. and {Guiriec}, S. and {Hays}, E. and {Hewitt}, J.~W. and {Horan}, D. and {J{\'o}hannesson}, G. and {Kocevski}, D. and {Kuss}, M. and {Latronico}, L. and {Li}, J. and {Longo}, F. and {Loparco}, F. and {Lovellette}, M.~N. and {Lubrano}, P. and {Maldera}, S. and {Manfreda}, A. and {Mart{\'\i}-Devesa}, G. and {Mazziotta}, M.~N. and {Mereu}, I. and {Meyer}, M. and {Michelson}, P.~F. and {Mirabal}, N. and {Mitthumsiri}, W. and {Mizuno}, T. and {Monzani}, M.~E. and {Moretti}, E. and {Morselli}, A. and {Moskalenko}, I.~V. and {Negro}, M. and {Nuss}, E. and {Ohno}, M. and {Omodei}, N. and {Orienti}, M. and {Orlando}, E. and {Palatiello}, M. and {Paliya}, V.~S. and {Paneque}, D. and {Persic}, M. and {Pesce-Rollins}, M. and {Petrosian}, V. and {Piron}, F. and {Poolakkil}, S. and {Poon}, H. and {Porter}, T.~A. and {Principe}, G. and {Racusin}, J.~L. and {Rain{\`o}}, S. and {Rando}, R. and {Razzano}, M. and {Razzaque}, S. and {Reimer}, A. and {Reimer}, O. and {Reposeur}, T. and {Ryde}, F. and {Serini}, D. and {Sgr{\`o}}, C. and {Siskind}, E.~J. and {Sonbas}, E. and {Spandre}, G. and {Spinelli}, P. and {Suson}, D.~J. and {Tajima}, H. and {Takahashi}, M. and {Tak}, D. and {Thayer}, J.~B. and {Torres}, D.~F. and {Troja}, E. and {Valverde}, J. and {Veres}, P. and {Vianello}, G. and {von Kienlin}, A. and {Wood}, K. and {Yassine}, M. and {Zhu}, S. and {Zimmer}, S.},
        title = "{A Decade of Gamma-Ray Bursts Observed by Fermi-LAT: The Second GRB Catalog}",
      journal = {\apj},
         year = 2019,
        month = jun,
       volume = {878},
       number = {1},
          eid = {52},
        pages = {52},
          doi = {10.3847/1538-4357/ab1d4e},
archivePrefix = {arXiv},
       eprint = {1906.11403},
 primaryClass = {astro-ph.HE},
       adsurl = {https://ui.adsabs.harvard.edu/abs/2019ApJ...878...52A}
}

@ARTICLE{2020ExA....50...91D,
       author = {{Dagoneau}, Nicolas and {Schanne}, St{\'e}phane and {Atteia}, Jean-Luc and {G{\"o}tz}, Diego and {Cordier}, Bertrand},
        title = "{Ultra-Long Gamma-Ray Bursts detection with SVOM/ECLAIRs}",
      journal = {Experimental Astronomy},
         year = 2020,
        month = jul,
       volume = {50},
       number = {1},
        pages = {91-123},
          doi = {10.1007/s10686-020-09665-w},
archivePrefix = {arXiv},
       eprint = {2005.12560},
 primaryClass = {astro-ph.HE},
       adsurl = {https://ui.adsabs.harvard.edu/abs/2020ExA....50...91D}
}

@ARTICLE{2020MNRAS.496.2910P,
       author = {{Petropoulou}, M. and {Beniamini}, P. and {Vasilopoulos}, G. and {Giannios}, D. and {Barniol Duran}, R.},
        title = "{Deciphering the properties of the central engine in GRB collapsars}",
      journal = {\mnras},
         year = 2020,
        month = aug,
       volume = {496},
       number = {3},
        pages = {2910-2921},
          doi = {10.1093/mnras/staa1695},
archivePrefix = {arXiv},
       eprint = {2006.07482},
 primaryClass = {astro-ph.HE},
       adsurl = {https://ui.adsabs.harvard.edu/abs/2020MNRAS.496.2910P}
}

@ARTICLE{2020A&A...641A...6P,
       author = {{Planck Collaboration} and {Aghanim}, N. and {Akrami}, Y. and {Ashdown}, M. and {Aumont}, J. and {Baccigalupi}, C. and {Ballardini}, M. and {Banday}, A.~J. and {Barreiro}, R.~B. and {Bartolo}, N. and {Basak}, S. and {Battye}, R. and {Benabed}, K. and {Bernard}, J. -P. and {Bersanelli}, M. and {Bielewicz}, P. and {Bock}, J.~J. and {Bond}, J.~R. and {Borrill}, J. and {Bouchet}, F.~R. and {Boulanger}, F. and {Bucher}, M. and {Burigana}, C. and {Butler}, R.~C. and {Calabrese}, E. and {Cardoso}, J. -F. and {Carron}, J. and {Challinor}, A. and {Chiang}, H.~C. and {Chluba}, J. and {Colombo}, L.~P.~L. and {Combet}, C. and {Contreras}, D. and {Crill}, B.~P. and {Cuttaia}, F. and {de Bernardis}, P. and {de Zotti}, G. and {Delabrouille}, J. and {Delouis}, J. -M. and {Di Valentino}, E. and {Diego}, J.~M. and {Dor{\'e}}, O. and {Douspis}, M. and {Ducout}, A. and {Dupac}, X. and {Dusini}, S. and {Efstathiou}, G. and {Elsner}, F. and {En{\ss}lin}, T.~A. and {Eriksen}, H.~K. and {Fantaye}, Y. and {Farhang}, M. and {Fergusson}, J. and {Fernandez-Cobos}, R. and {Finelli}, F. and {Forastieri}, F. and {Frailis}, M. and {Fraisse}, A.~A. and {Franceschi}, E. and {Frolov}, A. and {Galeotta}, S. and {Galli}, S. and {Ganga}, K. and {G{\'e}nova-Santos}, R.~T. and {Gerbino}, M. and {Ghosh}, T. and {Gonz{\'a}lez-Nuevo}, J. and {G{\'o}rski}, K.~M. and {Gratton}, S. and {Gruppuso}, A. and {Gudmundsson}, J.~E. and {Hamann}, J. and {Handley}, W. and {Hansen}, F.~K. and {Herranz}, D. and {Hildebrandt}, S.~R. and {Hivon}, E. and {Huang}, Z. and {Jaffe}, A.~H. and {Jones}, W.~C. and {Karakci}, A. and {Keih{\"a}nen}, E. and {Keskitalo}, R. and {Kiiveri}, K. and {Kim}, J. and {Kisner}, T.~S. and {Knox}, L. and {Krachmalnicoff}, N. and {Kunz}, M. and {Kurki-Suonio}, H. and {Lagache}, G. and {Lamarre}, J. -M. and {Lasenby}, A. and {Lattanzi}, M. and {Lawrence}, C.~R. and {Le Jeune}, M. and {Lemos}, P. and {Lesgourgues}, J. and {Levrier}, F. and {Lewis}, A. and {Liguori}, M. and {Lilje}, P.~B. and {Lilley}, M. and {Lindholm}, V. and {L{\'o}pez-Caniego}, M. and {Lubin}, P.~M. and {Ma}, Y. -Z. and {Mac{\'\i}as-P{\'e}rez}, J.~F. and {Maggio}, G. and {Maino}, D. and {Mandolesi}, N. and {Mangilli}, A. and {Marcos-Caballero}, A. and {Maris}, M. and {Martin}, P.~G. and {Martinelli}, M. and {Mart{\'\i}nez-Gonz{\'a}lez}, E. and {Matarrese}, S. and {Mauri}, N. and {McEwen}, J.~D. and {Meinhold}, P.~R. and {Melchiorri}, A. and {Mennella}, A. and {Migliaccio}, M. and {Millea}, M. and {Mitra}, S. and {Miville-Desch{\^e}nes}, M. -A. and {Molinari}, D. and {Montier}, L. and {Morgante}, G. and {Moss}, A. and {Natoli}, P. and {N{\o}rgaard-Nielsen}, H.~U. and {Pagano}, L. and {Paoletti}, D. and {Partridge}, B. and {Patanchon}, G. and {Peiris}, H.~V. and {Perrotta}, F. and {Pettorino}, V. and {Piacentini}, F. and {Polastri}, L. and {Polenta}, G. and {Puget}, J. -L. and {Rachen}, J.~P. and {Reinecke}, M. and {Remazeilles}, M. and {Renzi}, A. and {Rocha}, G. and {Rosset}, C. and {Roudier}, G. and {Rubi{\~n}o-Mart{\'\i}n}, J.~A. and {Ruiz-Granados}, B. and {Salvati}, L. and {Sandri}, M. and {Savelainen}, M. and {Scott}, D. and {Shellard}, E.~P.~S. and {Sirignano}, C. and {Sirri}, G. and {Spencer}, L.~D. and {Sunyaev}, R. and {Suur-Uski}, A. -S. and {Tauber}, J.~A. and {Tavagnacco}, D. and {Tenti}, M. and {Toffolatti}, L. and {Tomasi}, M. and {Trombetti}, T. and {Valenziano}, L. and {Valiviita}, J. and {Van Tent}, B. and {Vibert}, L. and {Vielva}, P. and {Villa}, F. and {Vittorio}, N. and {Wandelt}, B.~D. and {Wehus}, I.~K. and {White}, M. and {White}, S.~D.~M. and {Zacchei}, A. and {Zonca}, A.},
        title = "{Planck 2018 results. VI. Cosmological parameters}",
      journal = {\aap},
         year = 2020,
        month = sep,
       volume = {641},
          eid = {A6},
        pages = {A6},
          doi = {10.1051/0004-6361/201833910},
archivePrefix = {arXiv},
       eprint = {1807.06209},
 primaryClass = {astro-ph.CO},
       adsurl = {https://ui.adsabs.harvard.edu/abs/2020A&A...641A...6P}
}

@Article{         harris2020array,
 title         = {Array programming with {NumPy}},
 author        = {Charles R. Harris and K. Jarrod Millman and St{\'{e}}fan J.
                 van der Walt and Ralf Gommers and Pauli Virtanen and David
                 Cournapeau and Eric Wieser and Julian Taylor and Sebastian
                 Berg and Nathaniel J. Smith and Robert Kern and Matti Picus
                 and Stephan Hoyer and Marten H. van Kerkwijk and Matthew
                 Brett and Allan Haldane and Jaime Fern{\'{a}}ndez del
                 R{\'{i}}o and Mark Wiebe and Pearu Peterson and Pierre
                 G{\'{e}}rard-Marchant and Kevin Sheppard and Tyler Reddy and
                 Warren Weckesser and Hameer Abbasi and Christoph Gohlke and
                 Travis E. Oliphant},
 year          = {2020},
 month         = sep,
 journal       = {Nature},
 volume        = {585},
 number        = {7825},
 pages         = {357--362},
 doi           = {10.1038/s41586-020-2649-2},
 publisher     = {Springer Science and Business Media {LLC}},
 url           = {https://doi.org/10.1038/s41586-020-2649-2}
}

@ARTICLE{2020SciPy-NMeth,
  author  = {Virtanen, Pauli and Gommers, Ralf and Oliphant, Travis E. and
            Haberland, Matt and Reddy, Tyler and Cournapeau, David and
            Burovski, Evgeni and Peterson, Pearu and Weckesser, Warren and
            Bright, Jonathan and {van der Walt}, St{\'e}fan J. and
            Brett, Matthew and Wilson, Joshua and Millman, K. Jarrod and
            Mayorov, Nikolay and Nelson, Andrew R. J. and Jones, Eric and
            Kern, Robert and Larson, Eric and Carey, C J and
            Polat, {\.I}lhan and Feng, Yu and Moore, Eric W. and
            {VanderPlas}, Jake and Laxalde, Denis and Perktold, Josef and
            Cimrman, Robert and Henriksen, Ian and Quintero, E. A. and
            Harris, Charles R. and Archibald, Anne M. and
            Ribeiro, Ant{\^o}nio H. and Pedregosa, Fabian and
            {van Mulbregt}, Paul and {SciPy 1.0 Contributors}},
  title   = {{{SciPy} 1.0: Fundamental Algorithms for Scientific
            Computing in Python}},
  journal = {Nature Methods},
  year    = {2020},
  volume  = {17},
  pages   = {261--272},
  adsurl  = {https://rdcu.be/b08Wh},
  doi     = {10.1038/s41592-019-0686-2},
}

@INCOLLECTION{2022hxga.book...86Y,
       author = {{Yuan}, Weimin and {Zhang}, Chen and {Chen}, Yong and {Ling}, Zhixing},
        title = "{The Einstein Probe Mission}",
    booktitle = {Handbook of X-ray and Gamma-ray Astrophysics},
         year = 2022,
       editor = {{Bambi}, Cosimo and {Sangangelo}, Andrea},
          eid = {86},
        pages = {86},
          doi = {10.1007/978-981-16-4544-0_151-1},
       adsurl = {https://ui.adsabs.harvard.edu/abs/2022hxga.book...86Y}
}

@ARTICLE{2022ApJ...927..157M,
       author = {{Moss}, Michael and {Lien}, Amy and {Guiriec}, Sylvain and {Cenko}, S. Bradley and {Sakamoto}, Takanori},
        title = "{Instrumental Tip-of-the-iceberg Effects on the Prompt Emission of Swift/BAT Gamma-ray Bursts}",
      journal = {\apj},
         year = 2022,
        month = mar,
       volume = {927},
       number = {2},
          eid = {157},
        pages = {157},
          doi = {10.3847/1538-4357/ac4d94},
archivePrefix = {arXiv},
       eprint = {2111.13392},
 primaryClass = {astro-ph.HE},
       adsurl = {https://ui.adsabs.harvard.edu/abs/2022ApJ...927..157M}
}

\appendix
\section{Excluding Late-Mission GRBs} \label{sec:excluding_late_mission_grbs}

\begin{figure*}[ht!]
	\centering
	\includegraphics[width=0.3\linewidth]{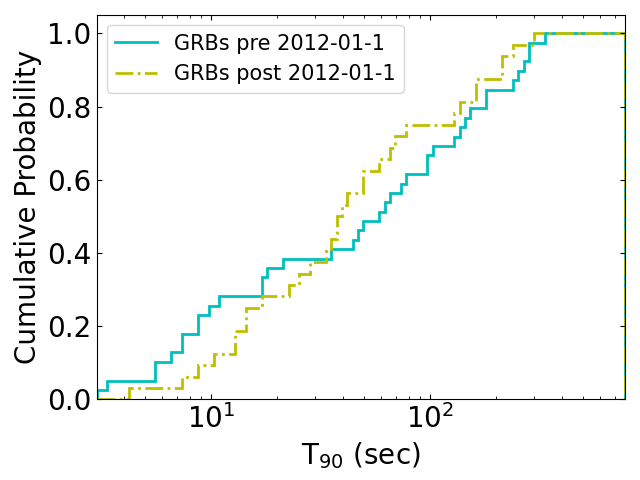}
	\includegraphics[width=0.3\linewidth]{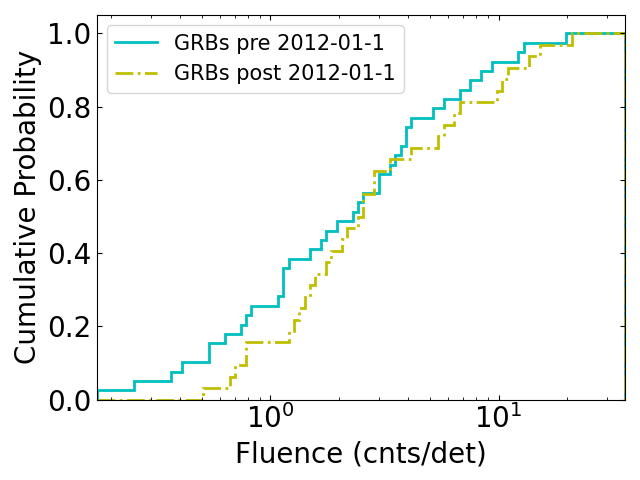}
	\includegraphics[width=0.3\linewidth]{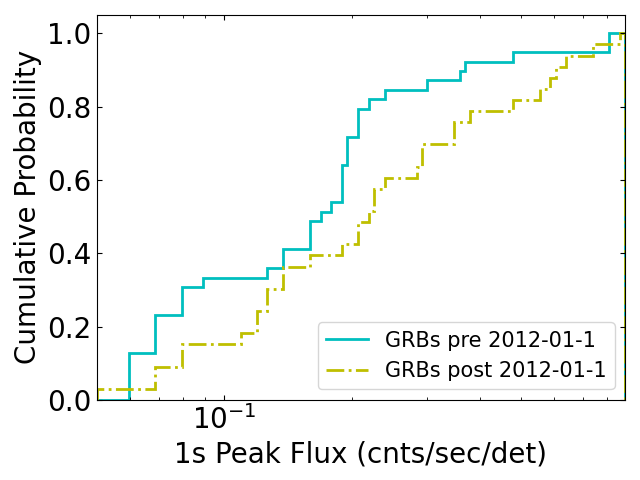}
	\caption{Displayed are the cumulative distributions of the measured durations (\textit{Top}), fluences (\textit{Center}), and peak fluxes (\textit{Bottom}) for Swift/BAT GRBs observed before 2012-01-01 (cyan) and those observed after (olive). The $p$-values obtained from KS-tests for these distributions can be found in Table }
	\label{fig: date comparisons}
\end{figure*}

In Figure \ref{fig: date comparisons} measurements of the pre-2012 and post-2012 samples are compared to one another and the respective KS-Test $p$-values calculated between the distributions are given in Table \ref{tab:p_vals_2012}. We can see that duration measurements of the pre-2012 and post-2012 samples are consistent with one another to within 1$\sigma$ (i.e., $p=0.61$), however, the fluence measurements of the two samples are only consistent within 2$\sigma$ (i.e., $p=0.24$) and the peak flux measurements are only just barely consistent at 2$\sigma$ (i.e., $p=0.05$). Naively, since these are all high-$z$ GRBs, these measurements are drawn from the same underlying distribution and a KS-Test should find $p\geq 0.32$ to indicate this fact. However, there is not a $>3\sigma$ difference between any of sample measurements so we cannot claim that the samples are being drawn from different distributions either. 

\begin{deluxetable}{lll}
	\tablewidth{0pt}
	\tablecaption{$p$-values obtained from KS-tests performed between Swift/BAT GRBs observed before and after 2012-01-01 \label{tab:p_vals_2012}}
	\tablehead
	{
		\colhead{Metric} & \colhead{$p$-value} & \colhead{Significance}
	}
	\startdata
		Durations & 0.61 & $< 1 \sigma$\\ 
		Fluences & 0.24 & $< 2 \sigma$\\ 
		Peak Flux & 0.05 & $= 2 \sigma$\\  
	\enddata
\end{deluxetable}

\begin{figure}[ht!]
	\centering
	\includegraphics[width=0.4\linewidth]{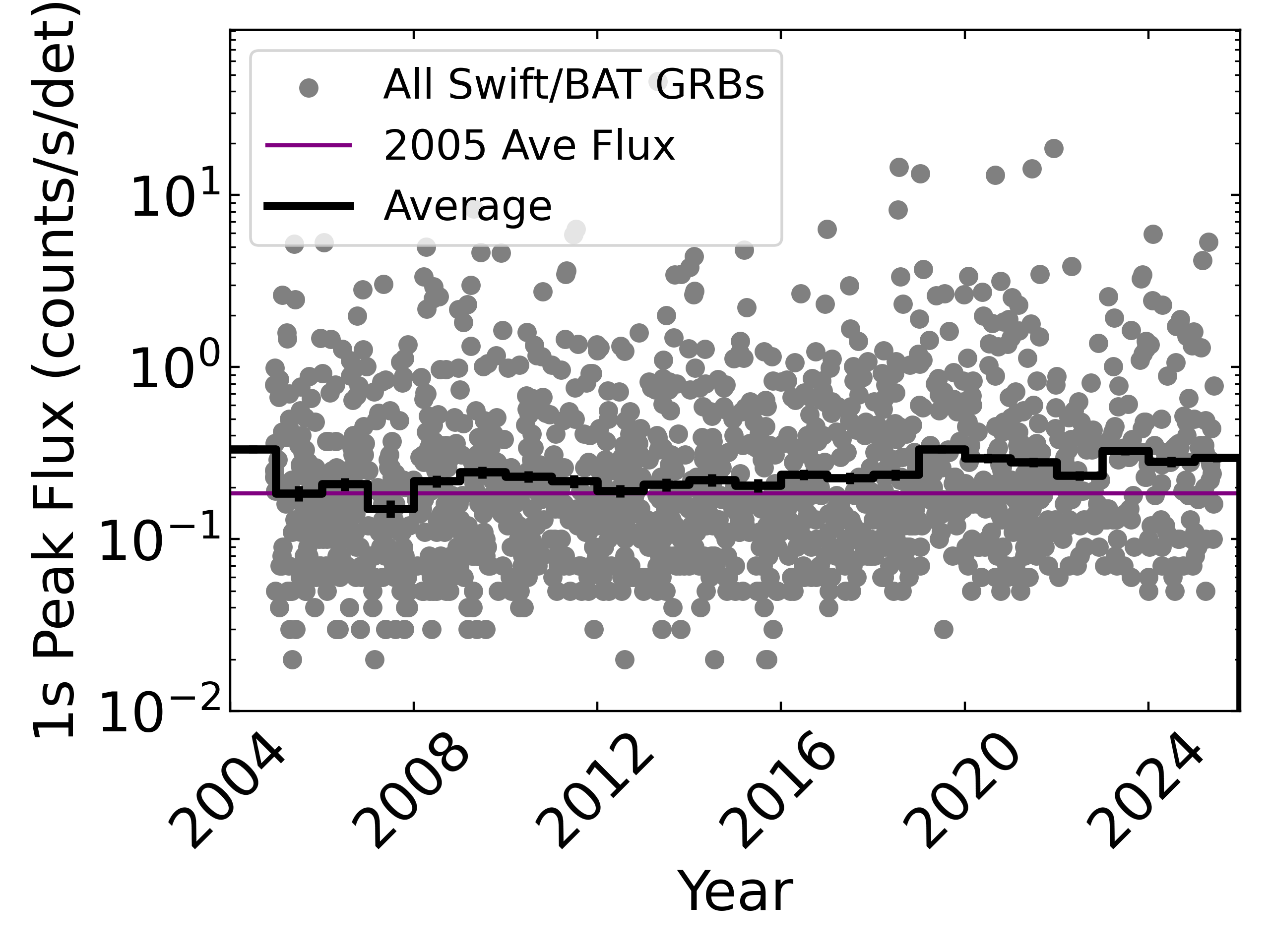}
	\includegraphics[width=0.4\linewidth]{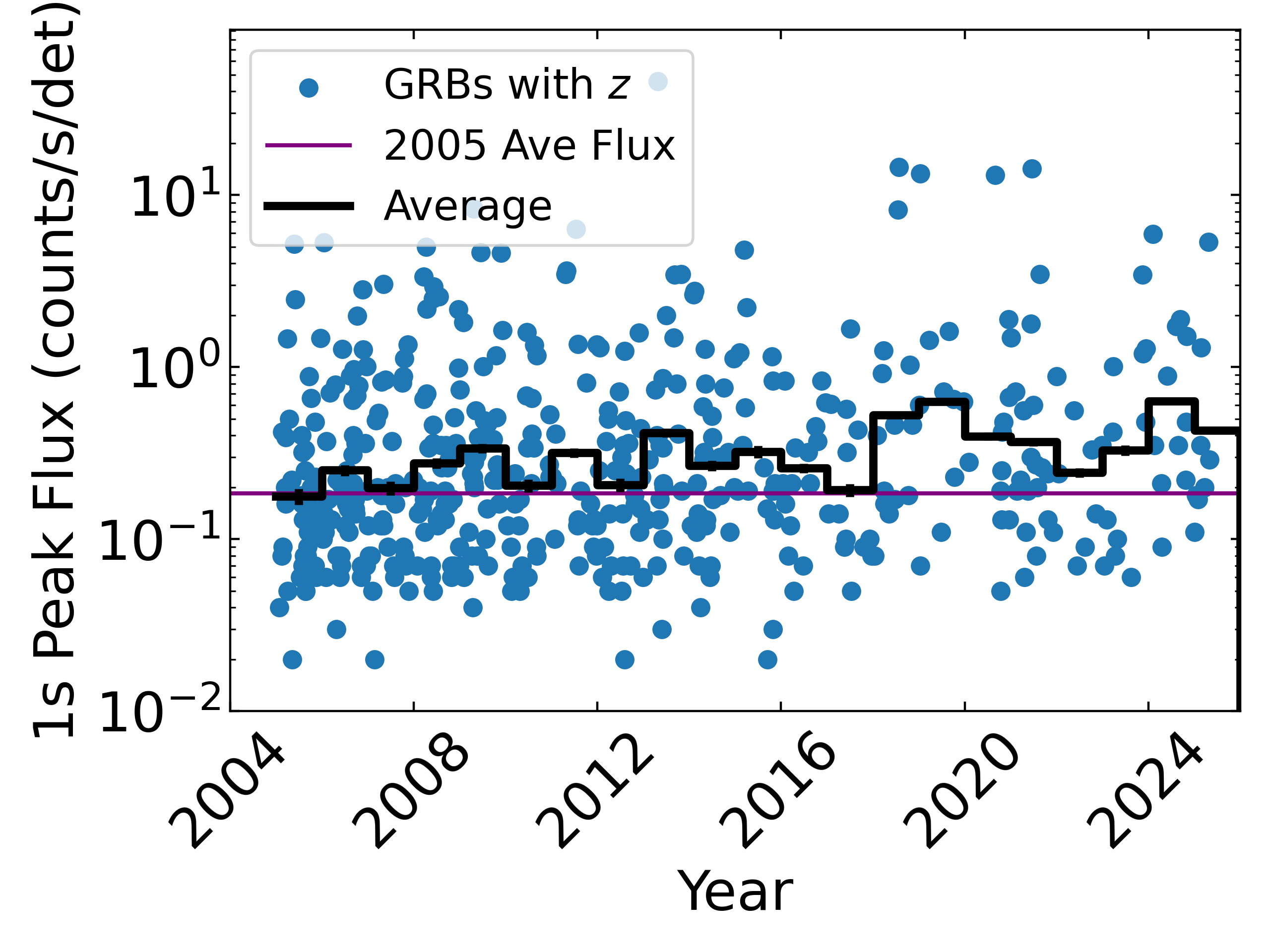}
	\includegraphics[width=0.4\linewidth]{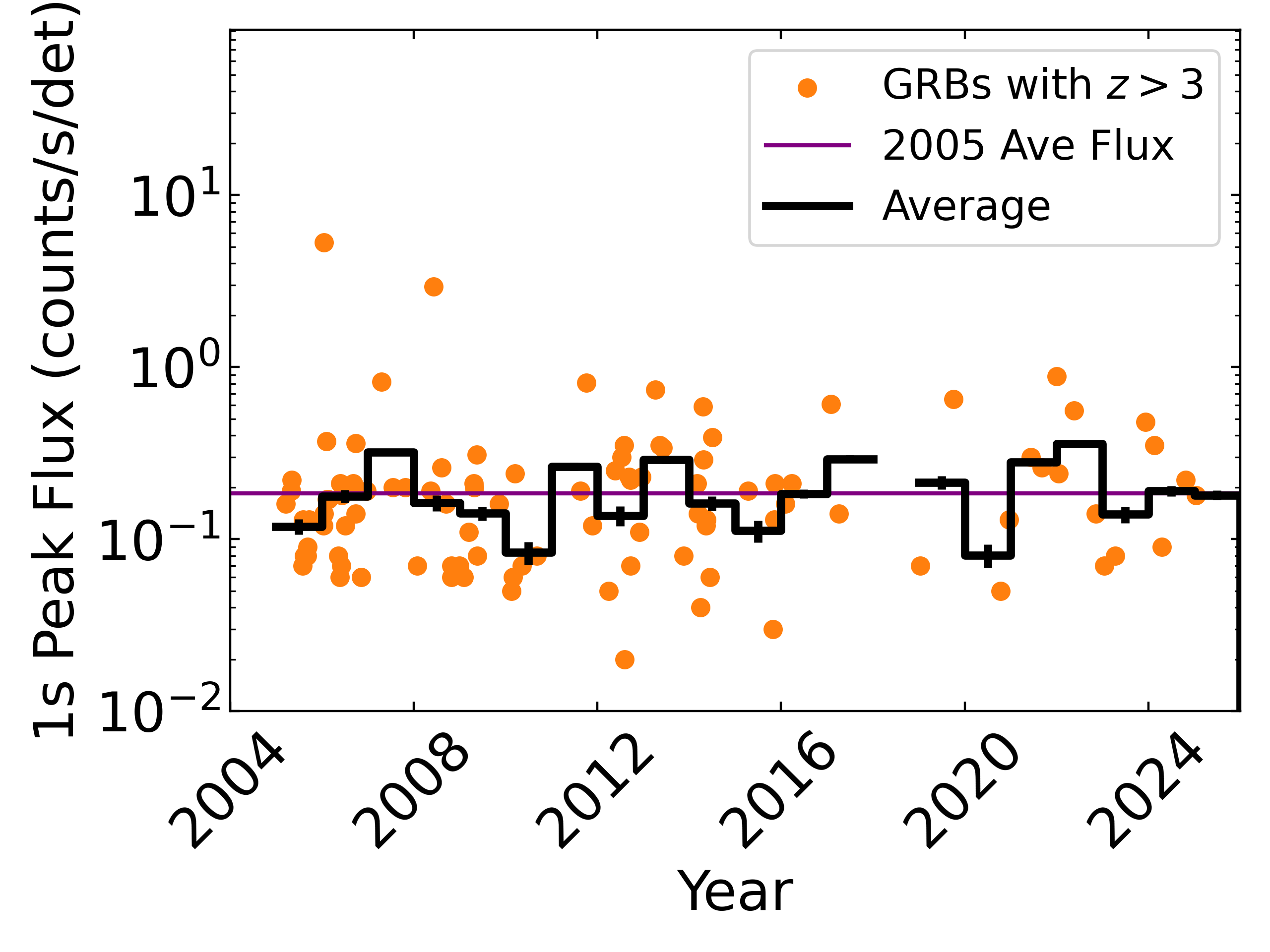}
	\includegraphics[width=0.4\linewidth]{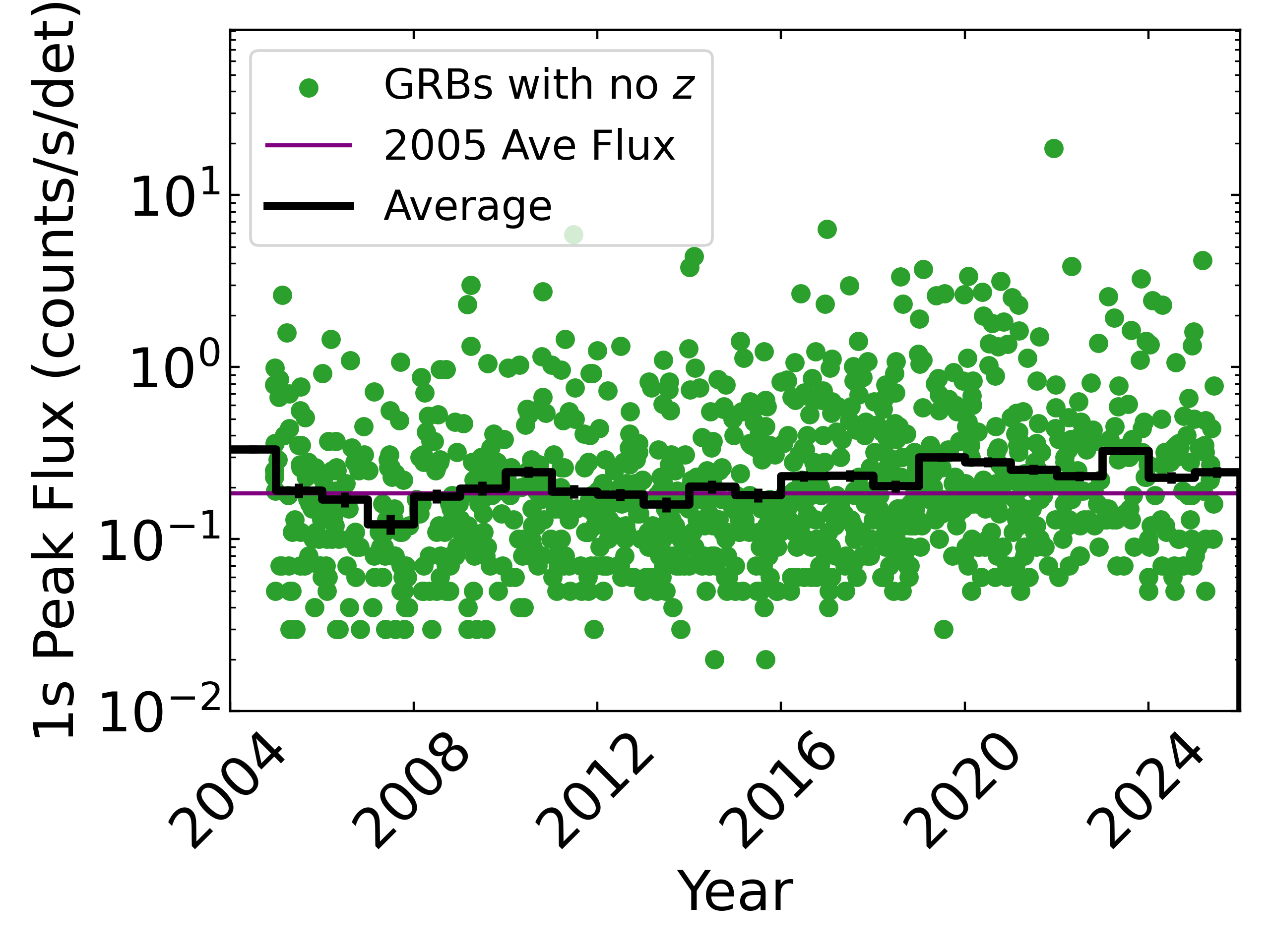}
	\caption{Displayed are the 1-second peak fluxes for \textit{a}: all Swift/BAT GRBs, \textit{b}: only GRBs with redshifts, \textit{c}: only GRBs with redshifts $z>3$, and \textit{d}: only GRBs without redshifts as a function of their observation date. The average peak flux in 6-month bins is shown with the solid lines in each plot. The thin purple line indicates the 2005 average peak flux.}
	\label{fig: peak fluxes vs date}
\end{figure}

We separated our observed high-$z$ GRB sample into GRBs observed before and after 2012-01-01, however, that date was simply an initial guess. We were not sure what date would be appropriate to make a cut at or even if an evolving bias was truly present at all. To investigate if there was any evolving bias in Swift/BAT GRBs and what date it may have become significant, we first looked into the average peak flux of Swift/BAT GRBs over the course of the mission's life (see Fig. \ref{fig: peak fluxes vs date} \textit{a}). There seems to be a slight increase in the average peak flux for the entire Swift/BAT GRB sample, however this increase is less than a factor of 2. Looking only at Swift/BAT GRBs that have measured redshifts, it seems this peak-flux increase may be more pronounced (see Fig. \ref{fig: peak fluxes vs date} \textit{b}). For GRBs with redshifts $z>3$, this bias does not seem to be present, but the sample size is quite low, especially so after 2016 (see Fig. \ref{fig: peak fluxes vs date} \textit{c}). But even for  GRBs without any redshifts, there seems to be a slight increase in the peak flux (see Fig. \ref{fig: peak fluxes vs date} \textit{d}). From this comparison, it looks as if $z>3$ GRBs actually stick close to the same average peak flux across the mission life time and that there is a possible increase in the average peak flux for $z<3$ GRBs. 

\begin{figure}[ht!]
	\centering
	\includegraphics[width=0.45\linewidth]{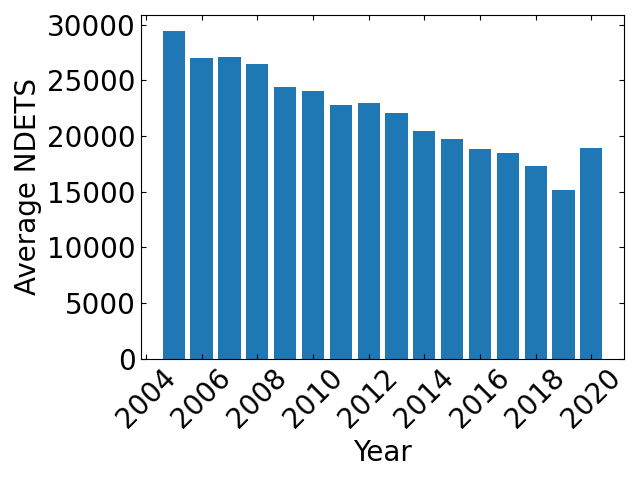}
	\caption{Average number of enabled/active detectors on the Swift/BAT detector plane. The total number of detectors at launch was 32,768. In early 2020, BAT recovered a significant number of detectors upon a spontaneous reboot.}
	\label{fig: ave ndets}
\end{figure}

This change in average peak flux could be due to a number of reasons. For instance, it may be due to the decrease in the number of active detectors on the BAT detector plane (see Fig. \ref{fig: ave ndets}). The sensitivity of BAT goes as $\propto\sqrt{N}$, where $N$ is the number of detectors active on the detector plane \citep{2008ApOpt..47.2739S,2013ApJS..207...19B}. The ratio between the number of active detectors on BAT at the beginning of Swift's lifetime (i.e., $N_0 = 32,768$) and the average number of active detectors in 2018 (i.e., $N_{18} \sim 15,000$) equates to $\sqrt{N_{18} / N_0 } = 0.676$, indicating that the sensitivity of BAT is expected to have decreased by $\sim 33\% $ between 2004 and 2018.

An alternative cause of the average peak flux change could be due an evolution in Swift's observing strategy; over Swift's lifetime, its average pointing interval has decreased from $\sim$990 sec in 2005 to $\sim$650 sec in 2025 and, consequently, its time spent slewing between pointings has increased. During a slew, Swift does not trigger and becomes less sensitive to dim and short events. 

\begin{figure}[ht!]
	\centering
	\includegraphics[width=0.31\linewidth]{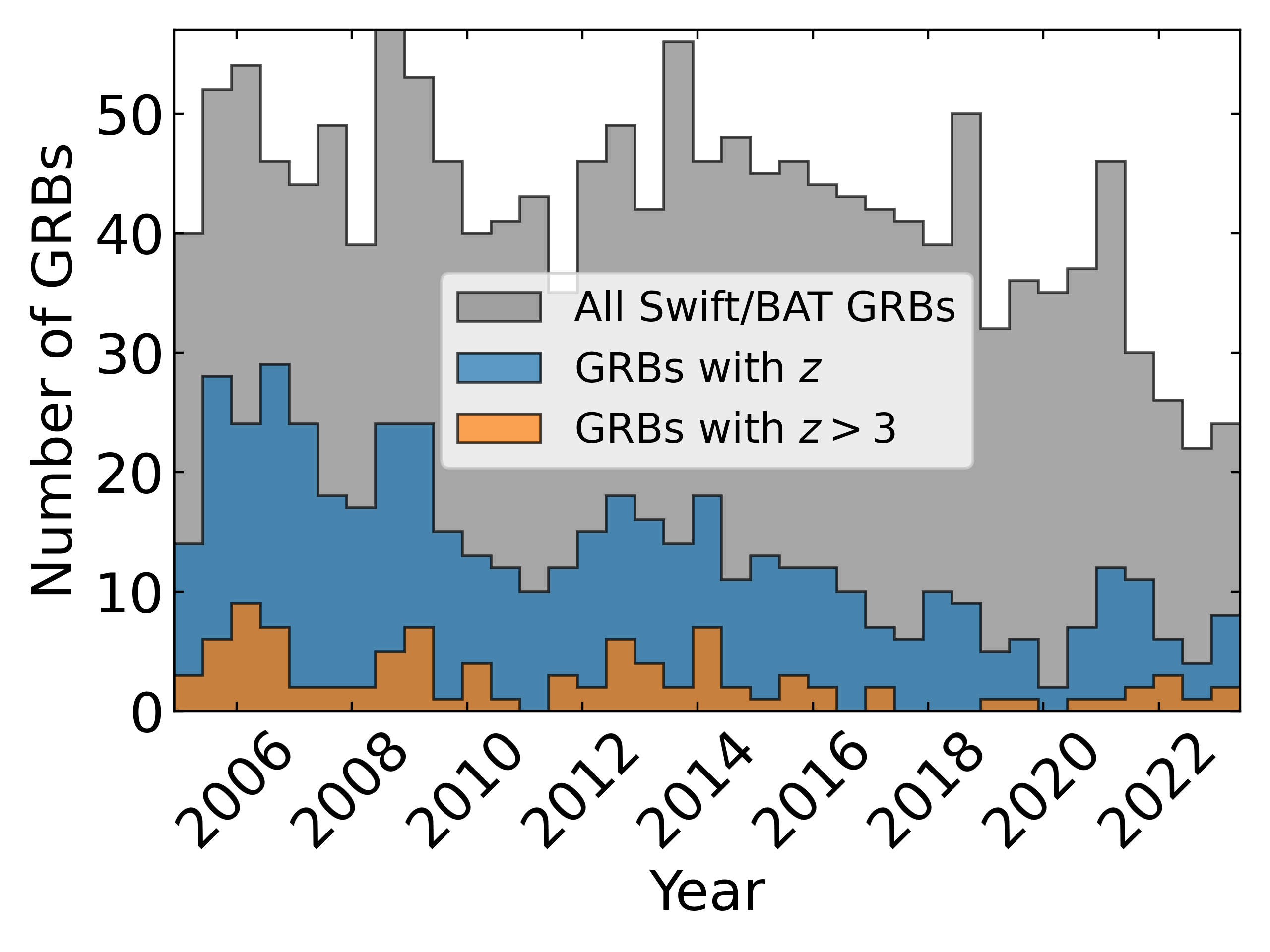}
	\includegraphics[width=0.31\linewidth]{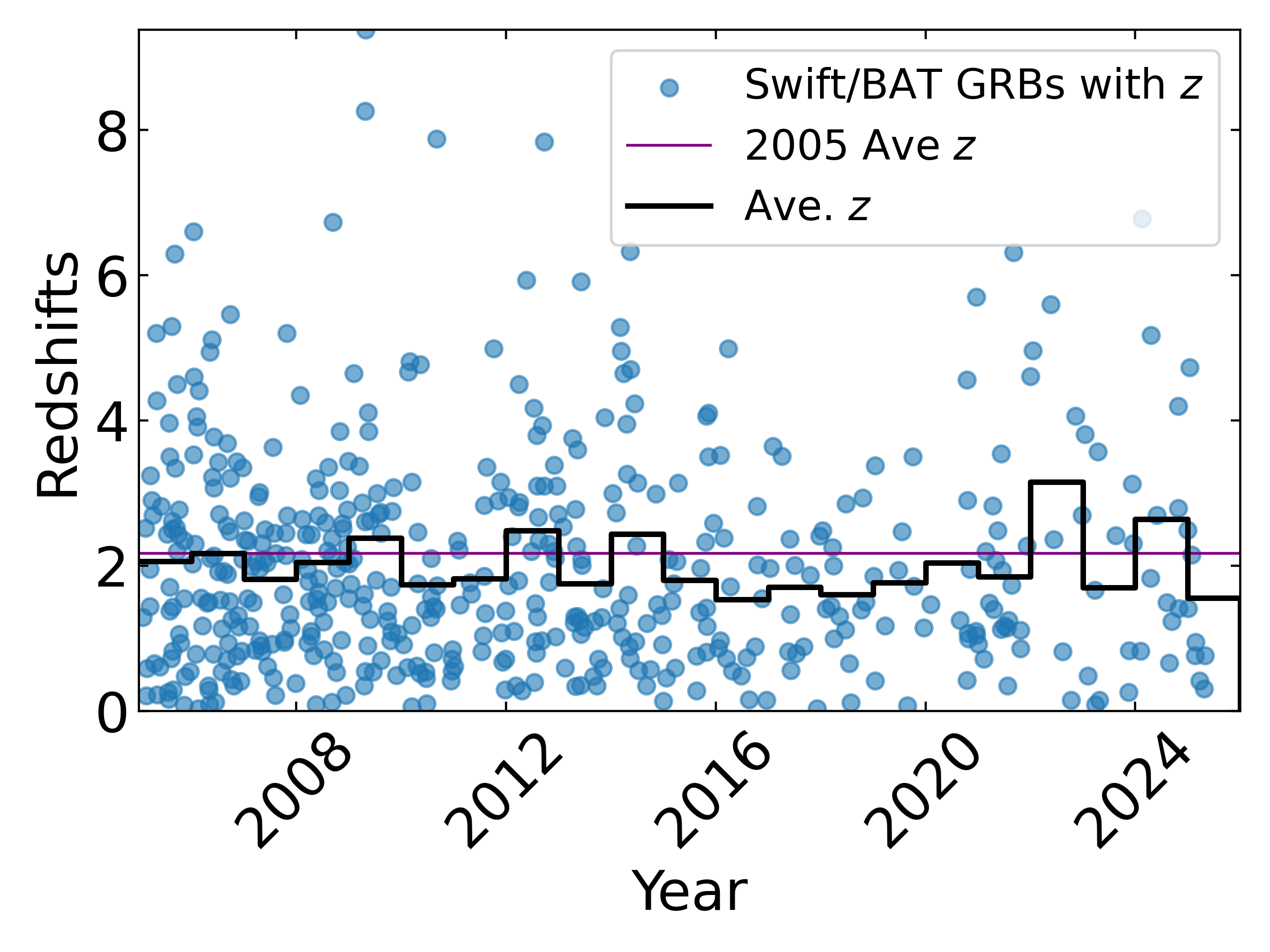}
	\includegraphics[width=0.31\linewidth]{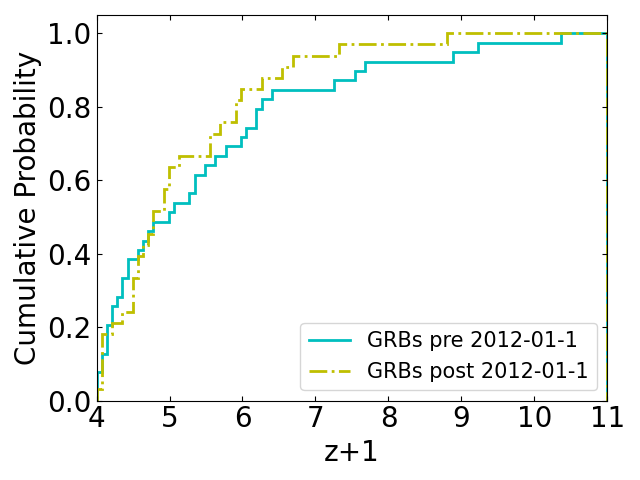}
	\caption{\textit{Left}: The number of redshift measurements obtained for Swift/BAT GRBs as a function of date since mission launch. All Swift/BAT GRBs are shown in gray, Swift/BAT GRBs with redshift in blue, and GRBs with redshifts $z>3$ in orange. \textit{Middle}: The average redshift measured for Swift/BAT GRBs as a function of their observation date. Swift/BAT GRBs are displayed in blue point. The black line indicates the average redshift since Swift's launch. Again, the thin purple line is just an extrapolation of the the 2005 average. \textit{Right}: Cumulative redshift distribution for GRBs observed before 2012-01-01 and those observed after. A KS-test performed between the two distributions obtains a $p$-value $=0.71$, indicating that they are not incompatible with being drawn from the same distribution.}
	\label{fig: z vs date}
\end{figure}

Lastly, over the last two decades, the fraction of GRBs observed by Swift that obtain redshift measurements from ground-based follow-up has decreased, leading to a lower redshift completeness and, possibly, also leading to some population bias (see Fig. \ref{fig: z vs date} \textit{Left}). \citet{2009ApJ...701..824N} found that a positive correlation exists for GRBs between their measured prompt emission fluence and their optical afterglow brightness measured after 11 hours, so it may be the case that ground-based facilities are choosing to perform follow-up observations of only GRBs with high-fluence prompt emission or, perhaps, performing less deep follow-up observations each year, leading to redshift measurements being obtained only for those GRBs with brighter afterglows and, consequently, higher fluence prompt emission. This could lead to either (i) an increase in the peak flux of high-$z$ GRBs -- since there exists a strong correlation that exists between GRB peak fluxes and fluences \citep{2019ApJ...878...52A} -- however, Figure \ref{fig: peak fluxes vs date} doesn't show any average increase for high-$z$ GRBs, or (ii) a decrease in the average redshift for Swift/BAT GRBs over time. In Figure \ref{fig: z vs date} \textit{Middle}, we show the average redshift of Swift/BAT triggered GRBs over time and find a possible small decrease of the average redshift starting in $\sim2016$, however the averages in 2022 and 2024 don't support this trend (although they are statistically limited). A KS-Test between the redshift distributions of the pre- and post- 2012 GRBs obtains $p=0.71$, meaning they are not incompatible with coming from the same underlying distribution (see Fig. \ref{fig: z vs date} \textit{Right}).

Although there are hints to an emerging bias, a conclusive answer on whether there is a bias developing for Swift/BAT GRBs, either due to BAT sensitivity degrading or observational strategies, may be difficult to uncover and is outside the scope of this work.

\newpage
\section{Data Tables} \label{app:data_tables}

In Tables \ref{tab: low-z} and \ref{tab: high-z} we display the samples of observed low-$z$ and high-$z$ GRBs are listed, respectively. These samples are both taken from a redshift complete sample of GRBs observed with the VLT/X-shooter \citep{2019A&A...623A..92S}, but includes a further cut of a 1-s peak flux $f_p \geq 2.6$ counts s$^{-1}$ cm$^{-2}$ in order to more closely follow the BAT6 sample \citep{2012ApJ...749...68S}.

\begin{deluxetable}{llllllc}
	\digitalasset
	\tablewidth{0pt}
	\tablecaption{GRBs in our low-$z$ sample (i.e., $z<1$) \label{tab: low-z}}
	\tablehead{
		\colhead{GRB Name} & \colhead{$z$} & \colhead{T$_{90, true}$} & \colhead{Fluence} & \colhead{$\alpha$} & \colhead{$E_p$} & \colhead{$N_0$} \\ 
		\colhead{} & \colhead{} & \colhead{(sec)} & \colhead{(cnts/det)} & \colhead{} & \colhead{(keV)} & \colhead{(cnts cm$^{-2}$ s$^{-1}$ keV$^{-1}$)}
		}
	\startdata
	050525A & 0.606 & 8.836 & 25.619 & -0.603 & 106.319 & 0.776 \\ 
	\hline
	060904B & 0.7029 & 189.98 & 2.926 & -1.061 & 342.243 & 0.024 \\ 
	\hline
	060912A & 0.937 & 5.028 & 2.253 & -1.196 & 86.386 & 0.112 \\ 
	\hline
	070508 & 0.82 & 20.9 & 29.116 & -0.679 & 213.491 & 0.307 \\ 
	\hline
	071010B & 0.947 & 36.124 & 8.2 & -1.001 & 68.532 & 0.133 \\ 
	\hline
	090424 & 0.544 & 49.46 & 33.035 & -0.984 & 183.397 & 0.826 \\ 
	\hline
	090510 & 0.903 & 5.664 & 0.846 & -0.567 & 264.487 & 0.053 \\ 
	\hline
	090618 & 0.54 & 113.34 & 177.209 & -1.235 & 364.469 & 0.363 \\ 
	\hline
	091018 & 0.971 & 4.368 & 3.092 & -1.298 & 35.062 & 0.191 \\ 
	\hline
	091127 & 0.49044 & 6.956 & 13.548 & -1.33 & 69.588 & 0.575 \\ 
	\hline
	100621A & 0.542 & 63.552 & 38.87 & -0.946 & 90.632 & 0.202 \\ 
	\hline
	100816A & 0.8049 & 2.884 & 2.693 & -0.494 & 153.042 & 0.171 \\ 
	\hline
	110715A & 0.8224 & 13.0 & 18.701 & -0.985 & 151.992 & 0.656 \\ 
	\hline
	111228A & 0.71627 & 101.244 & 16.039 & -1.644 & 93.714 & 0.109 \\ 
	\hline
	120907A & 0.97 & 6.08 & 0.954 & -1.301 & 411.019 & 0.026 \\ 
	\hline
	130427A & 0.3399 & 244.332 & 520.716 & -0.502 & 467.392 & 3.771 \\ 
	\hline
	130831A & 0.4791 & 30.192 & 10.745 & -1.545 & 149.325 & 0.122 \\ 
	\hline
	140506A & 0.889 & 111.104 & 4.09 & -0.53 & 100.377 & 0.217 \\ 
	\hline
	140512A & 0.725 & 154.112 & 20.406 & -1.007 & 224.843 & 0.066 \\ 
	\hline
	161001A & 0.891 & 2.6 & 0.978 & -0.572 & 375.703 & 0.037 \\ 
	\hline
	180720B & 0.654 & 108.4 & 121.571 & -0.664 & 214.739 & 0.853 \\ 
	\hline
	201024A & 0.999 & 5.0 & 1.5 & -1.832 & 25.887 & 0.034 \\ 
	\hline
	210104A & 0.92 & 32.048 & 14.183 & -1.047 & 164.371 & 0.136 \\ 
	\hline
	231118A & 0.8304 & 151.488 & 6.339 & -0.606 & 145.042 & 0.146 \\ 
	\hline
	240205B & 0.824 & 47.304 & 54.371 & -1.161 & 118.838 & 0.571 \\ 
	\hline
	250424A & 0.31 & 18.632 & 51.953 & -1.336 & 145.293 & 0.481 \\ 
	\enddata
\end{deluxetable}

\begin{deluxetable}{llll}
	\digitalasset
	\tablewidth{0pt}
	\tablecaption{GRBs in our high-$z$ sample (i.e., $z>3$) \label{tab: high-z}}
	\tablehead
	{
		\colhead{GRB Name} & \colhead{$z$} & \colhead{T$_{90, true}$} & \colhead{Fluence} \\
		\colhead{} & \colhead{} & \colhead{(sec)} & \colhead{(cnts/det)}
	}
	\startdata
		050319 & 3.2425 & 151.584 & 2.432 \\ 
		\hline
		050502B & 5.2 & 17.724 & 0.822 \\ 
		\hline
		050505 & 4.2748 & 58.852 & 3.422 \\ 
		\hline
		050814 & 5.3 & 142.852 & 3.06 \\ 
		\hline
		050904 & 6.295 & 181.576 & 7.806 \\ 
		\hline
		050908 & 3.3467 & 18.284 & 0.85 \\ 
		\hline
		050922B & 4.5 & 157.024 & 4.252 \\ 
		\hline
		060116 & 6.6 & 104.832 & 3.579 \\ 
		\hline
		060206 & 4.0559 & 7.552 & 1.559 \\ 
		\hline
		060223A & 4.41 & 11.32 & 1.099 \\ 
		\hline
		060522 & 5.11 & 69.124 & 1.753 \\ 
		\hline
		060526 & 3.2213 & 298.04 & 2.603 \\ 
		\hline
		060605 & 3.773 & 79.84 & 1.228 \\ 
		\hline
		060607A & 3.0749 & 103.032 & 4.069 \\ 
		\hline
		060906 & 3.6856 & 44.588 & 3.863 \\ 
		\hline
		060926 & 3.2086 & 8.824 & 0.56 \\ 
		\hline
		060927 & 5.4636 & 22.416 & 2.002 \\ 
		\hline
		061110B & 3.4344 & 135.248 & 1.845 \\ 
		\hline
		061222B & 3.355 & 37.248 & 4.041 \\ 
		\hline
		070420 & 3.01 & 77.024 & 20.501 \\ 
		\hline
		070721B & 3.6298 & 336.864 & 5.194 \\ 
		\hline
		071025 & 5.2 & 241.296 & 12.647 \\ 
		\hline
		080129 & 4.349 & 50.18 & 1.176 \\ 
		\hline
		080516 & 3.2 & 5.764 & 0.422 \\ 
		\hline
		080913 & 6.733 & 7.456 & 0.766 \\ 
		\hline
		081028A & 3.038 & 284.424 & 6.964 \\ 
		\hline
		081029 & 3.8479 & 275.104 & 3.158 \\ 
		\hline
		081228 & 3.44 & 3.0 & 0.173 \\ 
		\hline
		090205 & 4.6497 & 8.812 & 0.387 \\ 
		\hline
		090423 & 8.26 & 10.304 & 1.19 \\ 
		\hline
		090429B & 9.38 & 5.58 & 0.645 \\ 
		\hline
		090516A & 4.109 & 181.008 & 13.275 \\ 
		\hline
		090715B & 3.0 & 266.4 & 9.744 \\ 
		\hline
		091109A & 3.076 & 48.032 & 2.354 \\ 
		\hline
		100302A & 4.813 & 17.948 & 0.557 \\ 
		\hline
		100316A & 3.155 & 6.752 & 1.181 \\ 
		\hline
		100905A & 7.88 & 3.396 & 0.255 \\ 
		\hline
		110818A & 3.36 & 102.84 & 5.988 \\ 
		\hline
		111008A & 4.99005 & 62.848 & 8.51 \\ 
	\enddata
\end{deluxetable}

\begin{deluxetable}{llll}
	\digitalasset
	\tablewidth{0pt}
	\tablecaption{Continued: GRBs in our high-$z$ sample (i.e., $z>3$) \label{tab: high-z cont}}
	\tablehead
	{
		\colhead{GRB Name} & \colhead{$z$} & \colhead{T$_{90, true}$} & \colhead{Fluence} \\
		\colhead{} & \colhead{} & \colhead{(sec)} & \colhead{(cnts/det)}
	}
	\startdata
		120712A & 4.1745 & 14.808 & 2.685 \\ 
		\hline
		120802A & 3.796 & 50.288 & 2.911 \\ 
		\hline
		120909A & 3.93 & 220.596 & 11.349 \\ 
		\hline
		120922A & 3.1 & 168.224 & 9.912 \\ 
		\hline
		120923A & 7.84 & 26.076 & 0.687 \\ 
		\hline
		121201A & 3.385 & 38.0 & 1.399 \\ 
		\hline
		121217A & 3.1 & 778.092 & 10.497 \\ 
		\hline
		130408A & 3.758 & 4.24 & 2.226 \\ 
		\hline
		130514A & 3.6 & 214.192 & 15.499 \\ 
		\hline
		131117A & 4.042 & 10.88 & 0.528 \\ 
		\hline
		140114A & 3.0 & 139.948 & 6.482 \\ 
		\hline
		140304A & 5.283 & 14.784 & 1.659 \\ 
		\hline
		140311A & 4.954 & 70.48 & 3.372 \\ 
		\hline
		140419A & 3.956 & 80.076 & 21.973 \\ 
		\hline
		140423A & 3.26 & 134.144 & 14.079 \\ 
		\hline
		140515A & 6.33 & 23.416 & 1.253 \\ 
		\hline
		140518A & 4.707 & 60.524 & 2.114 \\ 
		\hline
		140703A & 3.14 & 68.644 & 5.917 \\ 
		\hline
		150413A & 3.139 & 243.596 & 6.824 \\ 
		\hline
		160203A & 3.52 & 17.44 & 1.788 \\ 
		\hline
		160327A & 4.99 & 33.744 & 2.509 \\ 
		\hline
		170202A & 3.645 & 37.76 & 5.6 \\ 
		\hline
		191004B & 3.503 & 300.128 & 4.383 \\ 
		\hline
		201014A & 4.56 & 36.196 & 0.706 \\ 
		\hline
		201221A & 5.7 & 44.316 & 2.699 \\ 
		\hline
		210610A & 3.54 & 13.616 & 1.55 \\ 
		\hline
		220101A & 4.61 & 161.888 & 35.848 \\ 
		\hline
		220117A & 4.961 & 50.56 & 2.974 \\ 
		\hline
		220521A & 5.6 & 13.548 & 1.46 \\ 
		\hline
		221110A & 4.06 & 8.98 & 0.798 \\ 
		\hline
		230116D & 3.81 & 41.0 & 1.332 \\ 
		\hline
		230414B & 3.568 & 29.06 & 0.82 \\ 
		\hline
		231210B & 3.13 & 7.472 & 1.921 \\ 
	\enddata
\end{deluxetable}

\figsetstart
\figsetnum{1}
\figsettitle{Low-$z$ GRB Light Curves and Simulation Results}

\end{document}